\documentclass[sigconf]{acmart}

\usepackage[detect-all=true,mode=text,binary-units=true]{siunitx}
\usepackage{paralist}
\usepackage{multirow}
\usepackage{multicol}
\usepackage{makecell}
\usepackage{colortbl}
\usepackage{tikz}
\usepackage{rotating}
\usepackage{hhline}
\usepackage{nicefrac}
\usepackage{tabu}
\usepackage{enumitem}
\usepackage{makecell}

\usepackage{caption}
\usepackage{subcaption}

\DeclareSIUnit{\million}{\text{M}}

\AtBeginDocument{%
  }

\copyrightyear{2023} 
\acmYear{2023} 
\setcopyright{acmlicensed}\acmConference[ASIA CCS 2023]{ACM ASIA Conference on Computer and Communications Security}{July 10--14, 2023}{Melbourne, VIC, Australia}
\acmBooktitle{ACM ASIA Conference on Computer and Communications Security (ASIA CCS 2023), July 10--14, 2023, Melbourne, VIC, Australia}
\acmPrice{15.00}
\acmDOI{10.1145/3579856.3590329}
\acmISBN{979-8-4007-0098-9/23/07}

\newif\ifdraft

\drafttrue

\ifdraft
\newcommand{\todo}[1]{{\textcolor{red}{\textbf{TODO:} #1}}}

\newcommand{\md}[1]{{\color{blue}{#1}}}
\newcommand{\cs}[1]{{\color{green}{#1}}}
\newcommand{\rd}[1]{{\color{purple}{#1}}}
\newcommand{\jp}[1]{{\color{brown}{#1}}}

\else
\newcommand{\todo}[1]{}

\newcommand{\md}[1]{{}}
\newcommand{\cs}[1]{{}}
\newcommand{\rd}[1]{{}}
\newcommand{\jp}[1]{{}}
\fi

\def\assert#1{\ifthenelse{#1}{}{\errmessage{ASSERT FAIL}}}

\newif\ifmanualheader

\manualheaderfalse

\ifmanualheader
  \makeatletter
  \if@ACM@anonymous
  \else
    \g@addto@macro\@subtitlenotes{\global\let\authors=\cleanauthors}
  \fi
  \makeatother
\fi

\newcommand*\circled[1]{\tikz[baseline=(char.base)]{
		\node[shape=circle,draw,thick,inner sep=1pt] (char) {#1};}}

\newcommand{\afblock}[1]{\textbf{#1:}}
\newcommand{\takeaway}[1]{\textit{\afblock{Takeaway} #1}}

\newcommand{\DefineRemark}[2]{%
  \expandafter\newcommand\csname rmk-#1\endcsname{#2}%
}
\DefineRemark{XX}{0}
\newcommand{\Remark}[1]{\csname rmk-#1\endcsname}

\DefineRemark{distinctlayersizeoverall}{50432883695509}
\DefineRemark{distinctnumlayersoverall}{1388414}
\DefineRemark{distinctnumrepooverall}{357898}
\DefineRemark{distinctsizelayersavgbyteoverall}{36324096}
\DefineRemark{distinctsizelayersavggigabyteoverall}{0}
\DefineRemark{distinctsizelayersavgkilobyteoverall}{36324}
\DefineRemark{distinctsizelayersavgmegabyteoverall}{36}
\DefineRemark{distinctsizelayersavgterabyteoverall}{0}
\DefineRemark{distinctsizelayersmaxbyteoverall}{30842192217}
\DefineRemark{distinctsizelayersmaxgigabyteoverall}{30}
\DefineRemark{distinctsizelayersmaxkilobyteoverall}{30842192}
\DefineRemark{distinctsizelayersmaxmegabyteoverall}{30842}
\DefineRemark{distinctsizelayersmaxterabyteoverall}{0}
\DefineRemark{distinctsizelayersminbyteoverall}{0}
\DefineRemark{distinctsizelayersmingigabyteoverall}{0}
\DefineRemark{distinctsizelayersminkilobyteoverall}{0}
\DefineRemark{distinctsizelayersminmegabyteoverall}{0}
\DefineRemark{distinctsizelayersminterabyteoverall}{0}
\DefineRemark{iiotavailabletagspergroup}{57914}
\DefineRemark{iiotavglastpulleddayspergroup}{1603}
\DefineRemark{iiotavglastpusheddayspergroup}{1603}
\DefineRemark{iiotavglastupdateddayspergroup}{922}
\DefineRemark{iiotavgrepopergroup}{2931}
\DefineRemark{iiotavgtagsperimagepergroup}{11}
\DefineRemark{iiotdistinctlayersizepergroup}{10975485215630}
\DefineRemark{iiotdistinctnumlayersgroup}{221844}
\DefineRemark{iiotdistinctnumlayerspergroup}{326298}
\DefineRemark{iiotdistinctnumrepopergrouponly}{61300}
\DefineRemark{iiotdistinctpctlayersgroup}{16}
\DefineRemark{iiotdistinctpctrepopergrouponly}{17}
\DefineRemark{iiotdistinctsizelayersavgbytepergroup}{33636385}
\DefineRemark{iiotdistinctsizelayersavggigabytepergroup}{0}
\DefineRemark{iiotdistinctsizelayersavgkilobytepergroup}{33636}
\DefineRemark{iiotdistinctsizelayersavgmegabytepergroup}{33}
\DefineRemark{iiotdistinctsizelayersavgterabytepergroup}{0}
\DefineRemark{iiotdistinctsizelayersmaxbytepergroup}{30842192217}
\DefineRemark{iiotdistinctsizelayersmaxgigabytepergroup}{30}
\DefineRemark{iiotdistinctsizelayersmaxkilobytepergroup}{30842192}
\DefineRemark{iiotdistinctsizelayersmaxmegabytepergroup}{30842}
\DefineRemark{iiotdistinctsizelayersmaxterabytepergroup}{0}
\DefineRemark{iiotdistinctsizelayersminbytepergroup}{23}
\DefineRemark{iiotdistinctsizelayersmingigabytepergroup}{0}
\DefineRemark{iiotdistinctsizelayersminkilobytepergroup}{0}
\DefineRemark{iiotdistinctsizelayersminmegabytepergroup}{0}
\DefineRemark{iiotdistinctsizelayersminterabytepergroup}{0}
\DefineRemark{iiotmaxlastpulleddayspergroup}{29892}
\DefineRemark{iiotmaxlastpusheddayspergroup}{29892}
\DefineRemark{iiotmaxlastupdateddayspergroup}{29891}
\DefineRemark{iiotmaxrepopergroup}{10000}
\DefineRemark{iiotmaxtagsperimagepergroup}{29357}
\DefineRemark{iiotminlastpulleddayspergroup}{0}
\DefineRemark{iiotminlastpusheddayspergroup}{0}
\DefineRemark{iiotminlastupdateddayspergroup}{0}
\DefineRemark{iiotminrepopergroup}{0}
\DefineRemark{iiotmintagsperimagepergroup}{0}
\DefineRemark{iiotnumlayerspergroup}{1485864}
\DefineRemark{iiotnumtaglatestpergroup}{44889}
\DefineRemark{iiotnumtagnonepergroup}{18872}
\DefineRemark{iiotnumtagspergroup}{1949976}
\DefineRemark{iiotq25lastpulleddayspergroup}{17}
\DefineRemark{iiotq25lastpusheddayspergroup}{17}
\DefineRemark{iiotq25lastupdateddayspergroup}{390}
\DefineRemark{iiotq25repopergroup}{0}
\DefineRemark{iiotq25tagsperimagepergroup}{1}
\DefineRemark{iiotq50lastpulleddayspergroup}{169}
\DefineRemark{iiotq50lastpusheddayspergroup}{169}
\DefineRemark{iiotq50lastupdateddayspergroup}{847}
\DefineRemark{iiotq50repopergroup}{117}
\DefineRemark{iiotq50tagsperimagepergroup}{1}
\DefineRemark{iiotq75lastpulleddayspergroup}{173}
\DefineRemark{iiotq75lastpusheddayspergroup}{173}
\DefineRemark{iiotq75lastupdateddayspergroup}{1401}
\DefineRemark{iiotq75repopergroup}{7183}
\DefineRemark{iiotq75tagsperimagepergroup}{2}
\DefineRemark{iiotrepopergroup}{76786}
\DefineRemark{iiotsizelayersavgbytepergroup}{29832022}
\DefineRemark{iiotsizelayersavggigabytepergroup}{0}
\DefineRemark{iiotsizelayersavgkilobytepergroup}{29832}
\DefineRemark{iiotsizelayersavgmegabytepergroup}{29}
\DefineRemark{iiotsizelayersavgterabytepergroup}{0}
\DefineRemark{iiotsizelayersmaxbytepergroup}{30842192217}
\DefineRemark{iiotsizelayersmaxgigabytepergroup}{30}
\DefineRemark{iiotsizelayersmaxkilobytepergroup}{30842192}
\DefineRemark{iiotsizelayersmaxmegabytepergroup}{30842}
\DefineRemark{iiotsizelayersmaxterabytepergroup}{0}
\DefineRemark{iiotsizelayersminbytepergroup}{23}
\DefineRemark{iiotsizelayersmingigabytepergroup}{0}
\DefineRemark{iiotsizelayersminkilobytepergroup}{0}
\DefineRemark{iiotsizelayersminmegabytepergroup}{0}
\DefineRemark{iiotsizelayersminterabytepergroup}{0}
\DefineRemark{iiotsizelayersq25bytepergroup}{602}
\DefineRemark{iiotsizelayersq25gigabytepergroup}{0}
\DefineRemark{iiotsizelayersq25kilobytepergroup}{0}
\DefineRemark{iiotsizelayersq25megabytepergroup}{0}
\DefineRemark{iiotsizelayersq25terabytepergroup}{0}
\DefineRemark{iiotsizelayersq50bytepergroup}{981699}
\DefineRemark{iiotsizelayersq50gigabytepergroup}{0}
\DefineRemark{iiotsizelayersq50kilobytepergroup}{981}
\DefineRemark{iiotsizelayersq50megabytepergroup}{0}
\DefineRemark{iiotsizelayersq50terabytepergroup}{0}
\DefineRemark{iiotsizelayersq75bytepergroup}{21350290}
\DefineRemark{iiotsizelayersq75gigabytepergroup}{0}
\DefineRemark{iiotsizelayersq75kilobytepergroup}{21350}
\DefineRemark{iiotsizelayersq75megabytepergroup}{21}
\DefineRemark{iiotsizelayersq75terabytepergroup}{0}
\DefineRemark{iiotsizelayerssumbytepergroup}{44326327744537}
\DefineRemark{iiotsizelayerssumgigabytepergroup}{44326}
\DefineRemark{iiotsizelayerssumkilobytepergroup}{44326327744}
\DefineRemark{iiotsizelayerssummegabytepergroup}{44326327}
\DefineRemark{iiotsizelayerssumterabytepergroup}{44}
\DefineRemark{iiotstddevrepopergroup}{4217}
\DefineRemark{iiotsumrepopergroup}{199311}
\DefineRemark{numtaglatestpergroup}{160}
\DefineRemark{numtagnonepergroup}{0}
\DefineRemark{standardavailabletagspergroup}{233649}
\DefineRemark{standardavglastpulleddayspergroup}{1568}
\DefineRemark{standardavglastpusheddayspergroup}{1568}
\DefineRemark{standardavglastupdateddayspergroup}{952}
\DefineRemark{standardavgrepopergroup}{4849}
\DefineRemark{standardavgtagsperimagepergroup}{19}
\DefineRemark{standarddistinctlayersizepergroup}{42400983099761}
\DefineRemark{standarddistinctnumlayersgroup}{1062112}
\DefineRemark{standarddistinctnumlayerspergroup}{1166570}
\DefineRemark{standarddistinctnumrepopergrouponly}{281112}
\DefineRemark{standarddistinctpctlayersgroup}{76}
\DefineRemark{standarddistinctpctrepopergrouponly}{79}
\DefineRemark{standarddistinctsizelayersavgbytepergroup}{36346711}
\DefineRemark{standarddistinctsizelayersavggigabytepergroup}{0}
\DefineRemark{standarddistinctsizelayersavgkilobytepergroup}{36346}
\DefineRemark{standarddistinctsizelayersavgmegabytepergroup}{36}
\DefineRemark{standarddistinctsizelayersavgterabytepergroup}{0}
\DefineRemark{standarddistinctsizelayersmaxbytepergroup}{27460444093}
\DefineRemark{standarddistinctsizelayersmaxgigabytepergroup}{27}
\DefineRemark{standarddistinctsizelayersmaxkilobytepergroup}{27460444}
\DefineRemark{standarddistinctsizelayersmaxmegabytepergroup}{27460}
\DefineRemark{standarddistinctsizelayersmaxterabytepergroup}{0}
\DefineRemark{standarddistinctsizelayersminbytepergroup}{0}
\DefineRemark{standarddistinctsizelayersmingigabytepergroup}{0}
\DefineRemark{standarddistinctsizelayersminkilobytepergroup}{0}
\DefineRemark{standarddistinctsizelayersminmegabytepergroup}{0}
\DefineRemark{standarddistinctsizelayersminterabytepergroup}{0}
\DefineRemark{standardmaxlastpulleddayspergroup}{29892}
\DefineRemark{standardmaxlastpusheddayspergroup}{29892}
\DefineRemark{standardmaxlastupdateddayspergroup}{29891}
\DefineRemark{standardmaxrepopergroup}{10000}
\DefineRemark{standardmaxtagsperimagepergroup}{69828}
\DefineRemark{standardminlastpulleddayspergroup}{0}
\DefineRemark{standardminlastpusheddayspergroup}{0}
\DefineRemark{standardminlastupdateddayspergroup}{0}
\DefineRemark{standardminrepopergroup}{0}
\DefineRemark{standardmintagsperimagepergroup}{0}
\DefineRemark{standardnumlayerspergroup}{3967848}
\DefineRemark{standardnumtaglatestpergroup}{171857}
\DefineRemark{standardnumtagnonepergroup}{62949}
\DefineRemark{standardnumtagspergroup}{7966650}
\DefineRemark{standardq25lastpulleddayspergroup}{16}
\DefineRemark{standardq25lastpusheddayspergroup}{16}
\DefineRemark{standardq25lastupdateddayspergroup}{409}
\DefineRemark{standardq25repopergroup}{684}
\DefineRemark{standardq25tagsperimagepergroup}{1}
\DefineRemark{standardq50lastpulleddayspergroup}{170}
\DefineRemark{standardq50lastpusheddayspergroup}{170}
\DefineRemark{standardq50lastupdateddayspergroup}{899}
\DefineRemark{standardq50repopergroup}{3563}
\DefineRemark{standardq50tagsperimagepergroup}{1}
\DefineRemark{standardq75lastpulleddayspergroup}{209}
\DefineRemark{standardq75lastpusheddayspergroup}{209}
\DefineRemark{standardq75lastupdateddayspergroup}{1469}
\DefineRemark{standardq75repopergroup}{10000}
\DefineRemark{standardq75tagsperimagepergroup}{3}
\DefineRemark{standardrepopergroup}{296598}
\DefineRemark{standardsizelayersavgbytepergroup}{33011688}
\DefineRemark{standardsizelayersavggigabytepergroup}{0}
\DefineRemark{standardsizelayersavgkilobytepergroup}{33011}
\DefineRemark{standardsizelayersavgmegabytepergroup}{33}
\DefineRemark{standardsizelayersavgterabytepergroup}{0}
\DefineRemark{standardsizelayersmaxbytepergroup}{27460444093}
\DefineRemark{standardsizelayersmaxgigabytepergroup}{27}
\DefineRemark{standardsizelayersmaxkilobytepergroup}{27460444}
\DefineRemark{standardsizelayersmaxmegabytepergroup}{27460}
\DefineRemark{standardsizelayersmaxterabytepergroup}{0}
\DefineRemark{standardsizelayersminbytepergroup}{0}
\DefineRemark{standardsizelayersmingigabytepergroup}{0}
\DefineRemark{standardsizelayersminkilobytepergroup}{0}
\DefineRemark{standardsizelayersminmegabytepergroup}{0}
\DefineRemark{standardsizelayersminterabytepergroup}{0}
\DefineRemark{standardsizelayersq25bytepergroup}{534}
\DefineRemark{standardsizelayersq25gigabytepergroup}{0}
\DefineRemark{standardsizelayersq25kilobytepergroup}{0}
\DefineRemark{standardsizelayersq25megabytepergroup}{0}
\DefineRemark{standardsizelayersq25terabytepergroup}{0}
\DefineRemark{standardsizelayersq50bytepergroup}{547590}
\DefineRemark{standardsizelayersq50gigabytepergroup}{0}
\DefineRemark{standardsizelayersq50kilobytepergroup}{547}
\DefineRemark{standardsizelayersq50megabytepergroup}{0}
\DefineRemark{standardsizelayersq50terabytepergroup}{0}
\DefineRemark{standardsizelayersq75bytepergroup}{20067699}
\DefineRemark{standardsizelayersq75gigabytepergroup}{0}
\DefineRemark{standardsizelayersq75kilobytepergroup}{20067}
\DefineRemark{standardsizelayersq75megabytepergroup}{20}
\DefineRemark{standardsizelayersq75terabytepergroup}{0}
\DefineRemark{standardsizelayerssumbytepergroup}{130985361636883}
\DefineRemark{standardsizelayerssumgigabytepergroup}{130985}
\DefineRemark{standardsizelayerssumkilobytepergroup}{130985361636}
\DefineRemark{standardsizelayerssummegabytepergroup}{130985361}
\DefineRemark{standardsizelayerssumterabytepergroup}{130}
\DefineRemark{standardstddevrepopergroup}{4283}
\DefineRemark{standardsumrepopergroup}{480065}

\DefineRemark{casignedcerts}{1060}
\DefineRemark{casignedcertspct}{4.8}
\DefineRemark{casignedparsedFindingextensionsextendedkeyusageany}{0}
\DefineRemark{casignedparsedFindingextensionsextendedkeyusageanypct}{0.0}
\DefineRemark{casignedparsedFindingextensionsextendedkeyusageclientauth}{1050}
\DefineRemark{casignedparsedFindingextensionsextendedkeyusageclientauthpct}{4.8}
\DefineRemark{casignedparsedFindingextensionsextendedkeyusagecodesigning}{3}
\DefineRemark{casignedparsedFindingextensionsextendedkeyusagecodesigningpct}{0.0}
\DefineRemark{casignedparsedFindingextensionsextendedkeyusageeapoverlan}{0}
\DefineRemark{casignedparsedFindingextensionsextendedkeyusageeapoverlanpct}{0.0}
\DefineRemark{casignedparsedFindingextensionsextendedkeyusageeapoverppp}{0}
\DefineRemark{casignedparsedFindingextensionsextendedkeyusageeapoverppppct}{0.0}
\DefineRemark{casignedparsedFindingextensionsextendedkeyusageemailprotection}{0}
\DefineRemark{casignedparsedFindingextensionsextendedkeyusageemailprotectionpct}{0.0}
\DefineRemark{casignedparsedFindingextensionsextendedkeyusageipsecendsystem}{0}
\DefineRemark{casignedparsedFindingextensionsextendedkeyusageipsecendsystempct}{0.0}
\DefineRemark{casignedparsedFindingextensionsextendedkeyusageipsecintermediatesystemusage}{0}
\DefineRemark{casignedparsedFindingextensionsextendedkeyusageipsecintermediatesystemusagepct}{0.0}
\DefineRemark{casignedparsedFindingextensionsextendedkeyusageipsectunnel}{0}
\DefineRemark{casignedparsedFindingextensionsextendedkeyusageipsectunnelpct}{0.0}
\DefineRemark{casignedparsedFindingextensionsextendedkeyusageipsecuser}{0}
\DefineRemark{casignedparsedFindingextensionsextendedkeyusageipsecuserpct}{0.0}
\DefineRemark{casignedparsedFindingextensionsextendedkeyusagemicrosoftcerttrustlistsigning}{0}
\DefineRemark{casignedparsedFindingextensionsextendedkeyusagemicrosoftcerttrustlistsigningpct}{0.0}
\DefineRemark{casignedparsedFindingextensionsextendedkeyusagemicrosoftdocumentsigning}{1}
\DefineRemark{casignedparsedFindingextensionsextendedkeyusagemicrosoftdocumentsigningpct}{0.0}
\DefineRemark{casignedparsedFindingextensionsextendedkeyusagemicrosoftefsrecovery}{0}
\DefineRemark{casignedparsedFindingextensionsextendedkeyusagemicrosoftefsrecoverypct}{0.0}
\DefineRemark{casignedparsedFindingextensionsextendedkeyusagemicrosoftencryptedfilesystem}{0}
\DefineRemark{casignedparsedFindingextensionsextendedkeyusagemicrosoftencryptedfilesystempct}{0.0}
\DefineRemark{casignedparsedFindingextensionsextendedkeyusagemicrosoftservergatedcrypto}{0}
\DefineRemark{casignedparsedFindingextensionsextendedkeyusagemicrosoftservergatedcryptopct}{0.0}
\DefineRemark{casignedparsedFindingextensionsextendedkeyusagemicrosoftsmartcardlogon}{0}
\DefineRemark{casignedparsedFindingextensionsextendedkeyusagemicrosoftsmartcardlogonpct}{0.0}
\DefineRemark{casignedparsedFindingextensionsextendedkeyusagenetscapeservergatedcrypto}{0}
\DefineRemark{casignedparsedFindingextensionsextendedkeyusagenetscapeservergatedcryptopct}{0.0}
\DefineRemark{casignedparsedFindingextensionsextendedkeyusageocspsigning}{0}
\DefineRemark{casignedparsedFindingextensionsextendedkeyusageocspsigningpct}{0.0}
\DefineRemark{casignedparsedFindingextensionsextendedkeyusageserverauth}{799}
\DefineRemark{casignedparsedFindingextensionsextendedkeyusageserverauthpct}{3.6}
\DefineRemark{casignedparsedFindingextensionsextendedkeyusagetimestamping}{0}
\DefineRemark{casignedparsedFindingextensionsextendedkeyusagetimestampingpct}{0.0}
\DefineRemark{casignedparsedFindingextensionsextendedkeyusageunknown}{1}
\DefineRemark{casignedparsedFindingextensionsextendedkeyusageunknownpct}{0.0}
\DefineRemark{casignedparsedFindingextensionskeyusagecertificatesign}{0}
\DefineRemark{casignedparsedFindingextensionskeyusagecertificatesignpct}{0.0}
\DefineRemark{casignedparsedFindingextensionskeyusagecontentcommitment}{3}
\DefineRemark{casignedparsedFindingextensionskeyusagecontentcommitmentpct}{0.0}
\DefineRemark{casignedparsedFindingextensionskeyusagecrlsign}{0}
\DefineRemark{casignedparsedFindingextensionskeyusagecrlsignpct}{0.0}
\DefineRemark{casignedparsedFindingextensionskeyusagedataencipherment}{1}
\DefineRemark{casignedparsedFindingextensionskeyusagedataenciphermentpct}{0.0}
\DefineRemark{casignedparsedFindingextensionskeyusagedecipheronly}{0}
\DefineRemark{casignedparsedFindingextensionskeyusagedecipheronlypct}{0.0}
\DefineRemark{casignedparsedFindingextensionskeyusagedigitalsignature}{1060}
\DefineRemark{casignedparsedFindingextensionskeyusagedigitalsignaturepct}{4.8}
\DefineRemark{casignedparsedFindingextensionskeyusageencipheronly}{0}
\DefineRemark{casignedparsedFindingextensionskeyusageencipheronlypct}{0.0}
\DefineRemark{casignedparsedFindingextensionskeyusagekeyagreement}{1}
\DefineRemark{casignedparsedFindingextensionskeyusagekeyagreementpct}{0.0}
\DefineRemark{casignedparsedFindingextensionskeyusagekeyencipherment}{791}
\DefineRemark{casignedparsedFindingextensionskeyusagekeyenciphermentpct}{3.6}
\DefineRemark{casignedvalidondownload}{141}
\DefineRemark{casignedvalidondownloadpct}{13}
\DefineRemark{casignedvalidonhistory}{631}
\DefineRemark{casignedvalidonhistorypct}{60}
\DefineRemark{knowncompromizedcerts}{22082}
\DefineRemark{knowncompromizedhostkeys}{11942}
\DefineRemark{privatecacerts}{7546}
\DefineRemark{privatecacertspct}{34}
\DefineRemark{privatecaparsedFindingextensionsextendedkeyusageany}{84}
\DefineRemark{privatecaparsedFindingextensionsextendedkeyusageanypct}{0.4}
\DefineRemark{privatecaparsedFindingextensionsextendedkeyusageclientauth}{2307}
\DefineRemark{privatecaparsedFindingextensionsextendedkeyusageclientauthpct}{10}
\DefineRemark{privatecaparsedFindingextensionsextendedkeyusagecodesigning}{22}
\DefineRemark{privatecaparsedFindingextensionsextendedkeyusagecodesigningpct}{0.1}
\DefineRemark{privatecaparsedFindingextensionsextendedkeyusageeapoverlan}{5}
\DefineRemark{privatecaparsedFindingextensionsextendedkeyusageeapoverlanpct}{0.0}
\DefineRemark{privatecaparsedFindingextensionsextendedkeyusageeapoverppp}{5}
\DefineRemark{privatecaparsedFindingextensionsextendedkeyusageeapoverppppct}{0.0}
\DefineRemark{privatecaparsedFindingextensionsextendedkeyusageemailprotection}{70}
\DefineRemark{privatecaparsedFindingextensionsextendedkeyusageemailprotectionpct}{0.3}
\DefineRemark{privatecaparsedFindingextensionsextendedkeyusageipsecendsystem}{6}
\DefineRemark{privatecaparsedFindingextensionsextendedkeyusageipsecendsystempct}{0.0}
\DefineRemark{privatecaparsedFindingextensionsextendedkeyusageipsecintermediatesystemusage}{9}
\DefineRemark{privatecaparsedFindingextensionsextendedkeyusageipsecintermediatesystemusagepct}{0.0}
\DefineRemark{privatecaparsedFindingextensionsextendedkeyusageipsectunnel}{6}
\DefineRemark{privatecaparsedFindingextensionsextendedkeyusageipsectunnelpct}{0.0}
\DefineRemark{privatecaparsedFindingextensionsextendedkeyusageipsecuser}{6}
\DefineRemark{privatecaparsedFindingextensionsextendedkeyusageipsecuserpct}{0.0}
\DefineRemark{privatecaparsedFindingextensionsextendedkeyusagemicrosoftcerttrustlistsigning}{3}
\DefineRemark{privatecaparsedFindingextensionsextendedkeyusagemicrosoftcerttrustlistsigningpct}{0.0}
\DefineRemark{privatecaparsedFindingextensionsextendedkeyusagemicrosoftdocumentsigning}{0}
\DefineRemark{privatecaparsedFindingextensionsextendedkeyusagemicrosoftdocumentsigningpct}{0.0}
\DefineRemark{privatecaparsedFindingextensionsextendedkeyusagemicrosoftefsrecovery}{2}
\DefineRemark{privatecaparsedFindingextensionsextendedkeyusagemicrosoftefsrecoverypct}{0.0}
\DefineRemark{privatecaparsedFindingextensionsextendedkeyusagemicrosoftencryptedfilesystem}{6}
\DefineRemark{privatecaparsedFindingextensionsextendedkeyusagemicrosoftencryptedfilesystempct}{0.0}
\DefineRemark{privatecaparsedFindingextensionsextendedkeyusagemicrosoftservergatedcrypto}{12}
\DefineRemark{privatecaparsedFindingextensionsextendedkeyusagemicrosoftservergatedcryptopct}{0.1}
\DefineRemark{privatecaparsedFindingextensionsextendedkeyusagemicrosoftsmartcardlogon}{14}
\DefineRemark{privatecaparsedFindingextensionsextendedkeyusagemicrosoftsmartcardlogonpct}{0.1}
\DefineRemark{privatecaparsedFindingextensionsextendedkeyusagenetscapeservergatedcrypto}{18}
\DefineRemark{privatecaparsedFindingextensionsextendedkeyusagenetscapeservergatedcryptopct}{0.1}
\DefineRemark{privatecaparsedFindingextensionsextendedkeyusageocspsigning}{40}
\DefineRemark{privatecaparsedFindingextensionsextendedkeyusageocspsigningpct}{0.2}
\DefineRemark{privatecaparsedFindingextensionsextendedkeyusageserverauth}{2271}
\DefineRemark{privatecaparsedFindingextensionsextendedkeyusageserverauthpct}{10}
\DefineRemark{privatecaparsedFindingextensionsextendedkeyusagetimestamping}{14}
\DefineRemark{privatecaparsedFindingextensionsextendedkeyusagetimestampingpct}{0.1}
\DefineRemark{privatecaparsedFindingextensionsextendedkeyusageunknown}{34}
\DefineRemark{privatecaparsedFindingextensionsextendedkeyusageunknownpct}{0.2}
\DefineRemark{privatecaparsedFindingextensionskeyusagecertificatesign}{797}
\DefineRemark{privatecaparsedFindingextensionskeyusagecertificatesignpct}{3.6}
\DefineRemark{privatecaparsedFindingextensionskeyusagecontentcommitment}{982}
\DefineRemark{privatecaparsedFindingextensionskeyusagecontentcommitmentpct}{4.4}
\DefineRemark{privatecaparsedFindingextensionskeyusagecrlsign}{307}
\DefineRemark{privatecaparsedFindingextensionskeyusagecrlsignpct}{1.4}
\DefineRemark{privatecaparsedFindingextensionskeyusagedataencipherment}{689}
\DefineRemark{privatecaparsedFindingextensionskeyusagedataenciphermentpct}{3.1}
\DefineRemark{privatecaparsedFindingextensionskeyusagedecipheronly}{6}
\DefineRemark{privatecaparsedFindingextensionskeyusagedecipheronlypct}{0.0}
\DefineRemark{privatecaparsedFindingextensionskeyusagedigitalsignature}{3503}
\DefineRemark{privatecaparsedFindingextensionskeyusagedigitalsignaturepct}{16}
\DefineRemark{privatecaparsedFindingextensionskeyusageencipheronly}{5}
\DefineRemark{privatecaparsedFindingextensionskeyusageencipheronlypct}{0.0}
\DefineRemark{privatecaparsedFindingextensionskeyusagekeyagreement}{592}
\DefineRemark{privatecaparsedFindingextensionskeyusagekeyagreementpct}{2.7}
\DefineRemark{privatecaparsedFindingextensionskeyusagekeyencipherment}{2818}
\DefineRemark{privatecaparsedFindingextensionskeyusagekeyenciphermentpct}{13}
\DefineRemark{privatecavalidondownload}{4970}
\DefineRemark{privatecavalidondownloadpct}{66}
\DefineRemark{privatecavalidonhistory}{6486}
\DefineRemark{privatecavalidonhistorypct}{86}
\DefineRemark{selfsignedcerts}{13476}
\DefineRemark{selfsignedcertspct}{61}
\DefineRemark{selfsignedparsedFindingextensionsextendedkeyusageany}{62}
\DefineRemark{selfsignedparsedFindingextensionsextendedkeyusageanypct}{0.3}
\DefineRemark{selfsignedparsedFindingextensionsextendedkeyusageclientauth}{479}
\DefineRemark{selfsignedparsedFindingextensionsextendedkeyusageclientauthpct}{2.2}
\DefineRemark{selfsignedparsedFindingextensionsextendedkeyusagecodesigning}{25}
\DefineRemark{selfsignedparsedFindingextensionsextendedkeyusagecodesigningpct}{0.1}
\DefineRemark{selfsignedparsedFindingextensionsextendedkeyusageeapoverlan}{1}
\DefineRemark{selfsignedparsedFindingextensionsextendedkeyusageeapoverlanpct}{0.0}
\DefineRemark{selfsignedparsedFindingextensionsextendedkeyusageeapoverppp}{1}
\DefineRemark{selfsignedparsedFindingextensionsextendedkeyusageeapoverppppct}{0.0}
\DefineRemark{selfsignedparsedFindingextensionsextendedkeyusageemailprotection}{24}
\DefineRemark{selfsignedparsedFindingextensionsextendedkeyusageemailprotectionpct}{0.1}
\DefineRemark{selfsignedparsedFindingextensionsextendedkeyusageipsecendsystem}{11}
\DefineRemark{selfsignedparsedFindingextensionsextendedkeyusageipsecendsystempct}{0.0}
\DefineRemark{selfsignedparsedFindingextensionsextendedkeyusageipsecintermediatesystemusage}{2}
\DefineRemark{selfsignedparsedFindingextensionsextendedkeyusageipsecintermediatesystemusagepct}{0.0}
\DefineRemark{selfsignedparsedFindingextensionsextendedkeyusageipsectunnel}{11}
\DefineRemark{selfsignedparsedFindingextensionsextendedkeyusageipsectunnelpct}{0.0}
\DefineRemark{selfsignedparsedFindingextensionsextendedkeyusageipsecuser}{11}
\DefineRemark{selfsignedparsedFindingextensionsextendedkeyusageipsecuserpct}{0.0}
\DefineRemark{selfsignedparsedFindingextensionsextendedkeyusagemicrosoftcerttrustlistsigning}{11}
\DefineRemark{selfsignedparsedFindingextensionsextendedkeyusagemicrosoftcerttrustlistsigningpct}{0.0}
\DefineRemark{selfsignedparsedFindingextensionsextendedkeyusagemicrosoftdocumentsigning}{0}
\DefineRemark{selfsignedparsedFindingextensionsextendedkeyusagemicrosoftdocumentsigningpct}{0.0}
\DefineRemark{selfsignedparsedFindingextensionsextendedkeyusagemicrosoftefsrecovery}{1}
\DefineRemark{selfsignedparsedFindingextensionsextendedkeyusagemicrosoftefsrecoverypct}{0.0}
\DefineRemark{selfsignedparsedFindingextensionsextendedkeyusagemicrosoftencryptedfilesystem}{11}
\DefineRemark{selfsignedparsedFindingextensionsextendedkeyusagemicrosoftencryptedfilesystempct}{0.0}
\DefineRemark{selfsignedparsedFindingextensionsextendedkeyusagemicrosoftservergatedcrypto}{11}
\DefineRemark{selfsignedparsedFindingextensionsextendedkeyusagemicrosoftservergatedcryptopct}{0.0}
\DefineRemark{selfsignedparsedFindingextensionsextendedkeyusagemicrosoftsmartcardlogon}{2}
\DefineRemark{selfsignedparsedFindingextensionsextendedkeyusagemicrosoftsmartcardlogonpct}{0.0}
\DefineRemark{selfsignedparsedFindingextensionsextendedkeyusagenetscapeservergatedcrypto}{11}
\DefineRemark{selfsignedparsedFindingextensionsextendedkeyusagenetscapeservergatedcryptopct}{0.0}
\DefineRemark{selfsignedparsedFindingextensionsextendedkeyusageocspsigning}{47}
\DefineRemark{selfsignedparsedFindingextensionsextendedkeyusageocspsigningpct}{0.2}
\DefineRemark{selfsignedparsedFindingextensionsextendedkeyusageserverauth}{731}
\DefineRemark{selfsignedparsedFindingextensionsextendedkeyusageserverauthpct}{3.3}
\DefineRemark{selfsignedparsedFindingextensionsextendedkeyusagetimestamping}{26}
\DefineRemark{selfsignedparsedFindingextensionsextendedkeyusagetimestampingpct}{0.1}
\DefineRemark{selfsignedparsedFindingextensionsextendedkeyusageunknown}{13}
\DefineRemark{selfsignedparsedFindingextensionsextendedkeyusageunknownpct}{0.1}
\DefineRemark{selfsignedparsedFindingextensionskeyusagecertificatesign}{1216}
\DefineRemark{selfsignedparsedFindingextensionskeyusagecertificatesignpct}{5.5}
\DefineRemark{selfsignedparsedFindingextensionskeyusagecontentcommitment}{766}
\DefineRemark{selfsignedparsedFindingextensionskeyusagecontentcommitmentpct}{3.5}
\DefineRemark{selfsignedparsedFindingextensionskeyusagecrlsign}{548}
\DefineRemark{selfsignedparsedFindingextensionskeyusagecrlsignpct}{2.5}
\DefineRemark{selfsignedparsedFindingextensionskeyusagedataencipherment}{367}
\DefineRemark{selfsignedparsedFindingextensionskeyusagedataenciphermentpct}{1.7}
\DefineRemark{selfsignedparsedFindingextensionskeyusagedecipheronly}{1}
\DefineRemark{selfsignedparsedFindingextensionskeyusagedecipheronlypct}{0.0}
\DefineRemark{selfsignedparsedFindingextensionskeyusagedigitalsignature}{1376}
\DefineRemark{selfsignedparsedFindingextensionskeyusagedigitalsignaturepct}{6.2}
\DefineRemark{selfsignedparsedFindingextensionskeyusageencipheronly}{1}
\DefineRemark{selfsignedparsedFindingextensionskeyusageencipheronlypct}{0.0}
\DefineRemark{selfsignedparsedFindingextensionskeyusagekeyagreement}{214}
\DefineRemark{selfsignedparsedFindingextensionskeyusagekeyagreementpct}{1.0}
\DefineRemark{selfsignedparsedFindingextensionskeyusagekeyencipherment}{1290}
\DefineRemark{selfsignedparsedFindingextensionskeyusagekeyenciphermentpct}{5.8}
\DefineRemark{selfsignedvalidondownload}{10629}
\DefineRemark{selfsignedvalidondownloadpct}{79}
\DefineRemark{selfsignedvalidonhistory}{12263}
\DefineRemark{selfsignedvalidonhistorypct}{91}

\DefineRemark{mostreoccurringdata}{AKIAIOSFODNN7EXAMPLE}
\DefineRemark{mostreoccurringgroup}{Cloud}
\DefineRemark{mostreoccurringnumlayer}{20497}
\DefineRemark{mostreoccurringnumocc}{291949}
\DefineRemark{mostreoccurringrule}{Amazon AWS API}
\DefineRemark{numconsidereddistinctlayersoverall}{1533104}
\DefineRemark{numconsideredlayersoverall}{1647300}
\DefineRemark{rawaccmaterialnummatchesimages}{70762873}
\DefineRemark{rawapicloudnummatchesimages}{6208995}
\DefineRemark{rawapifinancialnummatchesimages}{42901}
\DefineRemark{rawapiiotnummatchesimages}{297}
\DefineRemark{rawapisocialmedianummatchesimages}{6365854}
\DefineRemark{rawnummatchesimages}{181139838}
\DefineRemark{rawprivatekeynummatchesimages}{1377336}

\DefineRemark{frequencyNgramsTimeFactor}{29}
\DefineRemark{frequencyngrammax}{7}
\DefineRemark{frequencyngrammin}{4}
\DefineRemark{kompromatfilterednumdistincttotal}{1831}
\DefineRemark{kompromatfoundnumdistincttotal}{1914}
\DefineRemark{kompromatfoundrfcnumdistinct}{6}
\DefineRemark{kompromatfoundrfcnumdistinctmax}{6}
\DefineRemark{kompromatfoundrfcservicemax}{rfc}
\DefineRemark{kompromatfoundsoftwaretestsnumdistinct}{1820}
\DefineRemark{kompromatfoundsoftwaretestsnumdistinctmax}{951}
\DefineRemark{kompromatfoundsoftwaretestsservicemax}{openssl}
\DefineRemark{kompromatfoundtestvectorsnumdistinct}{3}
\DefineRemark{kompromatfoundtestvectorsnumdistinctmax}{3}
\DefineRemark{kompromatfoundtestvectorsservicemax}{wycheproof}
\DefineRemark{mostfiltered0matchingfpkompromat}{usr/local/src}
\DefineRemark{mostfiltered0secretsha256}{1473}
\DefineRemark{mostfiltered1matchingfpkompromat}{home/workhome/tlspuffin}
\DefineRemark{mostfiltered1secretsha256}{1008}
\DefineRemark{mostfiltered2matchingfpkompromat}{home/workhome/tlspuffin/deps}
\DefineRemark{mostfiltered2secretsha256}{1007}
\DefineRemark{mostfiltered3matchingfpkompromat}{home/workhome/tlspuffin/deps/rust-openssl}
\DefineRemark{mostfiltered3secretsha256}{990}
\DefineRemark{mostfiltered4matchingfpkompromat}{home/workhome/tlspuffin/deps/rust-openssl-src}
\DefineRemark{mostfiltered4secretsha256}{986}
\DefineRemark{mostfiltered5matchingfpkompromat}{home/workhome/tlspuffin/deps/rust-openssl-src/openssl}
\DefineRemark{mostfiltered5secretsha256}{986}
\DefineRemark{mostfiltered6matchingfpkompromat}{node/deps/openssl}
\DefineRemark{mostfiltered6secretsha256}{939}
\DefineRemark{mostfiltered7matchingfpkompromat}{node/deps/openssl/openssl}
\DefineRemark{mostfiltered7secretsha256}{939}
\DefineRemark{mostfiltered8matchingfpkompromat}{home/user/.local}
\DefineRemark{mostfiltered8secretsha256}{929}
\DefineRemark{mostfiltered9matchingfpkompromat}{usr/local/src/openssl-openssl-3.0.3}
\DefineRemark{mostfiltered9secretsha256}{924}
\DefineRemark{pkxmlnummatches}{67}
\DefineRemark{rawaccmaterialnumdistinctmatchesval}{18}
\DefineRemark{rawaccmaterialnummatchesval}{152}
\DefineRemark{rawapicloudnumdistinctmatchesval}{84}
\DefineRemark{rawapicloudnummatchesval}{416}
\DefineRemark{rawapifinancialnumdistinctmatchesval}{2}
\DefineRemark{rawapifinancialnummatchesval}{4}
\DefineRemark{rawapisocialmedianumdistinctmatchesval}{8}
\DefineRemark{rawapisocialmedianummatchesval}{14}
\DefineRemark{rawprivatekeynumdistinctmatchesval}{1}
\DefineRemark{rawprivatekeynummatchesval}{2}
\DefineRemark{sequencelimitnotsame}{4}
\DefineRemark{sequencelimitsame}{3}
\DefineRemark{topngramfilter0ngram}{....}
\DefineRemark{topngramfilter0occurrences}{63749}
\DefineRemark{topngramfilter1ngram}{acea}
\DefineRemark{topngramfilter1occurrences}{59112}
\DefineRemark{topngramfilter2ngram}{ceae}
\DefineRemark{topngramfilter2occurrences}{58782}
\DefineRemark{topngramfilter3ngram}{aceae}
\DefineRemark{topngramfilter3occurrences}{58744}
\DefineRemark{topngramfilter4ngram}{.....}
\DefineRemark{topngramfilter4occurrences}{57284}
\DefineRemark{topngramfilter5ngram}{......}
\DefineRemark{topngramfilter5occurrences}{51765}
\DefineRemark{topngramfilter6ngram}{spec}
\DefineRemark{topngramfilter6occurrences}{48095}
\DefineRemark{topngramfilter7ngram}{.......}
\DefineRemark{topngramfilter7occurrences}{46982}
\DefineRemark{topngramfilter8ngram}{atio}
\DefineRemark{topngramfilter8occurrences}{40739}
\DefineRemark{topngramfilter9ngram}{ation}
\DefineRemark{topngramfilter9occurrences}{40637}

\DefineRemark{affectediiotapirepositorynum}{706}
\DefineRemark{affectediiotapirepositorypct}{0.9}
\DefineRemark{affectediiotprivatekeyrepositorynum}{4159}
\DefineRemark{affectediiotprivatekeyrepositorypct}{5.4}
\DefineRemark{affectediiotrepositorynum}{4732}
\DefineRemark{affectediiotrepositorypct}{6.2}
\DefineRemark{affectedstandardapirepositorynum}{3100}
\DefineRemark{affectedstandardapirepositorypct}{1.0}
\DefineRemark{affectedstandardprivatekeyrepositorynum}{18693}
\DefineRemark{affectedstandardprivatekeyrepositorypct}{6.3}
\DefineRemark{affectedstandardrepositorynum}{21346}
\DefineRemark{affectedstandardrepositorypct}{7.2}
\DefineRemark{apicloud0numdistinctmatches}{1298}
\DefineRemark{apicloud0rule}{Google Service API}
\DefineRemark{apicloud1numdistinctmatches}{1213}
\DefineRemark{apicloud1rule}{Amazon AWS API}
\DefineRemark{apicloud2numdistinctmatches}{177}
\DefineRemark{apicloud2rule}{Alibaba API}
\DefineRemark{apicloud3numdistinctmatches}{146}
\DefineRemark{apicloud3rule}{Heroku API}
\DefineRemark{apicloud4numdistinctmatches}{36}
\DefineRemark{apicloud4rule}{MailGun API}
\DefineRemark{apicloud5numdistinctmatches}{29}
\DefineRemark{apicloud5rule}{MailChimp API}
\DefineRemark{apicloud6numdistinctmatches}{7}
\DefineRemark{apicloud6rule}{Twilio API}
\DefineRemark{apicloud7numdistinctmatches}{6}
\DefineRemark{apicloud7rule}{GitLab API}
\DefineRemark{apicloud8numdistinctmatches}{4}
\DefineRemark{apicloud8rule}{Github API}
\DefineRemark{apicloud9numdistinctmatches}{4}
\DefineRemark{apicloud9rule}{Google Cloud API}
\DefineRemark{apidownloadpct}{2.0}
\DefineRemark{apifinancial0numdistinctmatches}{18}
\DefineRemark{apifinancial0rule}{Stripe API}
\DefineRemark{apifinancial1numdistinctmatches}{6}
\DefineRemark{apifinancial1rule}{Amazon MWS API}
\DefineRemark{apifinancial2numdistinctmatches}{1}
\DefineRemark{apifinancial2rule}{Square API}
\DefineRemark{apigitpct}{4.0}
\DefineRemark{apiinstsshdpct}{0.3}
\DefineRemark{apinumdistinctmatches}{3158}
\DefineRemark{apisocialmedia0numdistinctmatches}{210}
\DefineRemark{apisocialmedia0rule}{Twitter API}
\DefineRemark{apisocialmedia1numdistinctmatches}{3}
\DefineRemark{apisocialmedia1rule}{Facebook API}
\DefineRemark{apisshkeygenpct}{0.1}
\DefineRemark{filteredimageapicloudNonenumdistinct}{2881}
\DefineRemark{filteredimageapicloudNonepctdistinct}{1.8}
\DefineRemark{filteredimageapicloudfilenumdistinct}{31844}
\DefineRemark{filteredimageapicloudfilepctdistinct}{20}
\DefineRemark{filteredimageapicloudrulenumdistinct}{67391}
\DefineRemark{filteredimageapicloudrulepctdistinct}{41}
\DefineRemark{filteredimageapicloudsequencenumdistinct}{59982}
\DefineRemark{filteredimageapicloudsequencepctdistinct}{37}
\DefineRemark{filteredimageapicloudunparsablenumdistinct}{704}
\DefineRemark{filteredimageapicloudunparsablepctdistinct}{0.4}
\DefineRemark{filteredimageapifinancialNonenumdistinct}{23}
\DefineRemark{filteredimageapifinancialNonepctdistinct}{2.1}
\DefineRemark{filteredimageapifinancialfilenumdistinct}{455}
\DefineRemark{filteredimageapifinancialfilepctdistinct}{42}
\DefineRemark{filteredimageapifinancialrulenumdistinct}{510}
\DefineRemark{filteredimageapifinancialrulepctdistinct}{48}
\DefineRemark{filteredimageapifinancialsequencenumdistinct}{84}
\DefineRemark{filteredimageapifinancialsequencepctdistinct}{7.8}
\DefineRemark{filteredimageapiiotfilenumdistinct}{94}
\DefineRemark{filteredimageapiiotfilepctdistinct}{39}
\DefineRemark{filteredimageapiiotrulenumdistinct}{116}
\DefineRemark{filteredimageapiiotrulepctdistinct}{48}
\DefineRemark{filteredimageapiiotsequencenumdistinct}{32}
\DefineRemark{filteredimageapiiotsequencepctdistinct}{13}
\DefineRemark{filteredimageapisocialmediaNonenumdistinct}{209}
\DefineRemark{filteredimageapisocialmediaNonepctdistinct}{0.0}
\DefineRemark{filteredimageapisocialmediafilenumdistinct}{432988}
\DefineRemark{filteredimageapisocialmediafilepctdistinct}{49}
\DefineRemark{filteredimageapisocialmediarulenumdistinct}{312117}
\DefineRemark{filteredimageapisocialmediarulepctdistinct}{35}
\DefineRemark{filteredimageapisocialmediasequencenumdistinct}{143886}
\DefineRemark{filteredimageapisocialmediasequencepctdistinct}{16}
\DefineRemark{filteredimageprivatekeyNonenumdistinct}{52150}
\DefineRemark{filteredimageprivatekeyNonepctdistinct}{81}
\DefineRemark{filteredimageprivatekeyfilenumdistinct}{9062}
\DefineRemark{filteredimageprivatekeyfilepctdistinct}{14}
\DefineRemark{filteredimageprivatekeykompromatnumdistinct}{1826}
\DefineRemark{filteredimageprivatekeykompromatpctdistinct}{2.8}
\DefineRemark{filteredimageprivatekeyunparsablenumdistinct}{1300}
\DefineRemark{filteredimageprivatekeyunparsablepctdistinct}{2.0}
\DefineRemark{filteredvalapicloudNonenumdistinct}{67}
\DefineRemark{filteredvalapicloudNonepctdistinct}{79}
\DefineRemark{filteredvalapicloudrulenumdistinct}{3}
\DefineRemark{filteredvalapicloudrulepctdistinct}{3.5}
\DefineRemark{filteredvalapicloudsequencenumdistinct}{15}
\DefineRemark{filteredvalapicloudsequencepctdistinct}{18}
\DefineRemark{filteredvalapifinancialNonenumdistinct}{2}
\DefineRemark{filteredvalapifinancialNonepctdistinct}{100}
\DefineRemark{filteredvalapisocialmediaNonenumdistinct}{4}
\DefineRemark{filteredvalapisocialmediaNonepctdistinct}{50}
\DefineRemark{filteredvalapisocialmediarulenumdistinct}{4}
\DefineRemark{filteredvalapisocialmediarulepctdistinct}{50}
\DefineRemark{filteredvalprivatekeyunparsablenumdistinct}{1}
\DefineRemark{filteredvalprivatekeyunparsablepctdistinct}{100}
\DefineRemark{imageapicloudnumdistinctmatches}{74460}
\DefineRemark{imageapifinancialnumdistinctmatches}{543}
\DefineRemark{imageapiiotnumdistinctmatches}{117}
\DefineRemark{imageapisocialmedianumdistinctmatches}{439822}
\DefineRemark{imageprivatekeynumdistinctmatches}{62282}
\DefineRemark{layerincludesecretapihubnum}{2437}
\DefineRemark{layerincludesecretapihubpct}{0.2}
\DefineRemark{layerincludesecretapiprivatenum}{744}
\DefineRemark{layerincludesecretapiprivatepct}{0.3}
\DefineRemark{layerincludesecrethubnum}{51515}
\DefineRemark{layerincludesecrethubpct}{3.7}
\DefineRemark{layerincludesecretprivatekeyhubnum}{49078}
\DefineRemark{layerincludesecretprivatekeyhubpct}{3.5}
\DefineRemark{layerincludesecretprivatekeyprivatenum}{3429}
\DefineRemark{layerincludesecretprivatekeyprivatepct}{1.3}
\DefineRemark{layerincludesecretprivatenum}{4173}
\DefineRemark{layerincludesecretprivatepct}{1.6}
\DefineRemark{numaffectedimages}{28621}
\DefineRemark{numaffectedimagesdockerhub}{24912}
\DefineRemark{numaffectedimagesprivate}{3709}
\DefineRemark{numnonemptyimages}{337171}
\DefineRemark{numnonemptyimagesdockerhub}{278026}
\DefineRemark{numnonemptyimagesprivate}{59145}
\DefineRemark{numsnakeoiletcssl}{6447}
\DefineRemark{pctaffectedimages}{8.5}
\DefineRemark{pctaffectedimagesdockerhub}{9.0}
\DefineRemark{pctaffectedimagesprivate}{6.3}
\DefineRemark{privatekey0numdistinctmatches}{52107}
\DefineRemark{privatekey0rule}{Private Key}
\DefineRemark{privatekeydownloadpct}{6.8}
\DefineRemark{privatekeygitpct}{2.6}
\DefineRemark{privatekeyinstsshdpct}{30}
\DefineRemark{privatekeynumdistinctmatches}{52107}
\DefineRemark{privatekeysshkeygenpct}{8.1}
\DefineRemark{totalvalidmatches}{55265}
\DefineRemark{valapicloudnumdistinctmatches}{84}
\DefineRemark{valapifinancialnumdistinctmatches}{2}
\DefineRemark{valapisocialmedianumdistinctmatches}{8}
\DefineRemark{validapicloudinvalidnumdistinctmatchestotal}{71591}
\DefineRemark{validapicloudinvalidpctdistinctmatchestotal}{96}
\DefineRemark{validapicloudvalidnumdistinctmatchestotal}{2920}
\DefineRemark{validapicloudvalidpctdistinctmatchestotal}{3.9}
\DefineRemark{validapifinancialinvalidnumdistinctmatchestotal}{520}
\DefineRemark{validapifinancialinvalidpctdistinctmatchestotal}{95}
\DefineRemark{validapifinancialvalidnumdistinctmatchestotal}{25}
\DefineRemark{validapifinancialvalidpctdistinctmatchestotal}{4.6}
\DefineRemark{validapiiotinvalidnumdistinctmatchestotal}{117}
\DefineRemark{validapiiotinvalidpctdistinctmatchestotal}{100}
\DefineRemark{validapisocialmediainvalidnumdistinctmatchestotal}{439617}
\DefineRemark{validapisocialmediainvalidpctdistinctmatchestotal}{100}
\DefineRemark{validapisocialmediavalidnumdistinctmatchestotal}{213}
\DefineRemark{validapisocialmediavalidpctdistinctmatchestotal}{0.0}
\DefineRemark{validimageapicloudinvalidnumdistinctmatches}{71580}
\DefineRemark{validimageapicloudinvalidpctdistinctmatches}{96}
\DefineRemark{validimageapicloudvalidnumdistinctmatches}{2880}
\DefineRemark{validimageapicloudvalidpctdistinctmatches}{3.9}
\DefineRemark{validimageapifinancialinvalidnumdistinctmatches}{520}
\DefineRemark{validimageapifinancialinvalidpctdistinctmatches}{96}
\DefineRemark{validimageapifinancialvalidnumdistinctmatches}{23}
\DefineRemark{validimageapifinancialvalidpctdistinctmatches}{4.2}
\DefineRemark{validimageapiiotinvalidnumdistinctmatches}{117}
\DefineRemark{validimageapiiotinvalidpctdistinctmatches}{100}
\DefineRemark{validimageapisocialmediainvalidnumdistinctmatches}{439613}
\DefineRemark{validimageapisocialmediainvalidpctdistinctmatches}{100}
\DefineRemark{validimageapisocialmediavalidnumdistinctmatches}{209}
\DefineRemark{validimageapisocialmediavalidpctdistinctmatches}{0.0}
\DefineRemark{validimageprivatekeyinvalidnumdistinctmatches}{10175}
\DefineRemark{validimageprivatekeyinvalidpctdistinctmatches}{16}
\DefineRemark{validimageprivatekeyvalidnumdistinctmatches}{52107}
\DefineRemark{validimageprivatekeyvalidpctdistinctmatches}{84}
\DefineRemark{validmatchmultiuseraccmaterialFalsenum}{5}
\DefineRemark{validmatchmultiuseraccmaterialFalsepct}{100}
\DefineRemark{validmatchmultiuserapiFalsenum}{2845}
\DefineRemark{validmatchmultiuserapiFalsepct}{90}
\DefineRemark{validmatchmultiuserapiTruenum}{314}
\DefineRemark{validmatchmultiuserapiTruepct}{9.9}
\DefineRemark{validmatchmultiuserlayer0apiFalsenum}{175}
\DefineRemark{validmatchmultiuserlayer0apiFalsepct}{90}
\DefineRemark{validmatchmultiuserlayer0apiTruenum}{19}
\DefineRemark{validmatchmultiuserlayer0apiTruepct}{9.8}
\DefineRemark{validmatchmultiuserlayer0privatekeyFalsenum}{1764}
\DefineRemark{validmatchmultiuserlayer0privatekeyFalsepct}{95}
\DefineRemark{validmatchmultiuserlayer0privatekeyTruenum}{95}
\DefineRemark{validmatchmultiuserlayer0privatekeyTruepct}{5.1}
\DefineRemark{validmatchmultiuserprivatekeyFalsenum}{49667}
\DefineRemark{validmatchmultiuserprivatekeyFalsepct}{95}
\DefineRemark{validmatchmultiuserprivatekeyTruenum}{2483}
\DefineRemark{validmatchmultiuserprivatekeyTruepct}{4.8}
\DefineRemark{validmatchnumdistinctcommands}{575}
\DefineRemark{validprivatekeyinvalidnumdistinctmatchestotal}{10175}
\DefineRemark{validprivatekeyinvalidpctdistinctmatchestotal}{16}
\DefineRemark{validprivatekeyvalidnumdistinctmatchestotal}{52107}
\DefineRemark{validprivatekeyvalidpctdistinctmatchestotal}{84}
\DefineRemark{validvalapicloudinvalidnumdistinctmatches}{17}
\DefineRemark{validvalapicloudinvalidpctdistinctmatches}{20}
\DefineRemark{validvalapicloudvalidnumdistinctmatches}{67}
\DefineRemark{validvalapicloudvalidpctdistinctmatches}{80}
\DefineRemark{validvalapifinancialvalidnumdistinctmatches}{2}
\DefineRemark{validvalapifinancialvalidpctdistinctmatches}{100}
\DefineRemark{validvalapisocialmediainvalidnumdistinctmatches}{4}
\DefineRemark{validvalapisocialmediainvalidpctdistinctmatches}{50}
\DefineRemark{validvalapisocialmediavalidnumdistinctmatches}{4}
\DefineRemark{validvalapisocialmediavalidpctdistinctmatches}{50}
\DefineRemark{validvalprivatekeyinvalidnumdistinctmatches}{1}
\DefineRemark{validvalprivatekeyinvalidpctdistinctmatches}{100}
\DefineRemark{valprivatekeynumdistinctmatches}{1}

\DefineRemark{private220801avgtagsperimageperdate}{75}
\DefineRemark{private220801maxtagsperimageperdate}{38638}
\DefineRemark{private220801mintagsperimageperdate}{0}
\DefineRemark{private220801numdistinctlayerspermeasurement}{516913}
\DefineRemark{private220801numlayerspermeasurement}{2480137}
\DefineRemark{private220801numselectedtagperdate}{55746}
\DefineRemark{private220801numtagsperdate}{6160615}
\DefineRemark{private220801q25layersmaxbytepermeasurement}{0}
\DefineRemark{private220801q25layersmaxgigabytepermeasurement}{0}
\DefineRemark{private220801q25layersmaxkilobytepermeasurement}{0}
\DefineRemark{private220801q25layersmaxmegabytepermeasurement}{0}
\DefineRemark{private220801q25layersmaxterabytepermeasurement}{0}
\DefineRemark{private220801q25tagsperimageperdate}{1}
\DefineRemark{private220801q50layersmaxbytepermeasurement}{0}
\DefineRemark{private220801q50layersmaxgigabytepermeasurement}{0}
\DefineRemark{private220801q50layersmaxkilobytepermeasurement}{0}
\DefineRemark{private220801q50layersmaxmegabytepermeasurement}{0}
\DefineRemark{private220801q50layersmaxterabytepermeasurement}{0}
\DefineRemark{private220801q50tagsperimageperdate}{1}
\DefineRemark{private220801q75layersmaxbytepermeasurement}{1569}
\DefineRemark{private220801q75layersmaxgigabytepermeasurement}{0}
\DefineRemark{private220801q75layersmaxkilobytepermeasurement}{1}
\DefineRemark{private220801q75layersmaxmegabytepermeasurement}{0}
\DefineRemark{private220801q75layersmaxterabytepermeasurement}{0}
\DefineRemark{private220801q75tagsperimageperdate}{3}
\DefineRemark{private220801sizelayersavgbytepermeasurement}{11692248}
\DefineRemark{private220801sizelayersavggigabytepermeasurement}{0}
\DefineRemark{private220801sizelayersavgkilobytepermeasurement}{11692}
\DefineRemark{private220801sizelayersavgmegabytepermeasurement}{11}
\DefineRemark{private220801sizelayersavgterabytepermeasurement}{0}
\DefineRemark{private220801sizelayersmaxbytepermeasurement}{138031576492}
\DefineRemark{private220801sizelayersmaxgigabytepermeasurement}{138}
\DefineRemark{private220801sizelayersmaxkilobytepermeasurement}{138031576}
\DefineRemark{private220801sizelayersmaxmegabytepermeasurement}{138031}
\DefineRemark{private220801sizelayersmaxterabytepermeasurement}{0}
\DefineRemark{private220801sizelayersminbytepermeasurement}{0}
\DefineRemark{private220801sizelayersmingigabytepermeasurement}{0}
\DefineRemark{private220801sizelayersminkilobytepermeasurement}{0}
\DefineRemark{private220801sizelayersminmegabytepermeasurement}{0}
\DefineRemark{private220801sizelayersminterabytepermeasurement}{0}
\DefineRemark{private220801sizelayerssumbytepermeasurement}{28998378512283}
\DefineRemark{private220801sizelayerssumgigabytepermeasurement}{28998}
\DefineRemark{private220801sizelayerssumkilobytepermeasurement}{28998378512}
\DefineRemark{private220801sizelayerssummegabytepermeasurement}{28998378}
\DefineRemark{private220801sizelayerssumterabytepermeasurement}{28}
\DefineRemark{private220806avgtagsperimageperdate}{81}
\DefineRemark{private220806maxtagsperimageperdate}{38638}
\DefineRemark{private220806mintagsperimageperdate}{0}
\DefineRemark{private220806numdistinctlayerspermeasurement}{491085}
\DefineRemark{private220806numlayerspermeasurement}{2445347}
\DefineRemark{private220806numselectedtagperdate}{55071}
\DefineRemark{private220806numtagsperdate}{6350311}
\DefineRemark{private220806q25layersmaxbytepermeasurement}{0}
\DefineRemark{private220806q25layersmaxgigabytepermeasurement}{0}
\DefineRemark{private220806q25layersmaxkilobytepermeasurement}{0}
\DefineRemark{private220806q25layersmaxmegabytepermeasurement}{0}
\DefineRemark{private220806q25layersmaxterabytepermeasurement}{0}
\DefineRemark{private220806q25tagsperimageperdate}{1}
\DefineRemark{private220806q50layersmaxbytepermeasurement}{0}
\DefineRemark{private220806q50layersmaxgigabytepermeasurement}{0}
\DefineRemark{private220806q50layersmaxkilobytepermeasurement}{0}
\DefineRemark{private220806q50layersmaxmegabytepermeasurement}{0}
\DefineRemark{private220806q50layersmaxterabytepermeasurement}{0}
\DefineRemark{private220806q50tagsperimageperdate}{1}
\DefineRemark{private220806q75layersmaxbytepermeasurement}{1376}
\DefineRemark{private220806q75layersmaxgigabytepermeasurement}{0}
\DefineRemark{private220806q75layersmaxkilobytepermeasurement}{1}
\DefineRemark{private220806q75layersmaxmegabytepermeasurement}{0}
\DefineRemark{private220806q75layersmaxterabytepermeasurement}{0}
\DefineRemark{private220806q75tagsperimageperdate}{3}
\DefineRemark{private220806sizelayersavgbytepermeasurement}{11528778}
\DefineRemark{private220806sizelayersavggigabytepermeasurement}{0}
\DefineRemark{private220806sizelayersavgkilobytepermeasurement}{11528}
\DefineRemark{private220806sizelayersavgmegabytepermeasurement}{11}
\DefineRemark{private220806sizelayersavgterabytepermeasurement}{0}
\DefineRemark{private220806sizelayersmaxbytepermeasurement}{138031576492}
\DefineRemark{private220806sizelayersmaxgigabytepermeasurement}{138}
\DefineRemark{private220806sizelayersmaxkilobytepermeasurement}{138031576}
\DefineRemark{private220806sizelayersmaxmegabytepermeasurement}{138031}
\DefineRemark{private220806sizelayersmaxterabytepermeasurement}{0}
\DefineRemark{private220806sizelayersminbytepermeasurement}{0}
\DefineRemark{private220806sizelayersmingigabytepermeasurement}{0}
\DefineRemark{private220806sizelayersminkilobytepermeasurement}{0}
\DefineRemark{private220806sizelayersminmegabytepermeasurement}{0}
\DefineRemark{private220806sizelayersminterabytepermeasurement}{0}
\DefineRemark{private220806sizelayerssumbytepermeasurement}{28191864283344}
\DefineRemark{private220806sizelayerssumgigabytepermeasurement}{28191}
\DefineRemark{private220806sizelayerssumkilobytepermeasurement}{28191864283}
\DefineRemark{private220806sizelayerssummegabytepermeasurement}{28191864}
\DefineRemark{private220806sizelayerssumterabytepermeasurement}{28}
\DefineRemark{privateavglayersbyteselected}{6750591}
\DefineRemark{privateavglayersgigabyteselected}{0}
\DefineRemark{privateavglayerskilobyteselected}{6750}
\DefineRemark{privateavglayersmegabyteselected}{6}
\DefineRemark{privateavglayersterabyteselected}{0}
\DefineRemark{privatemaxlayersbyteselected}{249809094}
\DefineRemark{privatemaxlayersgigabyteselected}{0}
\DefineRemark{privatemaxlayerskilobyteselected}{249809}
\DefineRemark{privatemaxlayersmegabyteselected}{249}
\DefineRemark{privatemaxlayersterabyteselected}{0}
\DefineRemark{privatemeasurement0date}{22-08-06}
\DefineRemark{privatemeasurement0registry}{95}
\DefineRemark{privatemeasurement0repository}{uhttpd}
\DefineRemark{privatemeasurement0sum}{2596}
\DefineRemark{privatemeasurement1date}{22-08-01}
\DefineRemark{privatemeasurement1registry}{2501}
\DefineRemark{privatemeasurement1repository}{uhttpd}
\DefineRemark{privatemeasurement1sum}{2596}
\DefineRemark{privatemeasurement220801date}{22-08-01}
\DefineRemark{privatemeasurement220801numtotal}{8076}
\DefineRemark{privatemeasurement220801numtotalnontls}{7593}
\DefineRemark{privatemeasurement220801numtotaltls}{483}
\DefineRemark{privatemeasurement220801repositorydistinctperdate}{51163}
\DefineRemark{privatemeasurement220801repositorymax}{100}
\DefineRemark{privatemeasurement220801repositorymean}{10}
\DefineRemark{privatemeasurement220801repositorymin}{1}
\DefineRemark{privatemeasurement220801repositorysum}{81570}
\DefineRemark{privatemeasurement220806date}{22-08-06}
\DefineRemark{privatemeasurement220806numtotal}{5656}
\DefineRemark{privatemeasurement220806numtotalnontls}{5184}
\DefineRemark{privatemeasurement220806numtotaltls}{472}
\DefineRemark{privatemeasurement220806repositorydistinctperdate}{50611}
\DefineRemark{privatemeasurement220806repositorymax}{100}
\DefineRemark{privatemeasurement220806repositorymean}{13}
\DefineRemark{privatemeasurement220806repositorymin}{1}
\DefineRemark{privatemeasurement220806repositorysum}{77900}
\DefineRemark{privatemeasurement2date}{22-08-06}
\DefineRemark{privatemeasurement2registry}{403}
\DefineRemark{privatemeasurement2repository}{nginx}
\DefineRemark{privatemeasurement2sum}{812}
\DefineRemark{privatemeasurement3date}{22-08-01}
\DefineRemark{privatemeasurement3registry}{409}
\DefineRemark{privatemeasurement3repository}{nginx}
\DefineRemark{privatemeasurement3sum}{812}
\DefineRemark{privatemeasurement4date}{22-08-06}
\DefineRemark{privatemeasurement4registry}{204}
\DefineRemark{privatemeasurement4repository}{redis}
\DefineRemark{privatemeasurement4sum}{409}
\DefineRemark{privatemeasurement5date}{22-08-01}
\DefineRemark{privatemeasurement5registry}{205}
\DefineRemark{privatemeasurement5repository}{redis}
\DefineRemark{privatemeasurement5sum}{409}
\DefineRemark{privatemeasurement6date}{22-08-06}
\DefineRemark{privatemeasurement6registry}{152}
\DefineRemark{privatemeasurement6repository}{busybox}
\DefineRemark{privatemeasurement6sum}{307}
\DefineRemark{privatemeasurement7date}{22-08-01}
\DefineRemark{privatemeasurement7registry}{155}
\DefineRemark{privatemeasurement7repository}{busybox}
\DefineRemark{privatemeasurement7sum}{307}
\DefineRemark{privatemeasurementdatemax}{22-08-01}
\DefineRemark{privatemeasurementdatemin}{22-08-06}
\DefineRemark{privatemeasurementmostreoccurringrepo}{uhttpd}
\DefineRemark{privatemeasurementmostreoccurringrepotimes}{2580}
\DefineRemark{privatemeasurementnumdistinctrepositoriesoverall}{53448}
\DefineRemark{privatemeasurementnumtotalmax}{8076}
\DefineRemark{privatemeasurementnumtotalmin}{5656}
\DefineRemark{privatemeasurementnumtotalnontlsmax}{7593}
\DefineRemark{privatemeasurementnumtotalnontlsmin}{5184}
\DefineRemark{privatemeasurementnumtotaltlsmax}{483}
\DefineRemark{privatemeasurementnumtotaltlsmin}{472}
\DefineRemark{privateminlayersbyteselected}{87}
\DefineRemark{privateminlayersgigabyteselected}{0}
\DefineRemark{privateminlayerskilobyteselected}{0}
\DefineRemark{privateminlayersmegabyteselected}{0}
\DefineRemark{privateminlayersterabyteselected}{0}
\DefineRemark{privatenumdistinctlayers}{551617}
\DefineRemark{privatenumdistinctlayersselected}{258889}
\DefineRemark{privatenumlayersselected}{5081615543739}
\DefineRemark{privateq25layersbyteselected}{341}
\DefineRemark{privateq25layersgigabyteselected}{0}
\DefineRemark{privateq25layerskilobyteselected}{0}
\DefineRemark{privateq25layersmegabyteselected}{0}
\DefineRemark{privateq25layersterabyteselected}{0}
\DefineRemark{privateq50layersbyteselected}{1732}
\DefineRemark{privateq50layersgigabyteselected}{0}
\DefineRemark{privateq50layerskilobyteselected}{1}
\DefineRemark{privateq50layersmegabyteselected}{0}
\DefineRemark{privateq50layersterabyteselected}{0}
\DefineRemark{privateq75layersbyteselected}{1967432}
\DefineRemark{privateq75layersgigabyteselected}{0}
\DefineRemark{privateq75layerskilobyteselected}{1967}
\DefineRemark{privateq75layersmegabyteselected}{1}
\DefineRemark{privateq75layersterabyteselected}{0}
\DefineRemark{tagspermeasurement220801numlatest}{34377}
\DefineRemark{tagspermeasurement220801numnone}{3704}
\DefineRemark{tagspermeasurement220806numlatest}{31997}
\DefineRemark{tagspermeasurement220806numnone}{3289}

\DefineRemark{20210501corrFingerprint}{535}
\DefineRemark{20210501numuniquehosts}{243419}
\DefineRemark{20210601corrFingerprint}{555}
\DefineRemark{20210601numuniquehosts}{245237}
\DefineRemark{20210701corrFingerprint}{614}
\DefineRemark{20210701numuniquehosts}{245538}
\DefineRemark{20210801corrFingerprint}{628}
\DefineRemark{20210801numuniquehosts}{253309}
\DefineRemark{20210901corrFingerprint}{641}
\DefineRemark{20210901numuniquehosts}{266174}
\DefineRemark{20211001corrFingerprint}{653}
\DefineRemark{20211001numuniquehosts}{285460}
\DefineRemark{20211101corrFingerprint}{641}
\DefineRemark{20211101numuniquehosts}{281342}
\DefineRemark{20211201corrFingerprint}{656}
\DefineRemark{20211201numuniquehosts}{292063}
\DefineRemark{20220101corrFingerprint}{650}
\DefineRemark{20220101numuniquehosts}{270333}
\DefineRemark{20220201corrFingerprint}{675}
\DefineRemark{20220201numuniquehosts}{272328}
\DefineRemark{20220301corrFingerprint}{701}
\DefineRemark{20220301numuniquehosts}{272583}
\DefineRemark{20220401corrFingerprint}{697}
\DefineRemark{20220401numuniquehosts}{275399}
\DefineRemark{20220501corrFingerprint}{711}
\DefineRemark{20220501numuniquehosts}{275380}
\DefineRemark{20220601corrFingerprint}{726}
\DefineRemark{20220601numuniquehosts}{274118}
\DefineRemark{20220701corrFingerprint}{732}
\DefineRemark{20220701numuniquehosts}{275066}
\DefineRemark{20220801corrFingerprint}{751}
\DefineRemark{20220801numuniquehosts}{276519}
\DefineRemark{20220901corrFingerprint}{740}
\DefineRemark{20220901numuniquehosts}{275269}
\DefineRemark{AMQP20210501corrFingerprint}{0}
\DefineRemark{AMQP20210501numuniquehosts}{0}
\DefineRemark{AMQP20210601corrFingerprint}{0}
\DefineRemark{AMQP20210601numuniquehosts}{0}
\DefineRemark{AMQP20210701corrFingerprint}{2}
\DefineRemark{AMQP20210701numuniquehosts}{21}
\DefineRemark{AMQP20210801corrFingerprint}{2}
\DefineRemark{AMQP20210801numuniquehosts}{21}
\DefineRemark{AMQP20210901corrFingerprint}{2}
\DefineRemark{AMQP20210901numuniquehosts}{22}
\DefineRemark{AMQP20211001corrFingerprint}{2}
\DefineRemark{AMQP20211001numuniquehosts}{22}
\DefineRemark{AMQP20211101corrFingerprint}{2}
\DefineRemark{AMQP20211101numuniquehosts}{26}
\DefineRemark{AMQP20211201corrFingerprint}{2}
\DefineRemark{AMQP20211201numuniquehosts}{27}
\DefineRemark{AMQP20220101corrFingerprint}{1}
\DefineRemark{AMQP20220101numuniquehosts}{12}
\DefineRemark{AMQP20220201corrFingerprint}{1}
\DefineRemark{AMQP20220201numuniquehosts}{12}
\DefineRemark{AMQP20220301corrFingerprint}{2}
\DefineRemark{AMQP20220301numuniquehosts}{18}
\DefineRemark{AMQP20220401corrFingerprint}{2}
\DefineRemark{AMQP20220401numuniquehosts}{18}
\DefineRemark{AMQP20220501corrFingerprint}{2}
\DefineRemark{AMQP20220501numuniquehosts}{20}
\DefineRemark{AMQP20220601corrFingerprint}{2}
\DefineRemark{AMQP20220601numuniquehosts}{17}
\DefineRemark{AMQP20220701corrFingerprint}{2}
\DefineRemark{AMQP20220701numuniquehosts}{22}
\DefineRemark{AMQP20220801corrFingerprint}{2}
\DefineRemark{AMQP20220801numuniquehosts}{21}
\DefineRemark{AMQP20220901corrFingerprint}{2}
\DefineRemark{AMQP20220901numuniquehosts}{19}
\DefineRemark{Elasticsearch20210501corrFingerprint}{1}
\DefineRemark{Elasticsearch20210501numuniquehosts}{4}
\DefineRemark{Elasticsearch20210601corrFingerprint}{1}
\DefineRemark{Elasticsearch20210601numuniquehosts}{4}
\DefineRemark{Elasticsearch20210701corrFingerprint}{1}
\DefineRemark{Elasticsearch20210701numuniquehosts}{4}
\DefineRemark{Elasticsearch20210801corrFingerprint}{1}
\DefineRemark{Elasticsearch20210801numuniquehosts}{4}
\DefineRemark{Elasticsearch20210901corrFingerprint}{1}
\DefineRemark{Elasticsearch20210901numuniquehosts}{5}
\DefineRemark{Elasticsearch20211001corrFingerprint}{1}
\DefineRemark{Elasticsearch20211001numuniquehosts}{5}
\DefineRemark{Elasticsearch20211101corrFingerprint}{1}
\DefineRemark{Elasticsearch20211101numuniquehosts}{6}
\DefineRemark{Elasticsearch20211201corrFingerprint}{1}
\DefineRemark{Elasticsearch20211201numuniquehosts}{5}
\DefineRemark{Elasticsearch20220101corrFingerprint}{1}
\DefineRemark{Elasticsearch20220101numuniquehosts}{5}
\DefineRemark{Elasticsearch20220201corrFingerprint}{1}
\DefineRemark{Elasticsearch20220201numuniquehosts}{5}
\DefineRemark{Elasticsearch20220301corrFingerprint}{1}
\DefineRemark{Elasticsearch20220301numuniquehosts}{5}
\DefineRemark{Elasticsearch20220401corrFingerprint}{1}
\DefineRemark{Elasticsearch20220401numuniquehosts}{4}
\DefineRemark{Elasticsearch20220501corrFingerprint}{1}
\DefineRemark{Elasticsearch20220501numuniquehosts}{5}
\DefineRemark{Elasticsearch20220601corrFingerprint}{1}
\DefineRemark{Elasticsearch20220601numuniquehosts}{4}
\DefineRemark{Elasticsearch20220701corrFingerprint}{1}
\DefineRemark{Elasticsearch20220701numuniquehosts}{5}
\DefineRemark{Elasticsearch20220801corrFingerprint}{1}
\DefineRemark{Elasticsearch20220801numuniquehosts}{4}
\DefineRemark{Elasticsearch20220901corrFingerprint}{1}
\DefineRemark{Elasticsearch20220901numuniquehosts}{3}
\DefineRemark{FTP20210501corrFingerprint}{15}
\DefineRemark{FTP20210501numuniquehosts}{8571}
\DefineRemark{FTP20210601corrFingerprint}{14}
\DefineRemark{FTP20210601numuniquehosts}{8572}
\DefineRemark{FTP20210701corrFingerprint}{16}
\DefineRemark{FTP20210701numuniquehosts}{8257}
\DefineRemark{FTP20210801corrFingerprint}{18}
\DefineRemark{FTP20210801numuniquehosts}{8352}
\DefineRemark{FTP20210901corrFingerprint}{19}
\DefineRemark{FTP20210901numuniquehosts}{8400}
\DefineRemark{FTP20211001corrFingerprint}{19}
\DefineRemark{FTP20211001numuniquehosts}{8799}
\DefineRemark{FTP20211101corrFingerprint}{19}
\DefineRemark{FTP20211101numuniquehosts}{8653}
\DefineRemark{FTP20211201corrFingerprint}{20}
\DefineRemark{FTP20211201numuniquehosts}{9153}
\DefineRemark{FTP20220101corrFingerprint}{15}
\DefineRemark{FTP20220101numuniquehosts}{8130}
\DefineRemark{FTP20220201corrFingerprint}{16}
\DefineRemark{FTP20220201numuniquehosts}{9696}
\DefineRemark{FTP20220301corrFingerprint}{22}
\DefineRemark{FTP20220301numuniquehosts}{7793}
\DefineRemark{FTP20220401corrFingerprint}{22}
\DefineRemark{FTP20220401numuniquehosts}{7760}
\DefineRemark{FTP20220501corrFingerprint}{22}
\DefineRemark{FTP20220501numuniquehosts}{7506}
\DefineRemark{FTP20220601corrFingerprint}{22}
\DefineRemark{FTP20220601numuniquehosts}{7194}
\DefineRemark{FTP20220701corrFingerprint}{23}
\DefineRemark{FTP20220701numuniquehosts}{7183}
\DefineRemark{FTP20220801corrFingerprint}{22}
\DefineRemark{FTP20220801numuniquehosts}{6882}
\DefineRemark{FTP20220901corrFingerprint}{23}
\DefineRemark{FTP20220901numuniquehosts}{6672}
\DefineRemark{HTTP20210501corrFingerprint}{432}
\DefineRemark{HTTP20210501numuniquehosts}{210250}
\DefineRemark{HTTP20210601corrFingerprint}{448}
\DefineRemark{HTTP20210601numuniquehosts}{211851}
\DefineRemark{HTTP20210701corrFingerprint}{452}
\DefineRemark{HTTP20210701numuniquehosts}{212158}
\DefineRemark{HTTP20210801corrFingerprint}{456}
\DefineRemark{HTTP20210801numuniquehosts}{208769}
\DefineRemark{HTTP20210901corrFingerprint}{456}
\DefineRemark{HTTP20210901numuniquehosts}{220757}
\DefineRemark{HTTP20211001corrFingerprint}{468}
\DefineRemark{HTTP20211001numuniquehosts}{238526}
\DefineRemark{HTTP20211101corrFingerprint}{465}
\DefineRemark{HTTP20211101numuniquehosts}{231970}
\DefineRemark{HTTP20211201corrFingerprint}{483}
\DefineRemark{HTTP20211201numuniquehosts}{240913}
\DefineRemark{HTTP20220101corrFingerprint}{481}
\DefineRemark{HTTP20220101numuniquehosts}{221056}
\DefineRemark{HTTP20220201corrFingerprint}{488}
\DefineRemark{HTTP20220201numuniquehosts}{218272}
\DefineRemark{HTTP20220301corrFingerprint}{501}
\DefineRemark{HTTP20220301numuniquehosts}{222180}
\DefineRemark{HTTP20220401corrFingerprint}{498}
\DefineRemark{HTTP20220401numuniquehosts}{224565}
\DefineRemark{HTTP20220501corrFingerprint}{506}
\DefineRemark{HTTP20220501numuniquehosts}{223449}
\DefineRemark{HTTP20220601corrFingerprint}{516}
\DefineRemark{HTTP20220601numuniquehosts}{221098}
\DefineRemark{HTTP20220701corrFingerprint}{519}
\DefineRemark{HTTP20220701numuniquehosts}{219351}
\DefineRemark{HTTP20220801corrFingerprint}{535}
\DefineRemark{HTTP20220801numuniquehosts}{216346}
\DefineRemark{HTTP20220901corrFingerprint}{515}
\DefineRemark{HTTP20220901numuniquehosts}{213573}
\DefineRemark{IMAP20210501corrFingerprint}{25}
\DefineRemark{IMAP20210501numuniquehosts}{589}
\DefineRemark{IMAP20210601corrFingerprint}{26}
\DefineRemark{IMAP20210601numuniquehosts}{669}
\DefineRemark{IMAP20210701corrFingerprint}{26}
\DefineRemark{IMAP20210701numuniquehosts}{1004}
\DefineRemark{IMAP20210801corrFingerprint}{26}
\DefineRemark{IMAP20210801numuniquehosts}{1339}
\DefineRemark{IMAP20210901corrFingerprint}{26}
\DefineRemark{IMAP20210901numuniquehosts}{1266}
\DefineRemark{IMAP20211001corrFingerprint}{25}
\DefineRemark{IMAP20211001numuniquehosts}{1237}
\DefineRemark{IMAP20211101corrFingerprint}{23}
\DefineRemark{IMAP20211101numuniquehosts}{1227}
\DefineRemark{IMAP20211201corrFingerprint}{25}
\DefineRemark{IMAP20211201numuniquehosts}{1192}
\DefineRemark{IMAP20220101corrFingerprint}{26}
\DefineRemark{IMAP20220101numuniquehosts}{1118}
\DefineRemark{IMAP20220201corrFingerprint}{26}
\DefineRemark{IMAP20220201numuniquehosts}{1214}
\DefineRemark{IMAP20220301corrFingerprint}{24}
\DefineRemark{IMAP20220301numuniquehosts}{1138}
\DefineRemark{IMAP20220401corrFingerprint}{24}
\DefineRemark{IMAP20220401numuniquehosts}{1316}
\DefineRemark{IMAP20220501corrFingerprint}{25}
\DefineRemark{IMAP20220501numuniquehosts}{1384}
\DefineRemark{IMAP20220601corrFingerprint}{24}
\DefineRemark{IMAP20220601numuniquehosts}{1566}
\DefineRemark{IMAP20220701corrFingerprint}{23}
\DefineRemark{IMAP20220701numuniquehosts}{1674}
\DefineRemark{IMAP20220801corrFingerprint}{23}
\DefineRemark{IMAP20220801numuniquehosts}{1751}
\DefineRemark{IMAP20220901corrFingerprint}{24}
\DefineRemark{IMAP20220901numuniquehosts}{1798}
\DefineRemark{Kubernetes20210501corrFingerprint}{0}
\DefineRemark{Kubernetes20210501numuniquehosts}{0}
\DefineRemark{Kubernetes20210601corrFingerprint}{0}
\DefineRemark{Kubernetes20210601numuniquehosts}{0}
\DefineRemark{Kubernetes20210701corrFingerprint}{0}
\DefineRemark{Kubernetes20210701numuniquehosts}{0}
\DefineRemark{Kubernetes20210801corrFingerprint}{0}
\DefineRemark{Kubernetes20210801numuniquehosts}{0}
\DefineRemark{Kubernetes20210901corrFingerprint}{0}
\DefineRemark{Kubernetes20210901numuniquehosts}{0}
\DefineRemark{Kubernetes20211001corrFingerprint}{0}
\DefineRemark{Kubernetes20211001numuniquehosts}{0}
\DefineRemark{Kubernetes20211101corrFingerprint}{0}
\DefineRemark{Kubernetes20211101numuniquehosts}{0}
\DefineRemark{Kubernetes20211201corrFingerprint}{0}
\DefineRemark{Kubernetes20211201numuniquehosts}{0}
\DefineRemark{Kubernetes20220101corrFingerprint}{0}
\DefineRemark{Kubernetes20220101numuniquehosts}{0}
\DefineRemark{Kubernetes20220201corrFingerprint}{0}
\DefineRemark{Kubernetes20220201numuniquehosts}{0}
\DefineRemark{Kubernetes20220301corrFingerprint}{0}
\DefineRemark{Kubernetes20220301numuniquehosts}{0}
\DefineRemark{Kubernetes20220401corrFingerprint}{0}
\DefineRemark{Kubernetes20220401numuniquehosts}{0}
\DefineRemark{Kubernetes20220501corrFingerprint}{0}
\DefineRemark{Kubernetes20220501numuniquehosts}{0}
\DefineRemark{Kubernetes20220601corrFingerprint}{0}
\DefineRemark{Kubernetes20220601numuniquehosts}{0}
\DefineRemark{Kubernetes20220701corrFingerprint}{0}
\DefineRemark{Kubernetes20220701numuniquehosts}{0}
\DefineRemark{Kubernetes20220801corrFingerprint}{0}
\DefineRemark{Kubernetes20220801numuniquehosts}{0}
\DefineRemark{Kubernetes20220901corrFingerprint}{12}
\DefineRemark{Kubernetes20220901numuniquehosts}{24}
\DefineRemark{LDAP20210501corrFingerprint}{0}
\DefineRemark{LDAP20210501numuniquehosts}{0}
\DefineRemark{LDAP20210601corrFingerprint}{0}
\DefineRemark{LDAP20210601numuniquehosts}{0}
\DefineRemark{LDAP20210701corrFingerprint}{0}
\DefineRemark{LDAP20210701numuniquehosts}{0}
\DefineRemark{LDAP20210801corrFingerprint}{0}
\DefineRemark{LDAP20210801numuniquehosts}{0}
\DefineRemark{LDAP20210901corrFingerprint}{0}
\DefineRemark{LDAP20210901numuniquehosts}{0}
\DefineRemark{LDAP20211001corrFingerprint}{0}
\DefineRemark{LDAP20211001numuniquehosts}{0}
\DefineRemark{LDAP20211101corrFingerprint}{0}
\DefineRemark{LDAP20211101numuniquehosts}{0}
\DefineRemark{LDAP20211201corrFingerprint}{0}
\DefineRemark{LDAP20211201numuniquehosts}{0}
\DefineRemark{LDAP20220101corrFingerprint}{0}
\DefineRemark{LDAP20220101numuniquehosts}{0}
\DefineRemark{LDAP20220201corrFingerprint}{1}
\DefineRemark{LDAP20220201numuniquehosts}{19}
\DefineRemark{LDAP20220301corrFingerprint}{1}
\DefineRemark{LDAP20220301numuniquehosts}{51}
\DefineRemark{LDAP20220401corrFingerprint}{1}
\DefineRemark{LDAP20220401numuniquehosts}{50}
\DefineRemark{LDAP20220501corrFingerprint}{1}
\DefineRemark{LDAP20220501numuniquehosts}{44}
\DefineRemark{LDAP20220601corrFingerprint}{1}
\DefineRemark{LDAP20220601numuniquehosts}{47}
\DefineRemark{LDAP20220701corrFingerprint}{1}
\DefineRemark{LDAP20220701numuniquehosts}{47}
\DefineRemark{LDAP20220801corrFingerprint}{1}
\DefineRemark{LDAP20220801numuniquehosts}{42}
\DefineRemark{LDAP20220901corrFingerprint}{1}
\DefineRemark{LDAP20220901numuniquehosts}{37}
\DefineRemark{MQTT20210501corrFingerprint}{12}
\DefineRemark{MQTT20210501numuniquehosts}{5164}
\DefineRemark{MQTT20210601corrFingerprint}{13}
\DefineRemark{MQTT20210601numuniquehosts}{5437}
\DefineRemark{MQTT20210701corrFingerprint}{15}
\DefineRemark{MQTT20210701numuniquehosts}{5659}
\DefineRemark{MQTT20210801corrFingerprint}{15}
\DefineRemark{MQTT20210801numuniquehosts}{5809}
\DefineRemark{MQTT20210901corrFingerprint}{16}
\DefineRemark{MQTT20210901numuniquehosts}{6170}
\DefineRemark{MQTT20211001corrFingerprint}{16}
\DefineRemark{MQTT20211001numuniquehosts}{6381}
\DefineRemark{MQTT20211101corrFingerprint}{16}
\DefineRemark{MQTT20211101numuniquehosts}{6688}
\DefineRemark{MQTT20211201corrFingerprint}{15}
\DefineRemark{MQTT20211201numuniquehosts}{7052}
\DefineRemark{MQTT20220101corrFingerprint}{14}
\DefineRemark{MQTT20220101numuniquehosts}{7076}
\DefineRemark{MQTT20220201corrFingerprint}{14}
\DefineRemark{MQTT20220201numuniquehosts}{7271}
\DefineRemark{MQTT20220301corrFingerprint}{14}
\DefineRemark{MQTT20220301numuniquehosts}{7195}
\DefineRemark{MQTT20220401corrFingerprint}{16}
\DefineRemark{MQTT20220401numuniquehosts}{7784}
\DefineRemark{MQTT20220501corrFingerprint}{15}
\DefineRemark{MQTT20220501numuniquehosts}{8075}
\DefineRemark{MQTT20220601corrFingerprint}{15}
\DefineRemark{MQTT20220601numuniquehosts}{8199}
\DefineRemark{MQTT20220701corrFingerprint}{15}
\DefineRemark{MQTT20220701numuniquehosts}{8353}
\DefineRemark{MQTT20220801corrFingerprint}{16}
\DefineRemark{MQTT20220801numuniquehosts}{8442}
\DefineRemark{MQTT20220901corrFingerprint}{17}
\DefineRemark{MQTT20220901numuniquehosts}{8674}
\DefineRemark{Misc20210501corrFingerprint}{105}
\DefineRemark{Misc20210501numuniquehosts}{5225}
\DefineRemark{Misc20210601corrFingerprint}{105}
\DefineRemark{Misc20210601numuniquehosts}{5165}
\DefineRemark{Misc20210701corrFingerprint}{105}
\DefineRemark{Misc20210701numuniquehosts}{4727}
\DefineRemark{Misc20210801corrFingerprint}{102}
\DefineRemark{Misc20210801numuniquehosts}{14950}
\DefineRemark{Misc20210901corrFingerprint}{97}
\DefineRemark{Misc20210901numuniquehosts}{16255}
\DefineRemark{Misc20211001corrFingerprint}{98}
\DefineRemark{Misc20211001numuniquehosts}{17206}
\DefineRemark{Misc20211101corrFingerprint}{81}
\DefineRemark{Misc20211101numuniquehosts}{19491}
\DefineRemark{Misc20211201corrFingerprint}{88}
\DefineRemark{Misc20211201numuniquehosts}{20412}
\DefineRemark{Misc20220101corrFingerprint}{76}
\DefineRemark{Misc20220101numuniquehosts}{20371}
\DefineRemark{Misc20220201corrFingerprint}{85}
\DefineRemark{Misc20220201numuniquehosts}{20988}
\DefineRemark{Misc20220301corrFingerprint}{83}
\DefineRemark{Misc20220301numuniquehosts}{21945}
\DefineRemark{Misc20220401corrFingerprint}{76}
\DefineRemark{Misc20220401numuniquehosts}{21609}
\DefineRemark{Misc20220501corrFingerprint}{81}
\DefineRemark{Misc20220501numuniquehosts}{23283}
\DefineRemark{Misc20220601corrFingerprint}{86}
\DefineRemark{Misc20220601numuniquehosts}{24835}
\DefineRemark{Misc20220701corrFingerprint}{85}
\DefineRemark{Misc20220701numuniquehosts}{27242}
\DefineRemark{Misc20220801corrFingerprint}{88}
\DefineRemark{Misc20220801numuniquehosts}{30975}
\DefineRemark{Misc20220901corrFingerprint}{122}
\DefineRemark{Misc20220901numuniquehosts}{33891}
\DefineRemark{MySQL20210501corrFingerprint}{1}
\DefineRemark{MySQL20210501numuniquehosts}{1}
\DefineRemark{MySQL20210601corrFingerprint}{1}
\DefineRemark{MySQL20210601numuniquehosts}{1}
\DefineRemark{MySQL20210701corrFingerprint}{1}
\DefineRemark{MySQL20210701numuniquehosts}{1}
\DefineRemark{MySQL20210801corrFingerprint}{1}
\DefineRemark{MySQL20210801numuniquehosts}{1}
\DefineRemark{MySQL20210901corrFingerprint}{2}
\DefineRemark{MySQL20210901numuniquehosts}{2}
\DefineRemark{MySQL20211001corrFingerprint}{2}
\DefineRemark{MySQL20211001numuniquehosts}{2}
\DefineRemark{MySQL20211101corrFingerprint}{3}
\DefineRemark{MySQL20211101numuniquehosts}{3}
\DefineRemark{MySQL20211201corrFingerprint}{2}
\DefineRemark{MySQL20211201numuniquehosts}{2}
\DefineRemark{MySQL20220101corrFingerprint}{1}
\DefineRemark{MySQL20220101numuniquehosts}{1}
\DefineRemark{MySQL20220201corrFingerprint}{1}
\DefineRemark{MySQL20220201numuniquehosts}{1}
\DefineRemark{MySQL20220301corrFingerprint}{2}
\DefineRemark{MySQL20220301numuniquehosts}{3}
\DefineRemark{MySQL20220401corrFingerprint}{2}
\DefineRemark{MySQL20220401numuniquehosts}{3}
\DefineRemark{MySQL20220501corrFingerprint}{1}
\DefineRemark{MySQL20220501numuniquehosts}{1}
\DefineRemark{MySQL20220601corrFingerprint}{1}
\DefineRemark{MySQL20220601numuniquehosts}{1}
\DefineRemark{MySQL20220701corrFingerprint}{3}
\DefineRemark{MySQL20220701numuniquehosts}{3}
\DefineRemark{MySQL20220801corrFingerprint}{3}
\DefineRemark{MySQL20220801numuniquehosts}{3}
\DefineRemark{MySQL20220901corrFingerprint}{3}
\DefineRemark{MySQL20220901numuniquehosts}{3}
\DefineRemark{POP320210501corrFingerprint}{11}
\DefineRemark{POP320210501numuniquehosts}{397}
\DefineRemark{POP320210601corrFingerprint}{13}
\DefineRemark{POP320210601numuniquehosts}{471}
\DefineRemark{POP320210701corrFingerprint}{12}
\DefineRemark{POP320210701numuniquehosts}{794}
\DefineRemark{POP320210801corrFingerprint}{13}
\DefineRemark{POP320210801numuniquehosts}{1053}
\DefineRemark{POP320210901corrFingerprint}{11}
\DefineRemark{POP320210901numuniquehosts}{883}
\DefineRemark{POP320211001corrFingerprint}{13}
\DefineRemark{POP320211001numuniquehosts}{845}
\DefineRemark{POP320211101corrFingerprint}{12}
\DefineRemark{POP320211101numuniquehosts}{837}
\DefineRemark{POP320211201corrFingerprint}{11}
\DefineRemark{POP320211201numuniquehosts}{794}
\DefineRemark{POP320220101corrFingerprint}{12}
\DefineRemark{POP320220101numuniquehosts}{712}
\DefineRemark{POP320220201corrFingerprint}{13}
\DefineRemark{POP320220201numuniquehosts}{843}
\DefineRemark{POP320220301corrFingerprint}{11}
\DefineRemark{POP320220301numuniquehosts}{757}
\DefineRemark{POP320220401corrFingerprint}{10}
\DefineRemark{POP320220401numuniquehosts}{938}
\DefineRemark{POP320220501corrFingerprint}{10}
\DefineRemark{POP320220501numuniquehosts}{1051}
\DefineRemark{POP320220601corrFingerprint}{11}
\DefineRemark{POP320220601numuniquehosts}{1276}
\DefineRemark{POP320220701corrFingerprint}{10}
\DefineRemark{POP320220701numuniquehosts}{1387}
\DefineRemark{POP320220801corrFingerprint}{12}
\DefineRemark{POP320220801numuniquehosts}{1464}
\DefineRemark{POP320220901corrFingerprint}{11}
\DefineRemark{POP320220901numuniquehosts}{1516}
\DefineRemark{PostgreSQL20210501corrFingerprint}{27}
\DefineRemark{PostgreSQL20210501numuniquehosts}{531}
\DefineRemark{PostgreSQL20210601corrFingerprint}{30}
\DefineRemark{PostgreSQL20210601numuniquehosts}{536}
\DefineRemark{PostgreSQL20210701corrFingerprint}{33}
\DefineRemark{PostgreSQL20210701numuniquehosts}{482}
\DefineRemark{PostgreSQL20210801corrFingerprint}{38}
\DefineRemark{PostgreSQL20210801numuniquehosts}{453}
\DefineRemark{PostgreSQL20210901corrFingerprint}{41}
\DefineRemark{PostgreSQL20210901numuniquehosts}{449}
\DefineRemark{PostgreSQL20211001corrFingerprint}{47}
\DefineRemark{PostgreSQL20211001numuniquehosts}{461}
\DefineRemark{PostgreSQL20211101corrFingerprint}{42}
\DefineRemark{PostgreSQL20211101numuniquehosts}{462}
\DefineRemark{PostgreSQL20211201corrFingerprint}{41}
\DefineRemark{PostgreSQL20211201numuniquehosts}{443}
\DefineRemark{PostgreSQL20220101corrFingerprint}{45}
\DefineRemark{PostgreSQL20220101numuniquehosts}{425}
\DefineRemark{PostgreSQL20220201corrFingerprint}{45}
\DefineRemark{PostgreSQL20220201numuniquehosts}{481}
\DefineRemark{PostgreSQL20220301corrFingerprint}{44}
\DefineRemark{PostgreSQL20220301numuniquehosts}{444}
\DefineRemark{PostgreSQL20220401corrFingerprint}{48}
\DefineRemark{PostgreSQL20220401numuniquehosts}{446}
\DefineRemark{PostgreSQL20220501corrFingerprint}{52}
\DefineRemark{PostgreSQL20220501numuniquehosts}{438}
\DefineRemark{PostgreSQL20220601corrFingerprint}{53}
\DefineRemark{PostgreSQL20220601numuniquehosts}{419}
\DefineRemark{PostgreSQL20220701corrFingerprint}{50}
\DefineRemark{PostgreSQL20220701numuniquehosts}{424}
\DefineRemark{PostgreSQL20220801corrFingerprint}{52}
\DefineRemark{PostgreSQL20220801numuniquehosts}{422}
\DefineRemark{PostgreSQL20220901corrFingerprint}{50}
\DefineRemark{PostgreSQL20220901numuniquehosts}{426}
\DefineRemark{SIP20210501corrFingerprint}{9}
\DefineRemark{SIP20210501numuniquehosts}{287}
\DefineRemark{SIP20210601corrFingerprint}{9}
\DefineRemark{SIP20210601numuniquehosts}{319}
\DefineRemark{SIP20210701corrFingerprint}{9}
\DefineRemark{SIP20210701numuniquehosts}{260}
\DefineRemark{SIP20210801corrFingerprint}{9}
\DefineRemark{SIP20210801numuniquehosts}{279}
\DefineRemark{SIP20210901corrFingerprint}{8}
\DefineRemark{SIP20210901numuniquehosts}{267}
\DefineRemark{SIP20211001corrFingerprint}{9}
\DefineRemark{SIP20211001numuniquehosts}{259}
\DefineRemark{SIP20211101corrFingerprint}{8}
\DefineRemark{SIP20211101numuniquehosts}{271}
\DefineRemark{SIP20211201corrFingerprint}{8}
\DefineRemark{SIP20211201numuniquehosts}{251}
\DefineRemark{SIP20220101corrFingerprint}{7}
\DefineRemark{SIP20220101numuniquehosts}{249}
\DefineRemark{SIP20220201corrFingerprint}{7}
\DefineRemark{SIP20220201numuniquehosts}{249}
\DefineRemark{SIP20220301corrFingerprint}{8}
\DefineRemark{SIP20220301numuniquehosts}{243}
\DefineRemark{SIP20220401corrFingerprint}{7}
\DefineRemark{SIP20220401numuniquehosts}{250}
\DefineRemark{SIP20220501corrFingerprint}{7}
\DefineRemark{SIP20220501numuniquehosts}{243}
\DefineRemark{SIP20220601corrFingerprint}{7}
\DefineRemark{SIP20220601numuniquehosts}{255}
\DefineRemark{SIP20220701corrFingerprint}{8}
\DefineRemark{SIP20220701numuniquehosts}{233}
\DefineRemark{SIP20220801corrFingerprint}{8}
\DefineRemark{SIP20220801numuniquehosts}{210}
\DefineRemark{SIP20220901corrFingerprint}{8}
\DefineRemark{SIP20220901numuniquehosts}{216}
\DefineRemark{SMTP20210501corrFingerprint}{48}
\DefineRemark{SMTP20210501numuniquehosts}{12400}
\DefineRemark{SMTP20210601corrFingerprint}{48}
\DefineRemark{SMTP20210601numuniquehosts}{12212}
\DefineRemark{SMTP20210701corrFingerprint}{49}
\DefineRemark{SMTP20210701numuniquehosts}{12020}
\DefineRemark{SMTP20210801corrFingerprint}{52}
\DefineRemark{SMTP20210801numuniquehosts}{12116}
\DefineRemark{SMTP20210901corrFingerprint}{54}
\DefineRemark{SMTP20210901numuniquehosts}{11488}
\DefineRemark{SMTP20211001corrFingerprint}{53}
\DefineRemark{SMTP20211001numuniquehosts}{11487}
\DefineRemark{SMTP20211101corrFingerprint}{49}
\DefineRemark{SMTP20211101numuniquehosts}{11501}
\DefineRemark{SMTP20211201corrFingerprint}{50}
\DefineRemark{SMTP20211201numuniquehosts}{11590}
\DefineRemark{SMTP20220101corrFingerprint}{48}
\DefineRemark{SMTP20220101numuniquehosts}{10910}
\DefineRemark{SMTP20220201corrFingerprint}{50}
\DefineRemark{SMTP20220201numuniquehosts}{11657}
\DefineRemark{SMTP20220301corrFingerprint}{50}
\DefineRemark{SMTP20220301numuniquehosts}{10534}
\DefineRemark{SMTP20220401corrFingerprint}{49}
\DefineRemark{SMTP20220401numuniquehosts}{10365}
\DefineRemark{SMTP20220501corrFingerprint}{50}
\DefineRemark{SMTP20220501numuniquehosts}{9587}
\DefineRemark{SMTP20220601corrFingerprint}{50}
\DefineRemark{SMTP20220601numuniquehosts}{8927}
\DefineRemark{SMTP20220701corrFingerprint}{46}
\DefineRemark{SMTP20220701numuniquehosts}{8853}
\DefineRemark{SMTP20220801corrFingerprint}{47}
\DefineRemark{SMTP20220801numuniquehosts}{8403}
\DefineRemark{SMTP20220901corrFingerprint}{43}
\DefineRemark{SMTP20220901numuniquehosts}{8165}
\DefineRemark{SSH20210501corrFingerprint}{0}
\DefineRemark{SSH20210501numuniquehosts}{0}
\DefineRemark{SSH20210601corrFingerprint}{0}
\DefineRemark{SSH20210601numuniquehosts}{0}
\DefineRemark{SSH20210701corrFingerprint}{44}
\DefineRemark{SSH20210701numuniquehosts}{151}
\DefineRemark{SSH20210801corrFingerprint}{49}
\DefineRemark{SSH20210801numuniquehosts}{162}
\DefineRemark{SSH20210901corrFingerprint}{57}
\DefineRemark{SSH20210901numuniquehosts}{201}
\DefineRemark{SSH20211001corrFingerprint}{54}
\DefineRemark{SSH20211001numuniquehosts}{221}
\DefineRemark{SSH20211101corrFingerprint}{60}
\DefineRemark{SSH20211101numuniquehosts}{200}
\DefineRemark{SSH20211201corrFingerprint}{59}
\DefineRemark{SSH20211201numuniquehosts}{219}
\DefineRemark{SSH20220101corrFingerprint}{67}
\DefineRemark{SSH20220101numuniquehosts}{266}
\DefineRemark{SSH20220201corrFingerprint}{73}
\DefineRemark{SSH20220201numuniquehosts}{1618}
\DefineRemark{SSH20220301corrFingerprint}{78}
\DefineRemark{SSH20220301numuniquehosts}{263}
\DefineRemark{SSH20220401corrFingerprint}{77}
\DefineRemark{SSH20220401numuniquehosts}{277}
\DefineRemark{SSH20220501corrFingerprint}{75}
\DefineRemark{SSH20220501numuniquehosts}{283}
\DefineRemark{SSH20220601corrFingerprint}{80}
\DefineRemark{SSH20220601numuniquehosts}{269}
\DefineRemark{SSH20220701corrFingerprint}{82}
\DefineRemark{SSH20220701numuniquehosts}{277}
\DefineRemark{SSH20220801corrFingerprint}{83}
\DefineRemark{SSH20220801numuniquehosts}{1542}
\DefineRemark{SSH20220901corrFingerprint}{77}
\DefineRemark{SSH20220901numuniquehosts}{240}
\DefineRemark{censysmaxoverallmonthsAMQPnumuniquehosts}{17}
\DefineRemark{censysmaxoverallmonthsANYCONNECTnumuniquehosts}{1}
\DefineRemark{censysmaxoverallmonthsELASTICSEARCHnumuniquehosts}{6}
\DefineRemark{censysmaxoverallmonthsFTPnumuniquehosts}{7939}
\DefineRemark{censysmaxoverallmonthsHTTPnumuniquehosts}{96348}
\DefineRemark{censysmaxoverallmonthsIMAPnumuniquehosts}{955}
\DefineRemark{censysmaxoverallmonthsKUBERNETESnumuniquehosts}{5}
\DefineRemark{censysmaxoverallmonthsLDAPnumuniquehosts}{51}
\DefineRemark{censysmaxoverallmonthsMQTTnumuniquehosts}{7501}
\DefineRemark{censysmaxoverallmonthsMYSQLnumuniquehosts}{2}
\DefineRemark{censysmaxoverallmonthsPOP3numuniquehosts}{934}
\DefineRemark{censysmaxoverallmonthsPOSTGRESnumuniquehosts}{302}
\DefineRemark{censysmaxoverallmonthsPROMETHEUSnumuniquehosts}{1}
\DefineRemark{censysmaxoverallmonthsSIPnumuniquehosts}{282}
\DefineRemark{censysmaxoverallmonthsSMTPnumuniquehosts}{11431}
\DefineRemark{censysmaxoverallmonthsSSHnumuniquehosts}{1355}
\DefineRemark{censysmaxoverallmonthsUNKNOWNnumuniquehosts}{29779}
\DefineRemark{censysmaxoverallmonthsXMPPnumuniquehosts}{11}

\DefineRemark{102102siemensdate}{05-29}
\DefineRemark{102102siemenspctconnsuccess}{0.9}
\DefineRemark{102102siemenspcttlsvalidoftransport}{0.0}
\DefineRemark{102102siemenspctvalidtls}{0.0}
\DefineRemark{102102siemenstotalas}{790}
\DefineRemark{102102siemenstotalcertcns}{0}
\DefineRemark{102102siemenstotalcerts}{0}
\DefineRemark{102102siemenstotalconnerr}{3091720}
\DefineRemark{102102siemenstotalconnsuccess}{29408}
\DefineRemark{102102siemenstotalhosts}{3121128}
\DefineRemark{102102siemenstotalsuccess}{6650}
\DefineRemark{102102siemenstotalsystemid}{1032}
\DefineRemark{102102siemenstotaltls}{0}
\DefineRemark{102102siemenstotaltlsauth}{0}
\DefineRemark{102102siemenstotaltlssuccess}{0}
\DefineRemark{102102tlssiemensdate}{05-29}
\DefineRemark{102102tlssiemenspctconnsuccess}{0.9}
\DefineRemark{102102tlssiemenspcttlsvalidoftransport}{35}
\DefineRemark{102102tlssiemenspctvalidtls}{0.0}
\DefineRemark{102102tlssiemenstotalas}{1}
\DefineRemark{102102tlssiemenstotalcertcns}{1}
\DefineRemark{102102tlssiemenstotalcerts}{1}
\DefineRemark{102102tlssiemenstotalconnerr}{3091720}
\DefineRemark{102102tlssiemenstotalconnsuccess}{29408}
\DefineRemark{102102tlssiemenstotalhosts}{3121128}
\DefineRemark{102102tlssiemenstotalsuccess}{1}
\DefineRemark{102102tlssiemenstotalsystemid}{0}
\DefineRemark{102102tlssiemenstotaltls}{10376}
\DefineRemark{102102tlssiemenstotaltlsauth}{10376}
\DefineRemark{102102tlssiemenstotaltlssuccess}{10281}
\DefineRemark{18831883mqttdate}{07-26}
\DefineRemark{18831883mqttpctconnsuccess}{15}
\DefineRemark{18831883mqttpcttlsvalidoftransport}{0.0}
\DefineRemark{18831883mqttpctvalidtls}{35}
\DefineRemark{18831883mqtttotalas}{6879}
\DefineRemark{18831883mqtttotalcertcns}{0}
\DefineRemark{18831883mqtttotalcerts}{0}
\DefineRemark{18831883mqtttotalconnerr}{3462525}
\DefineRemark{18831883mqtttotalconnsuccess}{603879}
\DefineRemark{18831883mqtttotalhosts}{4066404}
\DefineRemark{18831883mqtttotalsuccess}{602001}
\DefineRemark{18831883mqtttotalsystemid}{0}
\DefineRemark{18831883mqtttotaltls}{0}
\DefineRemark{18831883mqtttotaltlsauth}{0}
\DefineRemark{18831883mqtttotaltlssuccess}{0}
\DefineRemark{18831883tlsmqttdate}{07-26}
\DefineRemark{18831883tlsmqttpctconnsuccess}{15}
\DefineRemark{18831883tlsmqttpcttlsvalidoftransport}{16}
\DefineRemark{18831883tlsmqttpctvalidtls}{35}
\DefineRemark{18831883tlsmqtttotalas}{1904}
\DefineRemark{18831883tlsmqtttotalcertcns}{5043}
\DefineRemark{18831883tlsmqtttotalcerts}{8750}
\DefineRemark{18831883tlsmqtttotalconnerr}{3462525}
\DefineRemark{18831883tlsmqtttotalconnsuccess}{603879}
\DefineRemark{18831883tlsmqtttotalhosts}{4066404}
\DefineRemark{18831883tlsmqtttotalsuccess}{94660}
\DefineRemark{18831883tlsmqtttotalsystemid}{0}
\DefineRemark{18831883tlsmqtttotaltls}{94664}
\DefineRemark{18831883tlsmqtttotaltlsauth}{94664}
\DefineRemark{18831883tlsmqtttotaltlssuccess}{35686}
\DefineRemark{1999819998iec104date}{06-13}
\DefineRemark{1999819998iec104pctconnsuccess}{0.8}
\DefineRemark{1999819998iec104pcttlsvalidoftransport}{0.0}
\DefineRemark{1999819998iec104pctvalidtls}{0.0}
\DefineRemark{1999819998iec104totalas}{0}
\DefineRemark{1999819998iec104totalcertcns}{0}
\DefineRemark{1999819998iec104totalcerts}{0}
\DefineRemark{1999819998iec104totalconnerr}{4993055}
\DefineRemark{1999819998iec104totalconnsuccess}{42222}
\DefineRemark{1999819998iec104totalhosts}{5035277}
\DefineRemark{1999819998iec104totalsuccess}{0}
\DefineRemark{1999819998iec104totalsystemid}{0}
\DefineRemark{1999819998iec104totaltls}{0}
\DefineRemark{1999819998iec104totaltlsauth}{0}
\DefineRemark{1999819998iec104totaltlssuccess}{0}
\DefineRemark{1999819998tlsiec104date}{06-13}
\DefineRemark{1999819998tlsiec104pctconnsuccess}{0.8}
\DefineRemark{1999819998tlsiec104pcttlsvalidoftransport}{44}
\DefineRemark{1999819998tlsiec104pctvalidtls}{0.0}
\DefineRemark{1999819998tlsiec104totalas}{0}
\DefineRemark{1999819998tlsiec104totalcertcns}{0}
\DefineRemark{1999819998tlsiec104totalcerts}{0}
\DefineRemark{1999819998tlsiec104totalconnerr}{4993055}
\DefineRemark{1999819998tlsiec104totalconnsuccess}{42222}
\DefineRemark{1999819998tlsiec104totalhosts}{5035277}
\DefineRemark{1999819998tlsiec104totalsuccess}{0}
\DefineRemark{1999819998tlsiec104totalsystemid}{0}
\DefineRemark{1999819998tlsiec104totaltls}{18765}
\DefineRemark{1999819998tlsiec104totaltlsauth}{18762}
\DefineRemark{1999819998tlsiec104totaltlssuccess}{18695}
\DefineRemark{1999919999dnp3date}{07-17}
\DefineRemark{1999919999dnp3pctconnsuccess}{3.2}
\DefineRemark{1999919999dnp3pcttlsvalidoftransport}{0.0}
\DefineRemark{1999919999dnp3pctvalidtls}{0.7}
\DefineRemark{1999919999dnp3totalas}{0}
\DefineRemark{1999919999dnp3totalcertcns}{0}
\DefineRemark{1999919999dnp3totalcerts}{0}
\DefineRemark{1999919999dnp3totalconnerr}{4789043}
\DefineRemark{1999919999dnp3totalconnsuccess}{159623}
\DefineRemark{1999919999dnp3totalhosts}{4948666}
\DefineRemark{1999919999dnp3totalsuccess}{0}
\DefineRemark{1999919999dnp3totalsystemid}{0}
\DefineRemark{1999919999dnp3totaltls}{0}
\DefineRemark{1999919999dnp3totaltlsauth}{0}
\DefineRemark{1999919999dnp3totaltlssuccess}{0}
\DefineRemark{1999919999tlsdnp3date}{07-17}
\DefineRemark{1999919999tlsdnp3pctconnsuccess}{3.2}
\DefineRemark{1999919999tlsdnp3pcttlsvalidoftransport}{27}
\DefineRemark{1999919999tlsdnp3pctvalidtls}{0.7}
\DefineRemark{1999919999tlsdnp3totalas}{1}
\DefineRemark{1999919999tlsdnp3totalcertcns}{1}
\DefineRemark{1999919999tlsdnp3totalcerts}{1}
\DefineRemark{1999919999tlsdnp3totalconnerr}{4789043}
\DefineRemark{1999919999tlsdnp3totalconnsuccess}{159623}
\DefineRemark{1999919999tlsdnp3totalhosts}{4948666}
\DefineRemark{1999919999tlsdnp3totalsuccess}{1}
\DefineRemark{1999919999tlsdnp3totalsystemid}{0}
\DefineRemark{1999919999tlsdnp3totaltls}{43298}
\DefineRemark{1999919999tlsdnp3totaltlsauth}{43296}
\DefineRemark{1999919999tlsdnp3totaltlssuccess}{301}
\DefineRemark{2000020000dnp3date}{07-18}
\DefineRemark{2000020000dnp3pctconnsuccess}{6.4}
\DefineRemark{2000020000dnp3pcttlsvalidoftransport}{0.0}
\DefineRemark{2000020000dnp3pctvalidtls}{0.7}
\DefineRemark{2000020000dnp3totalas}{105}
\DefineRemark{2000020000dnp3totalcertcns}{0}
\DefineRemark{2000020000dnp3totalcerts}{0}
\DefineRemark{2000020000dnp3totalconnerr}{4110522}
\DefineRemark{2000020000dnp3totalconnsuccess}{279657}
\DefineRemark{2000020000dnp3totalhosts}{4390179}
\DefineRemark{2000020000dnp3totalsuccess}{585}
\DefineRemark{2000020000dnp3totalsystemid}{0}
\DefineRemark{2000020000dnp3totaltls}{0}
\DefineRemark{2000020000dnp3totaltlsauth}{0}
\DefineRemark{2000020000dnp3totaltlssuccess}{0}
\DefineRemark{2000020000tlsdnp3date}{07-18}
\DefineRemark{2000020000tlsdnp3pctconnsuccess}{6.4}
\DefineRemark{2000020000tlsdnp3pcttlsvalidoftransport}{18}
\DefineRemark{2000020000tlsdnp3pctvalidtls}{0.7}
\DefineRemark{2000020000tlsdnp3totalas}{2}
\DefineRemark{2000020000tlsdnp3totalcertcns}{2}
\DefineRemark{2000020000tlsdnp3totalcerts}{2}
\DefineRemark{2000020000tlsdnp3totalconnerr}{4110522}
\DefineRemark{2000020000tlsdnp3totalconnsuccess}{279657}
\DefineRemark{2000020000tlsdnp3totalhosts}{4390179}
\DefineRemark{2000020000tlsdnp3totalsuccess}{3}
\DefineRemark{2000020000tlsdnp3totalsystemid}{0}
\DefineRemark{2000020000tlsdnp3totaltls}{51584}
\DefineRemark{2000020000tlsdnp3totaltlsauth}{50906}
\DefineRemark{2000020000tlsdnp3totaltlssuccess}{1850}
\DefineRemark{220tlssshdate}{06-14}
\DefineRemark{220tlssshpctconnsuccess}{0.0}
\DefineRemark{220tlssshpcttlsvalidoftransport}{0}
\DefineRemark{220tlssshpctvalidtls}{0}
\DefineRemark{220tlssshtotalas}{0}
\DefineRemark{220tlssshtotalcertcns}{0}
\DefineRemark{220tlssshtotalcerts}{0}
\DefineRemark{220tlssshtotalconnerr}{22323287}
\DefineRemark{220tlssshtotalconnsuccess}{0}
\DefineRemark{220tlssshtotalhosts}{22323287}
\DefineRemark{220tlssshtotalsuccess}{0}
\DefineRemark{220tlssshtotalsystemid}{0}
\DefineRemark{220tlssshtotaltls}{0}
\DefineRemark{220tlssshtotaltlsauth}{0}
\DefineRemark{220tlssshtotaltlssuccess}{0}
\DefineRemark{22212221tlsenipdate}{06-20}
\DefineRemark{22212221tlsenippctconnsuccess}{0.2}
\DefineRemark{22212221tlsenippcttlsvalidoftransport}{100}
\DefineRemark{22212221tlsenippctvalidtls}{1.4}
\DefineRemark{22212221tlseniptotalas}{51}
\DefineRemark{22212221tlseniptotalcertcns}{2}
\DefineRemark{22212221tlseniptotalcerts}{75}
\DefineRemark{22212221tlseniptotalconnerr}{5089708}
\DefineRemark{22212221tlseniptotalconnsuccess}{10050}
\DefineRemark{22212221tlseniptotalhosts}{5099758}
\DefineRemark{22212221tlseniptotalsuccess}{90}
\DefineRemark{22212221tlseniptotalsystemid}{73}
\DefineRemark{22212221tlseniptotaltls}{10050}
\DefineRemark{22212221tlseniptotaltlsauth}{10050}
\DefineRemark{22212221tlseniptotaltlssuccess}{10050}
\DefineRemark{24042404iec104date}{06-12}
\DefineRemark{24042404iec104pctconnsuccess}{1.0}
\DefineRemark{24042404iec104pcttlsvalidoftransport}{0.0}
\DefineRemark{24042404iec104pctvalidtls}{0.0}
\DefineRemark{24042404iec104totalas}{460}
\DefineRemark{24042404iec104totalcertcns}{0}
\DefineRemark{24042404iec104totalcerts}{0}
\DefineRemark{24042404iec104totalconnerr}{4944814}
\DefineRemark{24042404iec104totalconnsuccess}{48233}
\DefineRemark{24042404iec104totalhosts}{4993047}
\DefineRemark{24042404iec104totalsuccess}{4818}
\DefineRemark{24042404iec104totalsystemid}{0}
\DefineRemark{24042404iec104totaltls}{0}
\DefineRemark{24042404iec104totaltlsauth}{0}
\DefineRemark{24042404iec104totaltlssuccess}{0}
\DefineRemark{24042404tlsiec104date}{06-12}
\DefineRemark{24042404tlsiec104pctconnsuccess}{1.0}
\DefineRemark{24042404tlsiec104pcttlsvalidoftransport}{22}
\DefineRemark{24042404tlsiec104pctvalidtls}{0.0}
\DefineRemark{24042404tlsiec104totalas}{0}
\DefineRemark{24042404tlsiec104totalcertcns}{0}
\DefineRemark{24042404tlsiec104totalcerts}{0}
\DefineRemark{24042404tlsiec104totalconnerr}{4944814}
\DefineRemark{24042404tlsiec104totalconnsuccess}{48233}
\DefineRemark{24042404tlsiec104totalhosts}{4993047}
\DefineRemark{24042404tlsiec104totalsuccess}{0}
\DefineRemark{24042404tlsiec104totalsystemid}{0}
\DefineRemark{24042404tlsiec104totaltls}{10846}
\DefineRemark{24042404tlsiec104totaltlsauth}{10841}
\DefineRemark{24042404tlsiec104totaltlssuccess}{10790}
\DefineRemark{30111911foxdate}{05-23}
\DefineRemark{30111911foxpctconnsuccess}{0.6}
\DefineRemark{30111911foxpcttlsvalidoftransport}{0.0}
\DefineRemark{30111911foxpctvalidtls}{0.8}
\DefineRemark{30111911foxtotalas}{1396}
\DefineRemark{30111911foxtotalcertcns}{0}
\DefineRemark{30111911foxtotalcerts}{0}
\DefineRemark{30111911foxtotalconnerr}{5601046}
\DefineRemark{30111911foxtotalconnsuccess}{34651}
\DefineRemark{30111911foxtotalhosts}{5635697}
\DefineRemark{30111911foxtotalsuccess}{20466}
\DefineRemark{30111911foxtotalsystemid}{10385}
\DefineRemark{30111911foxtotaltls}{0}
\DefineRemark{30111911foxtotaltlsauth}{0}
\DefineRemark{30111911foxtotaltlssuccess}{0}
\DefineRemark{30111911tlsfoxdate}{05-23}
\DefineRemark{30111911tlsfoxpctconnsuccess}{0.6}
\DefineRemark{30111911tlsfoxpcttlsvalidoftransport}{16}
\DefineRemark{30111911tlsfoxpctvalidtls}{0.8}
\DefineRemark{30111911tlsfoxtotalas}{0}
\DefineRemark{30111911tlsfoxtotalcertcns}{0}
\DefineRemark{30111911tlsfoxtotalcerts}{0}
\DefineRemark{30111911tlsfoxtotalconnerr}{5601046}
\DefineRemark{30111911tlsfoxtotalconnsuccess}{34651}
\DefineRemark{30111911tlsfoxtotalhosts}{5635697}
\DefineRemark{30111911tlsfoxtotalsuccess}{0}
\DefineRemark{30111911tlsfoxtotalsystemid}{0}
\DefineRemark{30111911tlsfoxtotaltls}{5547}
\DefineRemark{30111911tlsfoxtotaltlsauth}{5533}
\DefineRemark{30111911tlsfoxtotaltlssuccess}{5533}
\DefineRemark{30113011httpdate}{05-23}
\DefineRemark{30113011httppctconnsuccess}{3.9}
\DefineRemark{30113011httppcttlsvalidoftransport}{0.0}
\DefineRemark{30113011httppctvalidtls}{0.4}
\DefineRemark{30113011httptotalas}{1366}
\DefineRemark{30113011httptotalcertcns}{0}
\DefineRemark{30113011httptotalcerts}{0}
\DefineRemark{30113011httptotalconnerr}{5415489}
\DefineRemark{30113011httptotalconnsuccess}{220208}
\DefineRemark{30113011httptotalhosts}{5635697}
\DefineRemark{30113011httptotalsuccess}{18779}
\DefineRemark{30113011httptotalsystemid}{17772}
\DefineRemark{30113011httptotaltls}{0}
\DefineRemark{30113011httptotaltlsauth}{0}
\DefineRemark{30113011httptotaltlssuccess}{0}
\DefineRemark{30114911foxdate}{05-23}
\DefineRemark{30114911foxpctconnsuccess}{3.4}
\DefineRemark{30114911foxpcttlsvalidoftransport}{0.0}
\DefineRemark{30114911foxpctvalidtls}{0.8}
\DefineRemark{30114911foxtotalas}{0}
\DefineRemark{30114911foxtotalcertcns}{0}
\DefineRemark{30114911foxtotalcerts}{0}
\DefineRemark{30114911foxtotalconnerr}{5446123}
\DefineRemark{30114911foxtotalconnsuccess}{189574}
\DefineRemark{30114911foxtotalhosts}{5635697}
\DefineRemark{30114911foxtotalsuccess}{0}
\DefineRemark{30114911foxtotalsystemid}{0}
\DefineRemark{30114911foxtotaltls}{0}
\DefineRemark{30114911foxtotaltlsauth}{0}
\DefineRemark{30114911foxtotaltlssuccess}{0}
\DefineRemark{30114911tlsfoxdate}{05-23}
\DefineRemark{30114911tlsfoxpctconnsuccess}{3.4}
\DefineRemark{30114911tlsfoxpcttlsvalidoftransport}{85}
\DefineRemark{30114911tlsfoxpctvalidtls}{0.8}
\DefineRemark{30114911tlsfoxtotalas}{18}
\DefineRemark{30114911tlsfoxtotalcertcns}{19}
\DefineRemark{30114911tlsfoxtotalcerts}{22}
\DefineRemark{30114911tlsfoxtotalconnerr}{5446123}
\DefineRemark{30114911tlsfoxtotalconnsuccess}{189574}
\DefineRemark{30114911tlsfoxtotalhosts}{5635697}
\DefineRemark{30114911tlsfoxtotalsuccess}{25}
\DefineRemark{30114911tlsfoxtotalsystemid}{4}
\DefineRemark{30114911tlsfoxtotaltls}{160868}
\DefineRemark{30114911tlsfoxtotaltlsauth}{160868}
\DefineRemark{30114911tlsfoxtotaltlssuccess}{104679}
\DefineRemark{30115011httpdate}{05-23}
\DefineRemark{30115011httppctconnsuccess}{3.6}
\DefineRemark{30115011httppcttlsvalidoftransport}{0.0}
\DefineRemark{30115011httppctvalidtls}{0.4}
\DefineRemark{30115011httptotalas}{0}
\DefineRemark{30115011httptotalcertcns}{0}
\DefineRemark{30115011httptotalcerts}{0}
\DefineRemark{30115011httptotalconnerr}{5431225}
\DefineRemark{30115011httptotalconnsuccess}{204472}
\DefineRemark{30115011httptotalhosts}{5635697}
\DefineRemark{30115011httptotalsuccess}{0}
\DefineRemark{30115011httptotalsystemid}{0}
\DefineRemark{30115011httptotaltls}{0}
\DefineRemark{30115011httptotaltlsauth}{0}
\DefineRemark{30115011httptotaltlssuccess}{0}
\DefineRemark{30115011tlshttpdate}{05-23}
\DefineRemark{30115011tlshttppctconnsuccess}{3.6}
\DefineRemark{30115011tlshttppcttlsvalidoftransport}{79}
\DefineRemark{30115011tlshttppctvalidtls}{0.4}
\DefineRemark{30115011tlshttptotalas}{14}
\DefineRemark{30115011tlshttptotalcertcns}{16}
\DefineRemark{30115011tlshttptotalcerts}{18}
\DefineRemark{30115011tlshttptotalconnerr}{5431225}
\DefineRemark{30115011tlshttptotalconnsuccess}{204472}
\DefineRemark{30115011tlshttptotalhosts}{5635697}
\DefineRemark{30115011tlshttptotalsuccess}{20}
\DefineRemark{30115011tlshttptotalsystemid}{16}
\DefineRemark{30115011tlshttptotaltls}{162379}
\DefineRemark{30115011tlshttptotaltlsauth}{162379}
\DefineRemark{30115011tlshttptotaltlssuccess}{101779}
\DefineRemark{37823782siemensdate}{05-30}
\DefineRemark{37823782siemenspctconnsuccess}{0.9}
\DefineRemark{37823782siemenspcttlsvalidoftransport}{0.0}
\DefineRemark{37823782siemenspctvalidtls}{0.0}
\DefineRemark{37823782siemenstotalas}{0}
\DefineRemark{37823782siemenstotalcertcns}{0}
\DefineRemark{37823782siemenstotalcerts}{0}
\DefineRemark{37823782siemenstotalconnerr}{5409315}
\DefineRemark{37823782siemenstotalconnsuccess}{47895}
\DefineRemark{37823782siemenstotalhosts}{5457210}
\DefineRemark{37823782siemenstotalsuccess}{0}
\DefineRemark{37823782siemenstotalsystemid}{0}
\DefineRemark{37823782siemenstotaltls}{0}
\DefineRemark{37823782siemenstotaltlsauth}{0}
\DefineRemark{37823782siemenstotaltlssuccess}{0}
\DefineRemark{37823782tlssiemensdate}{05-30}
\DefineRemark{37823782tlssiemenspctconnsuccess}{0.9}
\DefineRemark{37823782tlssiemenspcttlsvalidoftransport}{23}
\DefineRemark{37823782tlssiemenspctvalidtls}{0.0}
\DefineRemark{37823782tlssiemenstotalas}{1}
\DefineRemark{37823782tlssiemenstotalcertcns}{1}
\DefineRemark{37823782tlssiemenstotalcerts}{1}
\DefineRemark{37823782tlssiemenstotalconnerr}{5409315}
\DefineRemark{37823782tlssiemenstotalconnsuccess}{47895}
\DefineRemark{37823782tlssiemenstotalhosts}{5457210}
\DefineRemark{37823782tlssiemenstotalsuccess}{1}
\DefineRemark{37823782tlssiemenstotalsystemid}{0}
\DefineRemark{37823782tlssiemenstotaltls}{11217}
\DefineRemark{37823782tlssiemenstotaltlsauth}{11217}
\DefineRemark{37823782tlssiemenstotaltlssuccess}{11162}
\DefineRemark{4481844818enipdate}{06-19}
\DefineRemark{4481844818enippctconnsuccess}{0.6}
\DefineRemark{4481844818enippcttlsvalidoftransport}{0.0}
\DefineRemark{4481844818enippctvalidtls}{1.4}
\DefineRemark{4481844818eniptotalas}{603}
\DefineRemark{4481844818eniptotalcertcns}{0}
\DefineRemark{4481844818eniptotalcerts}{0}
\DefineRemark{4481844818eniptotalconnerr}{4988796}
\DefineRemark{4481844818eniptotalconnsuccess}{29947}
\DefineRemark{4481844818eniptotalhosts}{5018743}
\DefineRemark{4481844818eniptotalsuccess}{6359}
\DefineRemark{4481844818eniptotalsystemid}{5996}
\DefineRemark{4481844818eniptotaltls}{0}
\DefineRemark{4481844818eniptotaltlsauth}{0}
\DefineRemark{4481844818eniptotaltlssuccess}{0}
\DefineRemark{4481844818tlsenipdate}{06-19}
\DefineRemark{4481844818tlsenippctconnsuccess}{0.6}
\DefineRemark{4481844818tlsenippcttlsvalidoftransport}{54}
\DefineRemark{4481844818tlsenippctvalidtls}{1.4}
\DefineRemark{4481844818tlseniptotalas}{0}
\DefineRemark{4481844818tlseniptotalcertcns}{0}
\DefineRemark{4481844818tlseniptotalcerts}{0}
\DefineRemark{4481844818tlseniptotalconnerr}{4988796}
\DefineRemark{4481844818tlseniptotalconnsuccess}{29947}
\DefineRemark{4481844818tlseniptotalhosts}{5018743}
\DefineRemark{4481844818tlseniptotalsuccess}{0}
\DefineRemark{4481844818tlseniptotalsystemid}{0}
\DefineRemark{4481844818tlseniptotaltls}{16098}
\DefineRemark{4481844818tlseniptotaltlsauth}{16098}
\DefineRemark{4481844818tlseniptotaltlssuccess}{16098}
\DefineRemark{48404840opcuadate}{08-14}
\DefineRemark{48404840opcuapctconnsuccess}{0.6}
\DefineRemark{48404840opcuapcttlsvalidoftransport}{0.0}
\DefineRemark{48404840opcuapctvalidtls}{0}
\DefineRemark{48404840opcuatotalas}{1}
\DefineRemark{48404840opcuatotalcertcns}{0}
\DefineRemark{48404840opcuatotalcerts}{1209} %
\DefineRemark{48404840opcuatotalconnerr}{5057934}
\DefineRemark{48404840opcuatotalconnsuccess}{30379}
\DefineRemark{48404840opcuatotalhosts}{5088313}
\DefineRemark{48404840opcuatotalsuccess}{1858}
\DefineRemark{48404840opcuatotalsystemid}{0}
\DefineRemark{48404840opcuatotaltls}{0}
\DefineRemark{48404840opcuatotaltlsauth}{0}
\DefineRemark{48404840opcuatotaltlssuccess}{0}
\DefineRemark{48404840tlsopcuadate}{08-14}
\DefineRemark{48404840tlsopcuapctconnsuccess}{1.1}
\DefineRemark{48404840tlsopcuapcttlsvalidoftransport}{13}
\DefineRemark{48404840tlsopcuapctvalidtls}{0.4}
\DefineRemark{48404840tlsopcuatotalas}{1}
\DefineRemark{48404840tlsopcuatotalcertcns}{6}
\DefineRemark{48404840tlsopcuatotalcerts}{6}
\DefineRemark{48404840tlsopcuatotalconnerr}{5038227}
\DefineRemark{48404840tlsopcuatotalconnsuccess}{54126}
\DefineRemark{48404840tlsopcuatotalhosts}{5092353}
\DefineRemark{48404840tlsopcuatotalsuccess}{6}
\DefineRemark{48404840tlsopcuatotalsystemid}{1}
\DefineRemark{48404840tlsopcuatotaltls}{7295}
\DefineRemark{48404840tlsopcuatotaltlsauth}{7293}
\DefineRemark{48404840tlsopcuatotaltlssuccess}{5310}
\DefineRemark{48434843opcuadate}{08-15}
\DefineRemark{48434843opcuapctconnsuccess}{1.0}
\DefineRemark{48434843opcuapcttlsvalidoftransport}{0.0}
\DefineRemark{48434843opcuapctvalidtls}{0.4}
\DefineRemark{48434843opcuatotalas}{0}
\DefineRemark{48434843opcuatotalcertcns}{0}
\DefineRemark{48434843opcuatotalcerts}{0}
\DefineRemark{48434843opcuatotalconnerr}{5089550}
\DefineRemark{48434843opcuatotalconnsuccess}{53887}
\DefineRemark{48434843opcuatotalhosts}{5143437}
\DefineRemark{48434843opcuatotalsuccess}{0}
\DefineRemark{48434843opcuatotalsystemid}{0}
\DefineRemark{48434843opcuatotaltls}{0}
\DefineRemark{48434843opcuatotaltlsauth}{0}
\DefineRemark{48434843opcuatotaltlssuccess}{0}
\DefineRemark{48434843tlsopcuadate}{08-15}
\DefineRemark{48434843tlsopcuapctconnsuccess}{1.0}
\DefineRemark{48434843tlsopcuapcttlsvalidoftransport}{14}
\DefineRemark{48434843tlsopcuapctvalidtls}{0.4}
\DefineRemark{48434843tlsopcuatotalas}{0}
\DefineRemark{48434843tlsopcuatotalcertcns}{0}
\DefineRemark{48434843tlsopcuatotalcerts}{0}
\DefineRemark{48434843tlsopcuatotalconnerr}{5089550}
\DefineRemark{48434843tlsopcuatotalconnsuccess}{53887}
\DefineRemark{48434843tlsopcuatotalhosts}{5143437}
\DefineRemark{48434843tlsopcuatotalsuccess}{0}
\DefineRemark{48434843tlsopcuatotalsystemid}{0}
\DefineRemark{48434843tlsopcuatotaltls}{7672}
\DefineRemark{48434843tlsopcuatotaltlsauth}{7672}
\DefineRemark{48434843tlsopcuatotaltlssuccess}{5467}
\DefineRemark{49111911foxdate}{05-24}
\DefineRemark{49111911foxpctconnsuccess}{0.4}
\DefineRemark{49111911foxpcttlsvalidoftransport}{0.0}
\DefineRemark{49111911foxpctvalidtls}{0.8}
\DefineRemark{49111911foxtotalas}{717}
\DefineRemark{49111911foxtotalcertcns}{0}
\DefineRemark{49111911foxtotalcerts}{0}
\DefineRemark{49111911foxtotalconnerr}{5505446}
\DefineRemark{49111911foxtotalconnsuccess}{21221}
\DefineRemark{49111911foxtotalhosts}{5526667}
\DefineRemark{49111911foxtotalsuccess}{6492}
\DefineRemark{49111911foxtotalsystemid}{722}
\DefineRemark{49111911foxtotaltls}{0}
\DefineRemark{49111911foxtotaltlsauth}{0}
\DefineRemark{49111911foxtotaltlssuccess}{0}
\DefineRemark{49111911tlsfoxdate}{05-24}
\DefineRemark{49111911tlsfoxpctconnsuccess}{0.4}
\DefineRemark{49111911tlsfoxpcttlsvalidoftransport}{26}
\DefineRemark{49111911tlsfoxpctvalidtls}{0.8}
\DefineRemark{49111911tlsfoxtotalas}{0}
\DefineRemark{49111911tlsfoxtotalcertcns}{0}
\DefineRemark{49111911tlsfoxtotalcerts}{0}
\DefineRemark{49111911tlsfoxtotalconnerr}{5505446}
\DefineRemark{49111911tlsfoxtotalconnsuccess}{21221}
\DefineRemark{49111911tlsfoxtotalhosts}{5526667}
\DefineRemark{49111911tlsfoxtotalsuccess}{0}
\DefineRemark{49111911tlsfoxtotalsystemid}{0}
\DefineRemark{49111911tlsfoxtotaltls}{5488}
\DefineRemark{49111911tlsfoxtotaltlsauth}{5474}
\DefineRemark{49111911tlsfoxtotaltlssuccess}{5472}
\DefineRemark{49113011httpdate}{05-24}
\DefineRemark{49113011httppctconnsuccess}{3.4}
\DefineRemark{49113011httppcttlsvalidoftransport}{0.0}
\DefineRemark{49113011httppctvalidtls}{0.4}
\DefineRemark{49113011httptotalas}{546}
\DefineRemark{49113011httptotalcertcns}{0}
\DefineRemark{49113011httptotalcerts}{0}
\DefineRemark{49113011httptotalconnerr}{5339456}
\DefineRemark{49113011httptotalconnsuccess}{187211}
\DefineRemark{49113011httptotalhosts}{5526667}
\DefineRemark{49113011httptotalsuccess}{4354}
\DefineRemark{49113011httptotalsystemid}{4154}
\DefineRemark{49113011httptotaltls}{0}
\DefineRemark{49113011httptotaltlsauth}{0}
\DefineRemark{49113011httptotaltlssuccess}{0}
\DefineRemark{49114911foxdate}{05-24}
\DefineRemark{49114911foxpctconnsuccess}{3.7}
\DefineRemark{49114911foxpcttlsvalidoftransport}{0.0}
\DefineRemark{49114911foxpctvalidtls}{0.8}
\DefineRemark{49114911foxtotalas}{0}
\DefineRemark{49114911foxtotalcertcns}{0}
\DefineRemark{49114911foxtotalcerts}{0}
\DefineRemark{49114911foxtotalconnerr}{5322821}
\DefineRemark{49114911foxtotalconnsuccess}{203846}
\DefineRemark{49114911foxtotalhosts}{5526667}
\DefineRemark{49114911foxtotalsuccess}{0}
\DefineRemark{49114911foxtotalsystemid}{0}
\DefineRemark{49114911foxtotaltls}{0}
\DefineRemark{49114911foxtotaltlsauth}{0}
\DefineRemark{49114911foxtotaltlssuccess}{0}
\DefineRemark{49114911tlsfoxdate}{05-24}
\DefineRemark{49114911tlsfoxpctconnsuccess}{3.7}
\DefineRemark{49114911tlsfoxpcttlsvalidoftransport}{82}
\DefineRemark{49114911tlsfoxpctvalidtls}{0.8}
\DefineRemark{49114911tlsfoxtotalas}{68}
\DefineRemark{49114911tlsfoxtotalcertcns}{76}
\DefineRemark{49114911tlsfoxtotalcerts}{79}
\DefineRemark{49114911tlsfoxtotalconnerr}{5322821}
\DefineRemark{49114911tlsfoxtotalconnsuccess}{203846}
\DefineRemark{49114911tlsfoxtotalhosts}{5526667}
\DefineRemark{49114911tlsfoxtotalsuccess}{188}
\DefineRemark{49114911tlsfoxtotalsystemid}{13}
\DefineRemark{49114911tlsfoxtotaltls}{167546}
\DefineRemark{49114911tlsfoxtotaltlsauth}{167545}
\DefineRemark{49114911tlsfoxtotaltlssuccess}{105082}
\DefineRemark{49115011httpdate}{05-24}
\DefineRemark{49115011httppctconnsuccess}{3.6}
\DefineRemark{49115011httppcttlsvalidoftransport}{0.0}
\DefineRemark{49115011httppctvalidtls}{0.4}
\DefineRemark{49115011httptotalas}{0}
\DefineRemark{49115011httptotalcertcns}{0}
\DefineRemark{49115011httptotalcerts}{0}
\DefineRemark{49115011httptotalconnerr}{5326954}
\DefineRemark{49115011httptotalconnsuccess}{199713}
\DefineRemark{49115011httptotalhosts}{5526667}
\DefineRemark{49115011httptotalsuccess}{0}
\DefineRemark{49115011httptotalsystemid}{0}
\DefineRemark{49115011httptotaltls}{0}
\DefineRemark{49115011httptotaltlsauth}{0}
\DefineRemark{49115011httptotaltlssuccess}{0}
\DefineRemark{49115011tlshttpdate}{05-24}
\DefineRemark{49115011tlshttppctconnsuccess}{3.6}
\DefineRemark{49115011tlshttppcttlsvalidoftransport}{82}
\DefineRemark{49115011tlshttppctvalidtls}{0.4}
\DefineRemark{49115011tlshttptotalas}{26}
\DefineRemark{49115011tlshttptotalcertcns}{30}
\DefineRemark{49115011tlshttptotalcerts}{32}
\DefineRemark{49115011tlshttptotalconnerr}{5326954}
\DefineRemark{49115011tlshttptotalconnsuccess}{199713}
\DefineRemark{49115011tlshttptotalhosts}{5526667}
\DefineRemark{49115011tlshttptotalsuccess}{67}
\DefineRemark{49115011tlshttptotalsystemid}{60}
\DefineRemark{49115011tlshttptotaltls}{163702}
\DefineRemark{49115011tlshttptotaltlsauth}{163701}
\DefineRemark{49115011tlshttptotaltlssuccess}{101276}
\DefineRemark{50005000httpdate}{08-06}
\DefineRemark{50005000httppctconnsuccess}{26}
\DefineRemark{50005000httppcttlsvalidoftransport}{0.0}
\DefineRemark{50005000httppctvalidtls}{0.4}
\DefineRemark{50005000httptotalas}{4}
\DefineRemark{50005000httptotalcertcns}{0}
\DefineRemark{50005000httptotalcerts}{0}
\DefineRemark{50005000httptotalconnerr}{3508308}
\DefineRemark{50005000httptotalconnsuccess}{1261001}
\DefineRemark{50005000httptotalhosts}{4769309}
\DefineRemark{50005000httptotalsuccess}{5}
\DefineRemark{50005000httptotalsystemid}{3}
\DefineRemark{50005000httptotaltls}{0}
\DefineRemark{50005000httptotaltlsauth}{0}
\DefineRemark{50005000httptotaltlssuccess}{0}
\DefineRemark{50005000tlshttpdate}{08-06}
\DefineRemark{50005000tlshttppctconnsuccess}{26}
\DefineRemark{50005000tlshttppcttlsvalidoftransport}{4.0}
\DefineRemark{50005000tlshttppctvalidtls}{0.4}
\DefineRemark{50005000tlshttptotalas}{0}
\DefineRemark{50005000tlshttptotalcertcns}{0}
\DefineRemark{50005000tlshttptotalcerts}{0}
\DefineRemark{50005000tlshttptotalconnerr}{3508308}
\DefineRemark{50005000tlshttptotalconnsuccess}{1261001}
\DefineRemark{50005000tlshttptotalhosts}{4769309}
\DefineRemark{50005000tlshttptotalsuccess}{0}
\DefineRemark{50005000tlshttptotalsystemid}{0}
\DefineRemark{50005000tlshttptotaltls}{50672}
\DefineRemark{50005000tlshttptotaltlsauth}{48782}
\DefineRemark{50005000tlshttptotaltlssuccess}{42441}
\DefineRemark{502502modbusdate}{06-05}
\DefineRemark{502502modbuspctconnsuccess}{1.8}
\DefineRemark{502502modbuspcttlsvalidoftransport}{0.0}
\DefineRemark{502502modbuspctvalidtls}{0.0}
\DefineRemark{502502modbustotalas}{2191}
\DefineRemark{502502modbustotalcertcns}{0}
\DefineRemark{502502modbustotalcerts}{0}
\DefineRemark{502502modbustotalconnerr}{3089711}
\DefineRemark{502502modbustotalconnsuccess}{56036}
\DefineRemark{502502modbustotalhosts}{3145747}
\DefineRemark{502502modbustotalsuccess}{30963}
\DefineRemark{502502modbustotalsystemid}{412}
\DefineRemark{502502modbustotaltls}{0}
\DefineRemark{502502modbustotaltlsauth}{0}
\DefineRemark{502502modbustotaltlssuccess}{0}
\DefineRemark{502502tlsmodbusdate}{06-05}
\DefineRemark{502502tlsmodbuspctconnsuccess}{1.8}
\DefineRemark{502502tlsmodbuspcttlsvalidoftransport}{20}
\DefineRemark{502502tlsmodbuspctvalidtls}{0.0}
\DefineRemark{502502tlsmodbustotalas}{0}
\DefineRemark{502502tlsmodbustotalcertcns}{0}
\DefineRemark{502502tlsmodbustotalcerts}{0}
\DefineRemark{502502tlsmodbustotalconnerr}{3089711}
\DefineRemark{502502tlsmodbustotalconnsuccess}{56036}
\DefineRemark{502502tlsmodbustotalhosts}{3145747}
\DefineRemark{502502tlsmodbustotalsuccess}{0}
\DefineRemark{502502tlsmodbustotalsystemid}{0}
\DefineRemark{502502tlsmodbustotaltls}{11020}
\DefineRemark{502502tlsmodbustotaltlsauth}{11020}
\DefineRemark{502502tlsmodbustotaltlssuccess}{10951}
\DefineRemark{56715671amqpdate}{08-08}
\DefineRemark{56715671amqppctconnsuccess}{4.4}
\DefineRemark{56715671amqppcttlsvalidoftransport}{0.0}
\DefineRemark{56715671amqppctvalidtls}{17}
\DefineRemark{56715671amqptotalas}{0}
\DefineRemark{56715671amqptotalcertcns}{0}
\DefineRemark{56715671amqptotalcerts}{0}
\DefineRemark{56715671amqptotalconnerr}{5583054}
\DefineRemark{56715671amqptotalconnsuccess}{259582}
\DefineRemark{56715671amqptotalhosts}{5842636}
\DefineRemark{56715671amqptotalsuccess}{0}
\DefineRemark{56715671amqptotalsystemid}{0}
\DefineRemark{56715671amqptotaltls}{0}
\DefineRemark{56715671amqptotaltlsauth}{0}
\DefineRemark{56715671amqptotaltlssuccess}{0}
\DefineRemark{56715671tlsamqpdate}{08-08}
\DefineRemark{56715671tlsamqppctconnsuccess}{4.4}
\DefineRemark{56715671tlsamqppcttlsvalidoftransport}{77}
\DefineRemark{56715671tlsamqppctvalidtls}{17}
\DefineRemark{56715671tlsamqptotalas}{443}
\DefineRemark{56715671tlsamqptotalcertcns}{2670}
\DefineRemark{56715671tlsamqptotalcerts}{3673}
\DefineRemark{56715671tlsamqptotalconnerr}{5583054}
\DefineRemark{56715671tlsamqptotalconnsuccess}{259582}
\DefineRemark{56715671tlsamqptotalhosts}{5842636}
\DefineRemark{56715671tlsamqptotalsuccess}{24828}
\DefineRemark{56715671tlsamqptotalsystemid}{8425}
\DefineRemark{56715671tlsamqptotaltls}{199844}
\DefineRemark{56715671tlsamqptotaltlsauth}{199742}
\DefineRemark{56715671tlsamqptotaltlssuccess}{149925}
\DefineRemark{56725672amqpdate}{08-09}
\DefineRemark{56725672amqppctconnsuccess}{5.8}
\DefineRemark{56725672amqppcttlsvalidoftransport}{0.0}
\DefineRemark{56725672amqppctvalidtls}{17}
\DefineRemark{56725672amqptotalas}{3688}
\DefineRemark{56725672amqptotalcertcns}{0}
\DefineRemark{56725672amqptotalcerts}{0}
\DefineRemark{56725672amqptotalconnerr}{5806018}
\DefineRemark{56725672amqptotalconnsuccess}{359090}
\DefineRemark{56725672amqptotalhosts}{6165108}
\DefineRemark{56725672amqptotalsuccess}{163174}
\DefineRemark{56725672amqptotalsystemid}{91628}
\DefineRemark{56725672amqptotaltls}{0}
\DefineRemark{56725672amqptotaltlsauth}{0}
\DefineRemark{56725672amqptotaltlssuccess}{0}
\DefineRemark{56725672tlsamqpdate}{08-09}
\DefineRemark{56725672tlsamqppctconnsuccess}{5.8}
\DefineRemark{56725672tlsamqppcttlsvalidoftransport}{46}
\DefineRemark{56725672tlsamqppctvalidtls}{17}
\DefineRemark{56725672tlsamqptotalas}{158}
\DefineRemark{56725672tlsamqptotalcertcns}{398}
\DefineRemark{56725672tlsamqptotalcerts}{692}
\DefineRemark{56725672tlsamqptotalconnerr}{5806018}
\DefineRemark{56725672tlsamqptotalconnsuccess}{359090}
\DefineRemark{56725672tlsamqptotalhosts}{6165108}
\DefineRemark{56725672tlsamqptotalsuccess}{9082}
\DefineRemark{56725672tlsamqptotalsystemid}{311}
\DefineRemark{56725672tlsamqptotaltls}{165452}
\DefineRemark{56725672tlsamqptotaltlsauth}{165054}
\DefineRemark{56725672tlsamqptotaltlssuccess}{117082}
\DefineRemark{56835683coapdate}{08-21}
\DefineRemark{56835683coappctconnsuccess}{84}
\DefineRemark{56835683coappcttlsvalidoftransport}{0.0}
\DefineRemark{56835683coappctvalidtls}{0.0}
\DefineRemark{56835683coaptotalas}{1}
\DefineRemark{56835683coaptotalcertcns}{0}
\DefineRemark{56835683coaptotalcerts}{0}
\DefineRemark{56835683coaptotalconnerr}{57523}
\DefineRemark{56835683coaptotalconnsuccess}{298437}
\DefineRemark{56835683coaptotalhosts}{355960}
\DefineRemark{56835683coaptotalsuccess}{294706}
\DefineRemark{56835683coaptotalsystemid}{321}
\DefineRemark{56835683coaptotaltls}{0}
\DefineRemark{56835683coaptotaltlsauth}{0}
\DefineRemark{56835683coaptotaltlssuccess}{0}
\DefineRemark{56835683tlscoapdate}{08-21}
\DefineRemark{56835683tlscoappctconnsuccess}{84}
\DefineRemark{56835683tlscoappcttlsvalidoftransport}{0.0}
\DefineRemark{56835683tlscoappctvalidtls}{0.0}
\DefineRemark{56835683tlscoaptotalas}{0}
\DefineRemark{56835683tlscoaptotalcertcns}{0}
\DefineRemark{56835683tlscoaptotalcerts}{0}
\DefineRemark{56835683tlscoaptotalconnerr}{57523}
\DefineRemark{56835683tlscoaptotalconnsuccess}{298437}
\DefineRemark{56835683tlscoaptotalhosts}{355960}
\DefineRemark{56835683tlscoaptotalsuccess}{0}
\DefineRemark{56835683tlscoaptotalsystemid}{0}
\DefineRemark{56835683tlscoaptotaltls}{2}
\DefineRemark{56835683tlscoaptotaltlsauth}{2}
\DefineRemark{56835683tlscoaptotaltlssuccess}{0}
\DefineRemark{802802modbusdate}{06-06}
\DefineRemark{802802modbuspctconnsuccess}{1.2}
\DefineRemark{802802modbuspcttlsvalidoftransport}{0.0}
\DefineRemark{802802modbuspctvalidtls}{0.0}
\DefineRemark{802802modbustotalas}{0}
\DefineRemark{802802modbustotalcertcns}{0}
\DefineRemark{802802modbustotalcerts}{0}
\DefineRemark{802802modbustotalconnerr}{3264403}
\DefineRemark{802802modbustotalconnsuccess}{38650}
\DefineRemark{802802modbustotalhosts}{3303053}
\DefineRemark{802802modbustotalsuccess}{0}
\DefineRemark{802802modbustotalsystemid}{0}
\DefineRemark{802802modbustotaltls}{0}
\DefineRemark{802802modbustotaltlsauth}{0}
\DefineRemark{802802modbustotaltlssuccess}{0}
\DefineRemark{802802tlsmodbusdate}{06-06}
\DefineRemark{802802tlsmodbuspctconnsuccess}{1.2}
\DefineRemark{802802tlsmodbuspcttlsvalidoftransport}{36}
\DefineRemark{802802tlsmodbuspctvalidtls}{0.0}
\DefineRemark{802802tlsmodbustotalas}{1}
\DefineRemark{802802tlsmodbustotalcertcns}{1}
\DefineRemark{802802tlsmodbustotalcerts}{1}
\DefineRemark{802802tlsmodbustotalconnerr}{3264403}
\DefineRemark{802802tlsmodbustotalconnsuccess}{38650}
\DefineRemark{802802tlsmodbustotalhosts}{3303053}
\DefineRemark{802802tlsmodbustotalsuccess}{1}
\DefineRemark{802802tlsmodbustotalsystemid}{0}
\DefineRemark{802802tlsmodbustotaltls}{13788}
\DefineRemark{802802tlsmodbustotaltlsauth}{13788}
\DefineRemark{802802tlsmodbustotaltlssuccess}{13707}
\DefineRemark{88838883mqttdate}{07-25}
\DefineRemark{88838883mqttpctconnsuccess}{10}
\DefineRemark{88838883mqttpcttlsvalidoftransport}{0.0}
\DefineRemark{88838883mqttpctvalidtls}{35}
\DefineRemark{88838883mqtttotalas}{5284}
\DefineRemark{88838883mqtttotalcertcns}{0}
\DefineRemark{88838883mqtttotalcerts}{0}
\DefineRemark{88838883mqtttotalconnerr}{5865668}
\DefineRemark{88838883mqtttotalconnsuccess}{671132}
\DefineRemark{88838883mqtttotalhosts}{6536800}
\DefineRemark{88838883mqtttotalsuccess}{651738}
\DefineRemark{88838883mqtttotalsystemid}{0}
\DefineRemark{88838883mqtttotaltls}{0}
\DefineRemark{88838883mqtttotaltlsauth}{0}
\DefineRemark{88838883mqtttotaltlssuccess}{0}
\DefineRemark{88838883tlsmqttdate}{07-25}
\DefineRemark{88838883tlsmqttpctconnsuccess}{10}
\DefineRemark{88838883tlsmqttpcttlsvalidoftransport}{87}
\DefineRemark{88838883tlsmqttpctvalidtls}{35}
\DefineRemark{88838883tlsmqtttotalas}{4550}
\DefineRemark{88838883tlsmqtttotalcertcns}{30810}
\DefineRemark{88838883tlsmqtttotalcerts}{46896}
\DefineRemark{88838883tlsmqtttotalconnerr}{5865668}
\DefineRemark{88838883tlsmqtttotalconnsuccess}{671132}
\DefineRemark{88838883tlsmqtttotalhosts}{6536800}
\DefineRemark{88838883tlsmqtttotalsuccess}{586229}
\DefineRemark{88838883tlsmqtttotalsystemid}{0}
\DefineRemark{88838883tlsmqtttotaltls}{586426}
\DefineRemark{88838883tlsmqtttotaltlsauth}{586424}
\DefineRemark{88838883tlsmqtttotaltlssuccess}{210636}
\DefineRemark{allfoxpctvalidtls}{0.8} %
\DefineRemark{allfoxtotalas}{1479} %
\DefineRemark{allfoxtotalcertcns}{0} %
\DefineRemark{allfoxtotalsuccess}{22134}
\DefineRemark{allhttppctvalidtls}{0.4} %
\DefineRemark{allhttptotalas}{1368} %
\DefineRemark{allhttptotalcertcns}{0} %
\DefineRemark{allhttptotalsuccess}{18975}
\DefineRemark{distinctcertificatesoverall}{156113}

\DefineRemark{numemaildisclosure}{1181}

\begin{document}

\title[An Internet-wide Study on Secrets in Container Images]{Secrets Revealed in Container Images:\\An Internet-wide Study on Occurrence and Impact}
\ifmanualheader
  \author{Markus Dahlmanns, Constantin Sander, Robin Decker, Klaus Wehrle}
  \def\cleanauthors{Markus Dahlmanns, Constantin Sander, Robin Decker, Klaus Wehrle}
  \affiliation{
  \textit{Communication and Distributed Systems}, RWTH Aachen University \city{Aachen} \country{Germany} \\
  \{dahlmanns, sander, decker, wehrle\}@comsys.rwth-aachen.de
  }
\else
  \author{Markus Dahlmanns}
  \email{dahlmanns@comsys.rwth-aachen.de}
  \affiliation{%
    \institution{RWTH Aachen University}
    \country{Germany}
  }

  \author{Constantin Sander}
  \email{sander@comsys.rwth-aachen.de}
  \affiliation{%
    \institution{RWTH Aachen University}
    \country{Germany}
  }

  \author{Robin Decker}
  \email{decker@comsys.rwth-aachen.de}
  \affiliation{%
    \institution{RWTH Aachen University}
    \country{Germany}
  }

  \author{Klaus Wehrle}
  \email{wehrle@comsys.rwth-aachen.de}
  \affiliation{%
    \institution{RWTH Aachen University}
    \country{Germany}
  }
\fi

\renewcommand{\shortauthors}{Dahlmanns et al.}

\begin{abstract}
Containerization allows bundling applications and their dependencies into a single image.
The containerization framework Docker eases the use of this concept and enables sharing images publicly, gaining high momentum.
However, it can lead to users creating and sharing images that include private keys or API secrets---either by mistake or out of negligence.
This leakage impairs the creator's security and that of everyone using the image.
Yet, the extent of this practice and how to counteract it remains unclear.

In this paper, we analyze \SI{\Remark{numnonemptyimages}}{}~images from Docker Hub and \SI{\Remark{privatemeasurementnumtotalmax}}{}~other private registries unveiling that \SI{\Remark{pctaffectedimages}}{\percent}~of images indeed include secrets.
Specifically, we find \SI{\Remark{validprivatekeyvalidnumdistinctmatchestotal}}{}~private keys and \SI{\Remark{apinumdistinctmatches}}{}~leaked API~secrets, both opening a large attack surface, i.e., putting authentication and confidentiality of privacy-sensitive data at stake and even allow active attacks.
We further document that those leaked keys are used in the wild:
While we discovered \SI{\Remark{casignedcerts}}{}~certificates relying on compromised keys being issued by public certificate authorities, based on further active Internet measurements, we find \SI{\Remark{20220901numuniquehosts}}{}~TLS and SSH hosts using leaked private keys for authentication.
To counteract this issue, we discuss how our methodology can be used to prevent secret leakage and reuse.
\end{abstract}

\ccsdesc[500]{Security and privacy~Network security}
\ccsdesc[500]{Security and privacy~Key management}

\keywords{network security, security configuration, secret leakage, container}

\maketitle

\section{Introduction}

While originally developed to isolate applications~\cite{docker-what-container}, containerization has become a new cornerstone of interconnected services as it significantly eases their deployment~\cite{Combe-2016, Pahl-2015, stackoverflow-survey-2021, kubernetes:online, abhishek:eurosys15:borg, liu-2022}.
To this end, Docker, the most prominent containerization framework~\cite{stackoverflow-survey-2021}, uses prebuilt images that include all software dependencies necessary to deploy an application~\cite{Combe-2016}.
Users only need to download an image from a registry or can derive their own image by adapting its configuration and included files.
These new images can then again be uploaded building a whole ecosystem of containerized applications.
For example, Docker Hub, the official Docker registry, comprises more than \SI{9000000}{}~images~\cite{DockerHub:online} anybody can use.

With this level of public exposure, any mistake during image creation can have drastic consequences.
Most notably, including confidential secrets such as cryptographic keys or API~secrets, by mistake or out of negligence, can introduce two security issues:
\begin{inparaenum}[(i)]
	\item attackers can misuse compromised secrets leading to potential loss of data, money, privacy, or control, and
	\item administrators instantiating images can rely on broken security, e.g., paving the way for Man-in-the-Middle attacks.
\end{inparaenum}
Aggravatingly, there is no easy tooling to show which files have been added---accidentally adding a secret is thus much easier than identifying such an incident.

Indeed, related work traced three reused private keys authenticating \SI{6000}{}~(Industrial)~Internet of Things services back to the occurrence in a Docker image~\cite{dahlmanns-2022}.
Additionally, blog entries produced anecdotal evidence that Docker images include further confidential security material~\cite{Lowhangi50:online,Turner95:online,Secretse30:online,Scanning55:online}.
However, comprehensive analyses on revealed security secrets at scale do not exist in this realm.
Instead, such analyses focus on GitHub repositories~\cite{PSADont65:online,Hundreds32:online,10000Git13:online,Sinha2015DetectingAM,Slackbot26:online,Rahman-2019,Rahman-2021,rahman-2019-2,rahman-2021-2,meli-2019}.
Hence, the extent for container images is unknown.

In this paper, we thus comprehensively study whether Docker images include confidential security material and whether administrators reuse these compromised secrets at large scale by
\begin{inparaenum}[(i)]
  \item scanning publicly available Docker images for confidential security material, and
  \item measure whether these secrets are used in practice on production deployments.
\end{inparaenum}
To this end, we analyze images available on the official and largest registry Docker Hub as well as examine the entire IPv4 address space for public registries and services relying their security on compromised secrets.

\newpage

\noindent\afblock{Contributions} Our main contributions are as follows.
\begin{itemize}[noitemsep,topsep=0pt,leftmargin=9pt]
  \item We found \SI{\Remark{privatemeasurementnumtotalmax}}{} Docker registries in the IPv4 address space that contain not only secrets but also potentially confidential software and likely allow attackers to replace images, e.g., with malware.
  \item After filtering test secrets, we identified \SI{\Remark{totalvalidmatches}}{}~leaked distinct secrets, i.e., \SI{\Remark{validprivatekeyvalidnumdistinctmatchestotal}}{}~private keys and \SI{\Remark{apinumdistinctmatches}}{}~API secrets, in \SI{\Remark{numaffectedimages}}{}~images (\SI{\Remark{pctaffectedimages}}{\percent}~of images we scanned are affected).
  \item We show that operators use \SI{\Remark{20220901corrFingerprint}}{} compromised private keys in practice affecting the authenticity of \SI{\Remark{20220901numuniquehosts}}{}~Internet-reachable hosts providing, i.a., HTTP, AMQP, MQTT, and LDAP services.
  \item We discuss improvements of the Docker paradigm to prevent secret leakage and reuse in the future as well as provide our software used to find and verify secrets~\cite{COMSYScode:online} to support mitigation.
\end{itemize}

\section{A Primer on the Docker Paradigm}

\begin{figure}[!t]
\centering
\includegraphics[width=\columnwidth]{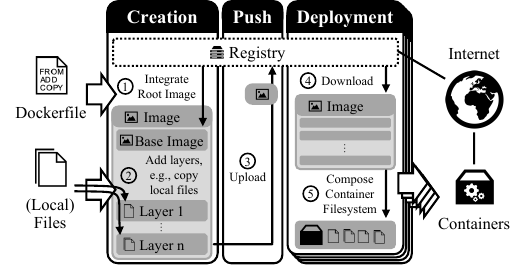}
\caption{The Docker Paradigm: Dockerfiles describe files and functions of images. Images are uploaded to a registry for sharing and potentially numerous administrators deploy containers based on a single image (according to~\cite{Combe-2016}).}%
\label{fig:docker-infrastructure}
\end{figure}

In contrast to other containerization frameworks, Docker~\cite{docker-what-container} does not only provide an isolated execution environment for applications.
Instead, Docker specifies an easy-to-use paradigm to create, share and deploy ready-to-run container images~\cite{Combe-2016}.
These images constitute the filesystems of the containers and include all dependencies necessary for the actual applications, i.e., they can include all kinds of files added during creation.
The completeness of these images allows to share them via (publicly accessible)~registries.
Figure~\ref{fig:docker-infrastructure} shows the structure and lifecycle of Docker images in detail, from creating images to sharing and running them.

\afblock{Image Creation}
To create an image, Docker uses a user-defined \emph{Dockerfile}~\cite{docker-dockerfile} to specify the image ingredients.
First~\circled{1}, the Dockerfile references another image, the base image, which is downloaded from a registry and comprises the initial file system of the new image.
Second~\circled{2}, image layers consisting of differential snapshots of the file system after running commands from the Dockerfile are created and stacked on each other~\cite{Combe-2016, docker-image-spec}.
These commands can include shell statements to, e.g., compile an application running in the container.
Furthermore, specific commands exist to embed environment variables or to add files from the host system into the image~\cite{Liu-2020, docker-dockerfile}.
While the files can be, e.g., source code or further dependencies, image creators can also easily and accidentally include (cryptographic) secrets into the image or its environment variables, putting the service's security at risk when leaked.
Once an image has been fully created, it exists as a self-containing unit, which is ready-to-run but also allows little insight on what has been added.

\afblock{Image Push}
After generating the image, creators can \emph{push} it to a registry~\cite{docker-registry-api}, e.g., the official and largest registry Docker Hub~\cite{DockerHub:online}, allowing to deploy containers among an own fleet of servers easily, but also to share it with other users~\cite{docker-what-container}.
To this end, the image layers are uploaded to the registry under a repository name and tag~\circled{3}.
Thereby, the repository name typically represents the application in the image, and the tag describes a version.
Conventionally, creators tag the newest image in a repository with \texttt{latest}.

\afblock{Container Deployment}
To run a Docker container, users \emph{pull} an image from a registry. %
When pulling, users first request an \emph{image manifest}~\cite{docker-image-spec} from the registry, including meta information about the image and its layers.
After downloading all layers~\circled{4}, Docker merges the content composing the file system for the new container~\circled{5}~\cite{Combe-2016}.
The application then finds an unchanged file system with all content provided by the image creator, i.e., all dependencies but also potentially added secrets, and can very likely provide services to the public Internet.
Since numerous containers of various users can base on a single image, included, and thus compromised, secrets could affect several deployments.

\takeaway{%
  The Docker paradigm eases distribution and deployment of applications.
  However, insight into what is added in images and up- or downloaded from a registry can be lost.
  Thus, secrets can be leaked and reused, impairing Internet-reachable services at scale.
}

\begin{table*}[!ht]
\renewcommand{\arraystretch}{1.2}%
\caption{%
Steps from our repository selection to the finally included and analyzed layers.
For found repositories, we give information on their distribution over our selected search terms in a query group, and for images on their distribution over found repositories.
For analyzed images, we report on the distribution of their age, and for layers, we show the distribution of their sizes.
By analyzing \SI{\Remark{distinctnumlayersoverall}}{} layers, we examine all files included in images of considered repositories for leaked secrets.
}%
\vspace{-0.75em}
\footnotesize
\centering
\setlength{\tabcolsep}{4.3pt}

\begin{tabular}{cccccccccccccccc}
\multicolumn{1}{c}{\multirow{2}{*}{\begin{tabular}[c]{@{}l@{}}\textbf{Query}\\\textbf{Group}\end{tabular}}} & & \multicolumn{4}{c}{\textbf{Repositories} \textit{(Section~\ref{sec:selectrepositories})}}                                                                                                                                                                                                                                                                                                                                                                                                                                                                                                                    & & \multicolumn{4}{c}{\textbf{Images} \textit{(Section~\ref{sec:selecttags})}}                                                                                                                                                                                                                                                                                                                                                                                                                                                                                                                                                                                                                                                                                                                                                                                                                                                                                                                                                                          & & \multicolumn{4}{c}{\textbf{Layers} \textit{(Section~\ref{sec:selectlayers})}}                                                                                                                                                                                                                                                                                                                                                                                                                                                                                                                                                                 \\
\cline{3-6}\cline{8-11}\cline{13-16}
\multicolumn{1}{c}{}                                                                      & & \multicolumn{1}{c}{\textbf{\#}}                                                                                                                                                                                                                                                                                                                            & \multicolumn{3}{c}{\textbf{distinct}}                                                                                                                                                                    & & \multicolumn{1}{c}{\textbf{\#}}                                                                                                                                                                                                                                                                                                                                                    & \multicolumn{1}{c}{\textbf{\texttt{latest}}}                                                                   & \multicolumn{1}{c}{\textbf{none}}                                                                              & \multicolumn{1}{c}{\textbf{analyzed [+age]}}                                                                                                                                                                                                                                                                                                                                                                          & & \multicolumn{1}{c}{\textbf{\# [+size]}}                                                                                                                                                                                                                                                                                                                                                                                                & \multicolumn{3}{c}{\textbf{distinct}}                                                                                                                                       \\
\hline\hline
Standard                                                                                  & & \multicolumn{1}{c}{\begin{tabular}[c]{@{}rl@{}} \multicolumn{2}{c}{\SI{\Remark{standardsumrepopergroup}}{}} \\ \textit{min:} & \SI{\Remark{standardminrepopergroup}}{} \\ $ p_{25} $: & \SI{\Remark{standardq25repopergroup}}{} \\ $ p_{50} $: & \SI{\Remark{standardq50repopergroup}}{} \\ $ p_{75} $: & \SI{\Remark{standardq75repopergroup}}{} \\ \textit{max:} & \SI{\Remark{standardmaxrepopergroup}}{} \end{tabular}}   & \SI{\Remark{standardrepopergroup}}{}  & $\xrightarrow[\SI{\Remark{standarddistinctpctrepopergrouponly}}{\percent}]{}$ & \multirow{8}{*}[0.5em]{\begin{tikzpicture} \draw (-0.1, 1.01) -- (0,1.01) -- (0,0.2); \node[yshift=-0.1em]{\SI{\Remark{distinctnumrepooverall}}{}}(0,0); \draw (-0.1, -1.01) -- (0, -1.01) -- (0,-0.2); \end{tikzpicture}} & & \multicolumn{1}{c}{\begin{tabular}[c]{@{}rl@{}} \multicolumn{2}{c}{\SI{\Remark{standardnumtagspergroup}}{}} \\ \textit{min:} & \SI{\Remark{standardmintagsperimagepergroup}}{} \\ $ p_{25} $: & \SI{\Remark{standardq25tagsperimagepergroup}}{} \\ $ p_{50} $: & \SI{\Remark{standardq50tagsperimagepergroup}}{} \\ $ p_{75} $: & \SI{\Remark{standardq75tagsperimagepergroup}}{} \\ \textit{max:} & \SI{\Remark{standardmaxtagsperimagepergroup}}{} \end{tabular}}   & \multicolumn{1}{c}{\SI{\Remark{standardnumtaglatestpergroup}}{}}  & \multicolumn{1}{c}{\SI{\Remark{standardnumtagnonepergroup}}{}}    & \multicolumn{1}{c}{\begin{tabular}[c]{@{}rl@{}} \multicolumn{2}{c}{\SI{\Remark{standardavailabletagspergroup}}{}} \\ \textit{min:} & \SI{\Remark{standardminlastupdateddayspergroup}}{\day} \\ $ p_{25} $: & \SI{\Remark{standardq25lastupdateddayspergroup}}{\day} \\ $ p_{50} $: & \SI{\Remark{standardq50lastupdateddayspergroup}}{\day} \\ $ p_{75} $: & \SI{\Remark{standardq75lastupdateddayspergroup}}{\day} \\ \textit{max:} & unset \end{tabular}}    & & \multicolumn{1}{c}{\begin{tabular}[c]{@{}rl@{}} \multicolumn{2}{c}{\SI{\Remark{standardnumlayerspergroup}}{}} \\ \textit{min:} & \SI{\Remark{standardsizelayersminbytepergroup}}{\byte} \\ $ p_{25} $: & \SI{\Remark{standardsizelayersq25bytepergroup}}{\byte} \\ $ p_{50} $: & \SI{\Remark{standardsizelayersq50kilobytepergroup}}{\kilo\byte} \\ $ p_{75} $: & \SI{\Remark{standardsizelayersq75megabytepergroup}}{\mega\byte} \\ \textit{max:} & \SI{\Remark{standardsizelayersmaxgigabytepergroup}}{\giga\byte} \end{tabular}}     & \SI{\Remark{standarddistinctnumlayerspergroup}}{} & $\xrightarrow[\SI{\Remark{standarddistinctpctlayersgroup}}{\percent}]{}$ & \multirow{8}{*}[0.5em]{\begin{tikzpicture} \draw (-0.1, 1.01) -- (0,1.01) -- (0,0.2); \node[yshift=-0.1em]{\SI{\Remark{distinctnumlayersoverall}}{}}(0,0); \draw (-0.1, -1.01) -- (0, -1.01) -- (0,-0.2); \end{tikzpicture}} \\
\cline{1-5}\cline{8-11}\cline{13-15}
IIoT                                                                                      & & \multicolumn{1}{c}{\begin{tabular}[c]{@{}rl@{}} \multicolumn{2}{c}{\SI{\Remark{iiotsumrepopergroup}}{}} \\ \textit{min:} & \SI{\Remark{iiotminrepopergroup}}{} \\ $ p_{25} $: & \SI{\Remark{iiotq25repopergroup}}{} \\ $ p_{50} $: & \SI{\Remark{iiotq50repopergroup}}{} \\ $ p_{75} $: & \SI{\Remark{iiotq75repopergroup}}{} \\ \textit{max:} & \SI{\Remark{iiotmaxrepopergroup}}{} \end{tabular}}                   & \SI{\Remark{iiotrepopergroup}}{}      & $\xrightarrow[\SI{\Remark{iiotdistinctpctrepopergrouponly}}{\percent}]{}$ &                                                                                                                                                                                                            & & \multicolumn{1}{c}{\begin{tabular}[c]{@{}rl@{}} \multicolumn{2}{c}{\SI{\Remark{iiotnumtagspergroup}}{}} \\ \textit{min:} & \SI{\Remark{iiotmintagsperimagepergroup}}{} \\ $ p_{25} $: & \SI{\Remark{iiotq25tagsperimagepergroup}}{} \\ $ p_{50} $: & \SI{\Remark{iiotq50tagsperimagepergroup}}{} \\ $ p_{75} $: & \SI{\Remark{iiotq75tagsperimagepergroup}}{} \\ \textit{max:} & \SI{\Remark{iiotmaxtagsperimagepergroup}}{} \end{tabular}}                   & \multicolumn{1}{c}{\SI{\Remark{iiotnumtaglatestpergroup}}{}}      & \multicolumn{1}{c}{\SI{\Remark{iiotnumtagnonepergroup}}{}}        & \multicolumn{1}{c}{\begin{tabular}[c]{@{}rl@{}} \multicolumn{2}{c}{\SI{\Remark{iiotavailabletagspergroup}}{}} \\ \textit{min:} & \SI{\Remark{iiotminlastupdateddayspergroup}}{\day} \\ $ p_{25} $: & \SI{\Remark{iiotq25lastupdateddayspergroup}}{\day} \\ $ p_{50} $: & \SI{\Remark{iiotq50lastupdateddayspergroup}}{\day} \\ $ p_{75} $: & \SI{\Remark{iiotq75lastupdateddayspergroup}}{\day} \\ \textit{max:} & unset \end{tabular}}                    & & \multicolumn{1}{c}{\begin{tabular}[c]{@{}rl@{}} \multicolumn{2}{c}{\SI{\Remark{iiotnumlayerspergroup}}{}} \\ \textit{min:} & \SI{\Remark{iiotsizelayersminbytepergroup}}{\byte} \\ $ p_{25} $: & \SI{\Remark{iiotsizelayersq25bytepergroup}}{\byte} \\ $ p_{50} $: & \SI{\Remark{iiotsizelayersq50kilobytepergroup}}{\kilo\byte} \\ $ p_{75} $: & \SI{\Remark{iiotsizelayersq75megabytepergroup}}{\mega\byte} \\ \textit{max:} & \SI{\Remark{iiotsizelayersmaxgigabytepergroup}}{\giga\byte} \end{tabular}}                       & \SI{\Remark{iiotdistinctnumlayerspergroup}}{}     & $\xrightarrow[\SI{\Remark{iiotdistinctpctlayersgroup}}{\percent}]{}$     &                                                                                                                                                                                    \\
\hline\hline
\end{tabular}

\vspace{-1.75em}
\label{tab:hub-overview}
\end{table*}

\section{Related Work}
Three streams of research motivate our analysis of confidential security material in Docker images: studies that detect leaked security material, research on publicly available Docker images, and Internet-wide scans disclosing security weaknesses at scale.

\afblock{Actively Leaked Security Material}
Currently, the search for leaked security material focuses on code repositories.
Several studies detected the leakage of passwords~\cite{PSADont65:online}, SSH private keys~\cite{Hundreds32:online}, Amazon Cloud API keys~\cite{10000Git13:online,Sinha2015DetectingAM}, and Slack API keys~\cite{Slackbot26:online}, using the built-in search of GitHub.
To allow broader searches, researchers entailed regular expressions but focused on specific file types~\cite{Rahman-2019,Rahman-2021,rahman-2021-2} or code snippets~\cite{rahman-2019-2}, i.e., the scale of this research was limited.
In contrast, Meli et al.\ performed a large scale study without focusing on specific file types, showing that $\sim$\SI{3.5}{\percent} of the \SI{4}{\million} analyzed code repositories on GitHub included leaked secrets~\cite{meli-2019}.
Further approaches use machine learning to improve the detection by relying on code semantics~\cite{Feng-2022}, false-positive detection~\cite{Saha-2020}, or both requiring further user input~\cite{KallT21,lrnto-icissp21}.
Away from GitHub, research proposed methods to investigate various platforms~\cite{Farinella-2021} and proved the presence of secrets in publicly available Android apps~\cite{Glanz-2020}.
A recent study underlines that most developers experienced secret leakage, and guidelines are insufficient for prevention~\cite{Krause-2022}.
While retroactively deleting leaked secrets does not help~\cite{WhyDelet25:online}, (non)-commercial approaches, e.g., GitGuardian~\cite{GitGuard70:online}, TruffleHog~\cite{truffles71:online}, or Gitrob~\cite{michenri37:online}, aim at preventing secret leakage for Git.

\afblock{Docker Images}
Besides Git, researchers and developers, early on without evidence, assumed leaked secrets in images for virtual machines or Docker and provided countermeasures~\cite{Balduzzi2012ASA,Combe-2016,Brady-2020,Howtokee46:online,Managese10:online,Huang-2019,Wei2009ManagingSO}.
Nevertheless, non-academic Web-blog studies~\cite{Lowhangi50:online,Turner95:online,Secretse30:online,Scanning55:online} still find leaked secrets in images on Docker Hub.
However, these studies either limit their scale~\cite{Lowhangi50:online,Turner95:online,Secretse30:online} to a few thousand images/secrets or restrict their methodology~\cite{Scanning55:online} to process large amounts of available images.
The latter study~\cite{Scanning55:online} finds \SI{46076}{}~affected images among \SI{6.3}{\million} images on Docker Hub, but only considers information available in Dockerfiles, e.g., specific file paths.
Meanwhile, SecretScanner~\cite{deepfenc70:online}, a smaller secret search tool, implements a function allowing users to find secrets in Docker images.

Still, a comprehensible, large-scale, and methodology-driven analysis on introduced security weaknesses by leaked security material is missing.
Instead, large-scale studies on Docker images focused on data compression~\cite{Zhao-2019-2,Zaho-2019}, software vulnerabilities~\cite{Zerouali-2019,Zerouali-2021,Jain-2021,Liu-2020}, or typosquatting of image names~\cite{liu-2022}.
Hence, as of now, it is unclear how widespread secret leakage is in images on Docker Hub as well as private Internet-reachable registries.
Moreover, it is unknown to what extent these compromised images are then used on the Internet and whether they weaken security at scale.

\afblock{Internet Measurements}
For understanding deployment security at scale, Internet-wide measurements have been a valuable tool in the past.
Internet scan services, such as Shodan~\cite{shodan} or Censys~\cite{durumeric2015search}, fetch and publish meta-information, e.g., security configurations, on Internet-reachable services.
Although these services often helped researchers analyzing the security of connected devices, e.g., cars~\cite{ueda_cars_2022} or (insecure)~Industrial IoT~(IIoT) deployments~\cite{hansson-analysisshodan-2018}, they usually do not see all deployments~\cite{barbieri2021assessing}.
Hence, researchers frequently conduct own active Internet measurement, e.g., using ZMap~\cite{durumeric-zmap-2013}. %
On the web, these measurements allowed to analyze the deployment of new TLS versions~\cite{2020-holz-tls13, lee-2021} and revealed wide security configuration mistakes~\cite{holz-tlscomm-2016,cui-insecureembedded-2010,holz-509pki-2011,chung-certificates-2016,kumar-certificates-2018,hiller20crosssigning,lee_spatial-tls_2021} or implementation deficits~\cite{heninger-mining-2012,adrian15dhfails,springall-tlsshortcuts-2016}.
Aside the web, researchers assessed the security of SSH services~\cite{gasser-ssh-2014,west-2022} and key-value stores leaking confidential data~\cite{Thousand61:online}.
For the IoT and IIoT, research revealed many deployments relying on vulnerable software~\cite{ceron2020online, kiravuo-vulnerabilitiesfinland-2015,genge-shovat-2016} and communicating without any security mechanism~\cite{mirian-icsmes-2016,xu2018increase,dahlmanns-2022,maggi2018fragility,nawrocki-passics-20}, e.g., access control.
Even with built-in security features, operators often configure such services insecurely~\cite{2020-dahlmanns-imc-opcua}.
For example, a massive reuse of certificates was traced back to a Docker image including certificates and corresponding private keys~\cite{dahlmanns-2022} jeopardizing the authenticity of numerous deployments.
Based on this, we claim that it is probable that there are further public Docker images that wrongly include confidential secrets and harm security on the Internet---especially when looking at the sheer size of Docker and Docker Hub.

\takeaway{
  Although the broad leakage of security secrets in code repositories is well understood, the spread of revealed secrets in Docker images and the introduced security risk for the Internet are unknown.
  However, known secret leakage detection techniques and Internet measurements are predestined to shed light on these issues.
}

\section{Composing our Dataset}

To answer whether Docker image creators actively compromise security secrets by publishing them in openly available Docker images, we set out and retrieve images from Docker Hub~(Section~\ref{sec:hub-dataset}) and publicly reachable private registries~(Section~\ref{sec:private-dataset}).

\subsection{Retrieving Images from Docker Hub}
\label{sec:hub-dataset}

Table~\ref{tab:hub-overview} guides through our composition process on Docker Hub, which has three tasks:
\begin{inparaenum}[(i)]
  \item composing a list of \emph{repositories},
  \item selecting one \emph{image} per repository to widely spread our analysis, and
  \item identifying \emph{layers} the images consist of.
\end{inparaenum}

\subsubsection{Repositories}
\label{sec:selectrepositories}

While Docker Hub limits the number of image downloads~\cite{docker-rate-limits} and we cannot download and analyze all \SI{15}{\peta\byte} of images available on Docker Hub~\cite{docker-storage} due to runtime and bandwidth restrictions, our analysis requires a selection of repositories of interest.
Furthermore, Docker Hub does not support listing all available images to choose from.
Hence, we use specific search terms to get images users retrieve when searching via the Web interface.
Our search terms~(which we elaborate in more detail in Appendix~\ref{sec:imageselectionhub}) build two query groups~(Table~\ref{tab:hub-overview}~(left));
\textbf{Standard} comprises mainstream communication protocol names~\cite{schumann-2022} and frequently used technologies~\cite{stackoverflow-survey-2021} for a wide analysis of images referencing current issues.
For comparison and more focusing on a specific area, we choose the Industrial Internet of Things~(IIoT) as past studies showed a great susceptibility to security faults~\cite{2020-dahlmanns-imc-opcua,dahlmanns-2022,nawrocki-passics-20,mirian-icsmes-2016,leverett-shodanclassification-2011,hansson-analysisshodan-2018}, i.e., \textbf{IIoT} includes protocol names from this area.

\begin{table*}[!ht]
\renewcommand{\arraystretch}{1.2}%
\caption{%
Overview of found private Docker registries, available image repositories, their tags, layers, and final layers included in our dataset.
We added \SI{\Remark{privatenumdistinctlayersselected}}{} randomly selected image layers, preferably from images tagged with \texttt{latest}, to our dataset.
}%
\vspace{-0.75em}
\footnotesize
\centering
\setlength{\tabcolsep}{4.1pt}

\begin{tabular}{ccccccccccccccccc}
\multicolumn{1}{c}{\multirow{2}{*}{\textbf{Date}}}  & & \multicolumn{1}{c}{\multirow{2}{*}{\textbf{\# Registries}}}                                                                                                                                                                                                                               & & \multicolumn{3}{c}{\textbf{Repositories}}                                                                                                                                                                                                                                                                                                                                                                                                                                                                                                                                                                                                                                     & & \multicolumn{4}{c}{\textbf{Images}}                                                                                                                                                                                                                                                                                                                                                                                                                                               & & \multicolumn{4}{c}{\textbf{Layers}}                                                                                                                                                                                                                                                                                                                                                                                                                                                                                                                                                                 \\
\cline{5-7}\cline{9-12}\cline{14-17}
                                                    & &                                                                                                                                                                                                                                                                                           & & \multicolumn{1}{c}{\textbf{\#}}                                                                                                                                                                                                                                                                                                                                         & \multicolumn{2}{c}{\textbf{distinct}}                                                                                                                                                                                                                                                               & & \multicolumn{1}{c}{\textbf{\#}}                                                                                                                                                                                                                                         & \multicolumn{1}{c}{\textbf{\texttt{latest}}}                               & \multicolumn{1}{c}{\textbf{none}}                                           & \multicolumn{1}{c}{\textbf{selected}}      & & \multicolumn{1}{c}{\textbf{\#}}          & \multicolumn{2}{c}{\textbf{distinct}}    & \multicolumn{1}{c}{\textbf{selected [+size]}}                                                                                                                        \\
\hline\hline
\Remark{privatemeasurement220801date}               & & \multicolumn{1}{c}{\begin{tabular}[c]{@{}rl@{}} \multicolumn{2}{c}{\SI{\Remark{privatemeasurement220801numtotal}}{}} \\ \textit{non-TLS:} & \SI{\Remark{privatemeasurement220801numtotalnontls}}{} \\ \textit{TLS:} & \SI{\Remark{privatemeasurement220801numtotaltls}}{} \end{tabular}}  & & \multicolumn{1}{c}{\begin{tabular}[c]{@{}rl@{}} \multicolumn{2}{c}{\SI{\Remark{privatemeasurement220801repositorysum}}{}} \\ \textit{min:} & \SI{\Remark{privatemeasurement220801repositorymin}}{} \\ \textit{avg:} & \SI{\Remark{privatemeasurement220801repositorymean}}{} \\ \textit{max:} & \SI{\Remark{privatemeasurement220801repositorymax}}{} \end{tabular}}    & \SI{\Remark{privatemeasurement220801repositorydistinctperdate}}{} & \multirow{6}{*}[0.5em]{\begin{tikzpicture}\draw (-0.1, 0.7) -- (0,0.7) -- (0,0.2); \node[yshift=-0.1em]{\SI{\Remark{privatemeasurementnumdistinctrepositoriesoverall}}{}}(0,0); \draw (-0.1, -0.7) -- (0, -0.7) -- (0,-0.2); \end{tikzpicture}}  & & \multicolumn{1}{c}{\begin{tabular}[c]{@{}rl@{}} \multicolumn{2}{c}{\SI{\Remark{private220801numtagsperdate}}{}} \\ \textit{min:} & \SI{\Remark{private220801mintagsperimageperdate}}{} \\ \textit{avg:} & \SI{\Remark{private220801avgtagsperimageperdate}}{} \\ \textit{max:} & \SI{\Remark{private220801maxtagsperimageperdate}}{} \end{tabular}}                   & \multicolumn{1}{c}{\SI{\Remark{tagspermeasurement220801numlatest}}{}}      & \multicolumn{1}{c}{\SI{\Remark{tagspermeasurement220801numnone}}{}}        & \SI{\Remark{private220801numselectedtagperdate}}{}                          & & \SI{\Remark{private220801numlayerspermeasurement}}{}                       & \SI{\Remark{private220801numdistinctlayerspermeasurement}}{}                       & \multirow{6}{*}[0.5em]{\begin{tikzpicture} \draw (-0.1, 0.7) -- (0,0.7) -- (0,0.2); \node[yshift=-0.1em]{\SI{\Remark{privatenumdistinctlayers}}{}}(0,0); \draw (-0.1, -0.7) -- (0, -0.7) -- (0,-0.2); \end{tikzpicture}}  & \multirow{6}{*}{\begin{tabular}[c]{@{}rl@{}} \multicolumn{2}{c}{\SI{\Remark{privatenumdistinctlayersselected}}{}} \\ \textit{min:} & \SI{\Remark{privateminlayersbyteselected}}{\byte} \\ \textit{avg:} & \SI{\Remark{privateavglayersmegabyteselected}}{\mega\byte} \\ \textit{max:} & \SI{\Remark{privatemaxlayersmegabyteselected}}{\mega\byte} \end{tabular}}                       \\
\cline{1-6}\cline{9-15}
\Remark{privatemeasurement220806date}               & & \multicolumn{1}{c}{\begin{tabular}[c]{@{}rl@{}} \multicolumn{2}{c}{\SI{\Remark{privatemeasurement220806numtotal}}{}} \\ \textit{non-TLS:} & \SI{\Remark{privatemeasurement220806numtotalnontls}}{} \\ \textit{TLS:} & \SI{\Remark{privatemeasurement220806numtotaltls}}{} \end{tabular}}  & & \multicolumn{1}{c}{\begin{tabular}[c]{@{}rl@{}} \multicolumn{2}{c}{\SI{\Remark{privatemeasurement220806repositorysum}}{}} \\ \textit{min:} & \SI{\Remark{privatemeasurement220806repositorymin}}{} \\ \textit{avg:} & \SI{\Remark{privatemeasurement220806repositorymean}}{} \\ \textit{max:} & \SI{\Remark{privatemeasurement220806repositorymax}}{} \end{tabular}}    & \SI{\Remark{privatemeasurement220806repositorydistinctperdate}}{} &                                                                                                                                                                                                                                 & & \multicolumn{1}{c}{\begin{tabular}[c]{@{}rl@{}} \multicolumn{2}{c}{\SI{\Remark{private220806numtagsperdate}}{}} \\ \textit{min:} & \SI{\Remark{private220806mintagsperimageperdate}}{} \\ \textit{avg:} & \SI{\Remark{private220806avgtagsperimageperdate}}{} \\ \textit{max:} & \SI{\Remark{private220806maxtagsperimageperdate}}{} \end{tabular}}                   & \multicolumn{1}{c}{\SI{\Remark{tagspermeasurement220806numlatest}}{}}      & \multicolumn{1}{c}{\SI{\Remark{tagspermeasurement220806numnone}}{}}        & \SI{\Remark{private220806numselectedtagperdate}}{}                          & & \SI{\Remark{private220806numlayerspermeasurement}}{}                       & \SI{\Remark{private220806numdistinctlayerspermeasurement}}{}                       &                                                                                                                                                                                                     &                        \\
\hline\hline
\end{tabular}

\vspace{-1.75em}
\label{tab:private-overview}
\end{table*}

We list the number of repositories covered by our analysis per query group, i.e., the sum of found repositories of all search terms of a group, in Table~\ref{tab:hub-overview}~(column~Repositories-\#).
To further convey the prevalence of our search terms, we indicate the minimum, maximum, and 25-, 50-, and 75-percentiles of search results for included terms, i.e., higher values of lower percentiles would imply a higher prevalence. %
While both query groups contain terms that lead to no results~(\textit{min}), i.e., the term is not mentioned in any repository name or description, terms in the standard group generate more results due to their closer correlation to frequently used technologies than IIoT protocols~($p_{25}$, $p_{50}$, $p_{75}$).
Docker Hub's API limits the number of results to \SI{10000}{}~(\textit{max}).

As different search terms lead to overlapping repositories, we further report on the distinct number of repositories gradually, i.e., per query group, and overall.
In total, we gathered \SI{\Remark{distinctnumrepooverall}}{}~distinct repositories subject to our study of which \SI{\Remark{standarddistinctpctrepopergrouponly}}{\percent} are uniquely added by our standard search terms and \SI{\Remark{iiotdistinctpctrepopergrouponly}}{\percent} by IIoT related search queries.

\subsubsection{Images}
\label{sec:selecttags}

Table~\ref{tab:hub-overview}~(column~Images-\#) indicates how many images were available in total over the distinct repositories of a search group.
While repositories mostly contain different images, including the same software in other versions and thereby comprising similar files, we choose to analyze one tag per repository to spread our analysis as widely as possible.
Here, we select images tagged with \texttt{latest} which is used as Docker's default and typically includes the newest version of an image.
However, not all repositories contain images tagged with \texttt{latest}~(as shown in Table~\ref{tab:hub-overview}~(column~Images-\texttt{latest}).
Here, we select the image with the latest changes~(as reported by Docker Hub's API).
Empty repositories~(Table~\ref{tab:hub-overview}~(column~Images-none)), i.e., have no image layers available, cannot include any secrets.
Besides the number of images that are covered by our study~(column~Images-analyzed), we also report on the age of the images to analyze how long they are already available on Docker Hub.
The ages of images included in both query groups roughly have the same distribution indicating that although the number of images found by our IIoT-related queries is lower image creators update their images in the same frequency as image creators of images included in our Standard group.

\subsubsection{Layers}
\label{sec:selectlayers}

While we report on the number of layers included in all images~(Table~\ref{tab:hub-overview} column~Layers-\#), different images often share the same layers, e.g., layers from frequently used base images.
Hence, to speed up our search for leaked secrets, we analyze each distinct layer only once.
We show the distinct number of layers gradually, i.e., per query group, and overall.
To cover all \SI{\Remark{distinctnumrepooverall}}{} repositories, we analyze \SI{\Remark{distinctnumlayersoverall}}{}~layers.
~(\SI{\Remark{standarddistinctpctlayersgroup}}{\percent} uniquely added by Standard-related, \SI{\Remark{iiotdistinctpctlayersgroup}}{\percent} by IIoT-related repositories).

\subsection{Images from Private Docker Registries}
\label{sec:private-dataset}

Since image creators might upload sensitive images preferably to private registries, we want to include images from these registries in our analysis.
Table~\ref{tab:private-overview} shows our steps taken to extend our dataset with images from private registries, i.e., we search private registries, and, subsequently, include a subset of available layers.

\subsubsection{Find Private Registries and Repositories}
\label{sec:findprivate}

To find publicly reachable Docker registries, we scan the complete IPv4 address space for services running on the standard port for Docker registries, i.e., TCP port~5000, under comprehensive ethical measures~(cf.\ Appendix~\ref{sec:ethics}) twice to analyze short-term fluctuations~(Table~\ref{tab:private-overview}~(left)).
Both times, we perform a TCP~SYN scan using \texttt{zmap}~\cite{durumeric-zmap-2013}, identifying hosts running a service behind this port and subsequently send an HTTP request as defined by Docker's Registry API~\cite{docker-registry-api} for verification.
Whenever we do not receive a valid HTTP response, we retry via HTTPS.
While we found up to \SI{\Remark{privatemeasurementnumtotalmax}}{} private registries on \Remark{privatemeasurementdatemax}, the difference in found registries in comparison to our scan on \Remark{privatemeasurementdatemin} is due to registries in Amazon AWS-related ASes that do not reply after our first scan anymore.
Since these registries only contain the same and single image~(uhttpd), they might relate to another research project, e.g., implementing a registry honeypot.

Contrarily to Docker Hub's API, the API of private registries allows listing available repositories without search terms.
However, we limit our requests to receive a maximum of 100~repositories per registry to prevent any overloads.
As such, the found private registries provide \SI{\Remark{privatemeasurement220801repositorysum}}{} resp.\ \SI{\Remark{privatemeasurement220806repositorysum}}{}~repositories.
Since the registries do not implement access control for read access, clients are able to download all included images.
Notably, by default also write access is not restricted~\cite{dockerdeployregistry}, i.e., attackers might be able to inject malware. %

\assert{\equal{\Remark{privatemeasurement0repository}}{uhttpd}}
\assert{\equal{\Remark{privatemeasurement2repository}}{nginx}}
\assert{\equal{\Remark{privatemeasurement4repository}}{redis}}

\begin{table*}[!t]
\renewcommand{\arraystretch}{1.2}%
\caption{%
Domains of secrets covered in our analysis, their potential threats~(left), the number of (distinct) matches we found with our corresponding regular expressions in images on Docker Hub and private registries~(center), as well as the number of matches we validated~(right). We excluded rules with too arbitrary, thus unverifiable, matches~(\textcolor{gray}{gray}) from validation process.
}%
\vspace{-0.75em}
\footnotesize
\centering
\setlength{\tabcolsep}{2.1pt}
\arrayrulecolor[rgb]{0.753,0.753,0.753}
\begin{tabular}{cccccccccc}
\multicolumn{3}{c}{\textbf{Regular Expressions}~\textit{(Section~\ref{sec:regexselection} / Appendix~\ref{sec:regularexpressions})}}                                                                                                                                                        & & \multicolumn{2}{c}{\textbf{(Distinct) Matches} \textit{(Sec.~\ref{sec:matching})}}  & & \multicolumn{3}{c}{\textbf{Valid Secrets} \textit{(Section~\ref{sec:match-validation})}}                                                                \\ 
\arrayrulecolor{black}\cline{1-3}\cline{5-6}\cline{8-10}
\multicolumn{2}{c}{\textbf{Domain}}                             & \textbf{Potential Threat / \textit{(Service) Type}}                                                                                                                                                                                      & & \textbf{Images}                 & \textbf{Variables}   & & \textbf{Images}                 & \textbf{Variables}   & \textbf{Total}                      \\
\arrayrulecolor{black}\hline\hline
\multicolumn{2}{c}{\multirow{2}{*}{Private Key}}                & Perform man-in-the-middle attacks, fake identity, \dots                                                                                                                              & & \multirow{2}{*}{\begin{tabular}[c]{@{}cc@{}} \SI{\Remark{rawprivatekeynummatchesimages}}{} \\ (\SI{\Remark{imageprivatekeynumdistinctmatches}}{}) \end{tabular}} & \multirow{2}{*}{\begin{tabular}[c]{@{}cc@{}} \SI{\Remark{rawprivatekeynummatchesval}}{} \\ (\SI{\Remark{valprivatekeynumdistinctmatches}}{}) \end{tabular}} & & \multirow{2}{*}{\SI{\Remark{validimageprivatekeyvalidnumdistinctmatches}}{}} & \multirow{2}{*}{\SI{0}{}} & \multirow{2}{*}{\SI{\Remark{validprivatekeyvalidnumdistinctmatchestotal}}{}}   \\ 
\arrayrulecolor[rgb]{0.753,0.753,0.753}\cline{3-3}
\multicolumn{2}{c}{}                                            & \begin{tabular}[c]{@{}c@{}}\textit{PEM Private Key, PEM Private Key Block, PEM PKCS7, XML Private Key}\end{tabular}                                                         & &                                     &                               & &                                     &                               &                                     \\ 
\arrayrulecolor{black}\hline\hline
\parbox[t]{2mm}{\multirow{12}{*}{\rotatebox[origin=c]{90}{\textbf{API}}}} & \multirow{4}{*}{Cloud}        & \begin{tabular}[c]{@{}c@{}}Manage services, create new API keys, reconfigure DNS, access emails / SMS,\\control voice calls, read / alter private repositories, \dots\end{tabular}                                                                               & & \multirow{4}{*}{\begin{tabular}[c]{@{}cc@{}} \SI{\Remark{rawapicloudnummatchesimages}}{} \\ (\SI{\Remark{imageapicloudnumdistinctmatches}}{}) \end{tabular}} & \multirow{4}{*}{\begin{tabular}[c]{@{}cc@{}} \SI{\Remark{rawapicloudnummatchesval}}{} \\ (\SI{\Remark{valapicloudnumdistinctmatches}}{}) \end{tabular}} & & \multirow{4}{*}{\SI{\Remark{validimageapicloudvalidnumdistinctmatches}}{}} & \multirow{4}{*}{\SI{\Remark{validvalapicloudvalidnumdistinctmatches}}{}} & \multirow{4}{*}{\SI{\Remark{validapicloudvalidnumdistinctmatchestotal}}{}}   \\ 
\arrayrulecolor[rgb]{0.753,0.753,0.753}\cline{3-3}
                                &                               & \begin{tabular}[c]{@{}c@{}}\textit{Alibaba\textsuperscript{\cite{truffles71:online}}, Amazon AWS\textsuperscript{\cite{truffles71:online}}, \textcolor{gray}{Azure\textsuperscript{\cite{truffles71:online}}}, \textcolor{gray}{DigitalOcean\textsuperscript{\cite{truffles71:online}}}, Github\textsuperscript{\cite{truffles71:online}}, Gitlab (\textcolor{gray}{v1}, v2)\textsuperscript{\cite{truffles71:online}}}, \\ \textit{Google Cloud\textsuperscript{\cite{truffles71:online}}, Google Services\textsuperscript{\cite{meli-2019}}, Heroku\textsuperscript{\cite{truffles71:online}}, IBM Cloud Identity Service\textsuperscript{\cite{truffles71:online}},}\\\textit{\textcolor{gray}{Login Radius\textsuperscript{\cite{truffles71:online}}}, MailChimp\textsuperscript{\cite{meli-2019}}, MailGun\textsuperscript{\cite{meli-2019}}, Microsoft Teams\textsuperscript{\cite{truffles71:online}}, \textcolor{gray}{Netlify\textsuperscript{\cite{truffles71:online}}}, Twilio\textsuperscript{\cite{meli-2019}}}\end{tabular}       & &                                     &                               & &                                     &                               &                                     \\ 
\arrayrulecolor{black}\cline{2-10}
                                & \multirow{3}{*}{Financial}    & List / perform payments, inspect / alter invoices, \dots                                                                                                                             & & \multirow{3}{*}{\begin{tabular}[c]{@{}cc@{}} \SI{\Remark{rawapifinancialnummatchesimages}}{} \\ (\SI{\Remark{imageapifinancialnumdistinctmatches}}{}) \end{tabular}} & \multirow{3}{*}{\begin{tabular}[c]{@{}cc@{}} \SI{\Remark{rawapifinancialnummatchesval}}{} \\ (\SI{\Remark{valapifinancialnumdistinctmatches}}{}) \end{tabular}} & & \multirow{3}{*}{\SI{\Remark{validimageapifinancialvalidnumdistinctmatches}}{}} & \multirow{3}{*}{\SI{\Remark{validvalapifinancialvalidnumdistinctmatches}}{}} & \multirow{3}{*}{\SI{\Remark{validapifinancialvalidnumdistinctmatchestotal}}{}}   \\ 
\arrayrulecolor[rgb]{0.753,0.753,0.753}\cline{3-3}
                                &                               & \begin{tabular}[c]{@{}c@{}}\textit{Amazon MWS\textsuperscript{\cite{meli-2019}}, \textcolor{gray}{Bitfinex\textsuperscript{\cite{truffles71:online}}}, \textcolor{gray}{Coinbase\textsuperscript{\cite{truffles71:online}}}, \textcolor{gray}{Currency Cloud\textsuperscript{\cite{truffles71:online}}}, \textcolor{gray}{Paydirt\textsuperscript{\cite{truffles71:online}}}, \textcolor{gray}{Paymo\textsuperscript{\cite{truffles71:online}}}, }\\\textit{\textcolor{gray}{Paymongo\textsuperscript{\cite{truffles71:online}}}, Paypal Braintree\textsuperscript{\cite{meli-2019}}, Picatic\textsuperscript{\cite{meli-2019}}, Stripe\textsuperscript{\cite{meli-2019}}, Square\textsuperscript{\cite{meli-2019}}, \textcolor{gray}{Ticketmaster\textsuperscript{\cite{truffles71:online}}}, \textcolor{gray}{WePay\textsuperscript{\cite{truffles71:online}}}}\end{tabular}   & &                                     &                               & &                                     &                               &                                     \\ 
\arrayrulecolor{black}\cline{2-10}
                                & \multirow{2}{*}{Social Media} & Tweet, access direct messages, retrieve relationships, \dots                                                                                                                         & & \multirow{2}{*}{\begin{tabular}[c]{@{}cc@{}} \SI{\Remark{rawapisocialmedianummatchesimages}}{} \\ (\SI{\Remark{imageapisocialmedianumdistinctmatches}}{}) \end{tabular}} & \multirow{2}{*}{\begin{tabular}[c]{@{}cc@{}} \SI{\Remark{rawapisocialmedianummatchesval}}{} \\ (\SI{\Remark{valapisocialmedianumdistinctmatches}}{}) \end{tabular}} & & \multirow{2}{*}{\SI{\Remark{validimageapisocialmediavalidnumdistinctmatches}}{}} & \multirow{2}{*}{\SI{\Remark{validvalapisocialmediavalidnumdistinctmatches}}{}} & \multirow{2}{*}{\SI{\Remark{validapisocialmediavalidnumdistinctmatchestotal}}{}}   \\ 
\arrayrulecolor[rgb]{0.753,0.753,0.753}\cline{3-3}
                                &                               & \textit{Facebook\textsuperscript{\textcolor{gray}{\cite{truffles71:online}}, \cite{meli-2019}}, Twitter\textsuperscript{\textcolor{gray}{\cite{meli-2019}}}}                                                                                                                                                            & &                                     &                               & &                                     &                               &                                     \\
\arrayrulecolor{black}\cline{2-10}
                                & \multirow{2}{*}{IoT}          & Retrieve (privacy-sensitive) IoT data, e.g., track cars, \dots                                                                                                                           & & \multirow{2}{*}{\begin{tabular}[c]{@{}cc@{}} \SI{\Remark{rawapiiotnummatchesimages}}{} \\ (\SI{\Remark{imageapiiotnumdistinctmatches}}{}) \end{tabular}} & \multirow{2}{*}{\begin{tabular}[c]{@{}cc@{}} \SI{0}{} \\ (\SI{0}{}) \end{tabular}} & & \multirow{2}{*}{\SI{\Remark{XX}}{}} & \multirow{2}{*}{\SI{\Remark{XX}}{}} & \multirow{2}{*}{\SI{\Remark{XX}}{}}   \\ 
\arrayrulecolor[rgb]{0.753,0.753,0.753}\cline{3-3}
                                &                               & \textit{\textcolor{gray}{Accuweather\textsuperscript{\cite{truffles71:online}}}, Adafruit IO\textsuperscript{\cite{truffles71:online}}, \textcolor{gray}{OpenUV\textsuperscript{\cite{truffles71:online}}}, \textcolor{gray}{Tomtom\textsuperscript{\cite{truffles71:online}}}}                                                                                                                                     & &                                     &                               & &                                     &                               &                                     \\
\arrayrulecolor{black}\hline\hline
\end{tabular}
\vspace{-1.75em}
\label{tab:matchtargets}
\end{table*}

While being publicly available on private registries but not filtered by any search terms, the content of these images is of special interest.
Here, often the repository name indicates the image's content and thus allows conclusions on widely distributed applications, i.e., over both measurements, \texttt{\Remark{privatemeasurement0repository}} is the most reoccurring repository name~(reoccurring \SI{\Remark{privatemeasurement0sum}}{}~times, but only during our first scan).
Repository names on the second and third place, i.e., \texttt{\Remark{privatemeasurement2repository}} and \texttt{\Remark{privatemeasurement4repository}}, indicate proxy and cloud services where image creators might have included security secrets before uploading it to their registry.
Beyond the scope of security secrets, other repository names occurring less often, e.g., \texttt{api-payments-gateway\_prod} or \texttt{smarthome\_web}, imply that image creators might include confidential software, source code, private data, or information on systems especially worthy of protection in openly available Docker images.

\subsubsection{Image and Layer Selection}
\label{sec:imageprivate}
For all found repositories, we collect the lists of available images and their tags~(Table~\ref{tab:private-overview}~(center)).
Although private registries typically do not implement any rate limiting like Docker~Hub, we do not want to overload found registries or their Internet connections.
Hence, to spread our analysis as far as possible but limit the load on each registry, we choose one tag per image.
Similar to our selection process on Docker Hub, typically, in each repository, we select images tagged as \texttt{latest} to download the corresponding manifest.
Whenever no \texttt{latest} image is available, we sort all available images naturally by their tag~(to account for version numbers as tags), and select the maximum~(i.e., the newest version), as the API does not provide any information on the latest changes.
Subsequently, we download the corresponding image manifests to retrieve accompanying layers.
To further limit load on Internet connections of found registries, we do not download all available layers for included secrets.
Instead, we randomly select layers of chosen images such that the sum of their sizes does not exceed \SI{250}{\mega\byte} per registry and per measurement.
All in all, we added \SI{\Remark{privatenumdistinctlayersselected}}{}~layers from private registries to our dataset.

\takeaway{
In parallel to Docker Hub numerous private registries exist providing images to the public.
Overall, we assemble a dataset of \SI{\Remark{numconsideredlayersoverall}}{}~layers from \SI{\Remark{numnonemptyimages}}{}~images subject to our future research.
Furthermore, private registries might allow attackers to, e.g., inject malware, potentially infecting container deployments at scale as well.
}

\section{Leaked Secrets in Docker Images}

Next, we search in considered images for included secrets~(Section~\ref{sec:layer-analysis}), discuss the origin of affected images to later evaluate remedies~(Section~\ref{sec:origin-analysis}), and analyze also found certificates compromised due to private key leakage to estimate arising risks~(Section~\ref{sec:accmaterial}).

\subsection{Searching for Secrets}
\label{sec:layer-analysis}

To analyze available images for included secrets, we align our approach to established methods~\cite{meli-2019,truffles71:online}, i.e., we choose and extend regular expressions identifying specific secrets and match these on files and environment variables.
Additionally, we extensively filter our matches to exclude false positives.

\subsubsection{Regular Expression Selection}
\label{sec:regexselection}

We base our selection of regular expressions on previous work to find secrets in code repositories~\cite{meli-2019,truffles71:online}~(we further elaborate on our election process and expressions in Appendix~\ref{sec:regularexpressions}). %
Table~\ref{tab:matchtargets}~(left) names the domains of secrets that our selected expressions match and indicates how attackers could misuse these secrets.
We start with regular expressions composed by Meli~et~al.~\cite{meli-2019} due to their selection of unambiguous expressions~(reducing false positives) matching secrets with a high threat when leaked.
We extend their expressions for private keys to match a larger variety, e.g., also OpenSSH private keys.
Moreover, we widen the set by expressions matching API secrets of trending technologies~\cite{stackoverflow-survey-2021} based on match rules from TruffleHog~\cite{truffles71:online}.
However, TruffleHog's rules are relatively ambiguous and incur many false positives, which TruffleHog filters by validating the API secrets against their respective endpoints.
As our ethical considerations do not allow for any further use of the secrets~(cf.\ Appendix~\ref{sec:ethics}), we focus on rules which expect at least one fixed character and later add further filtering and verification steps.

\subsubsection{Matching Potential Secrets}
\label{sec:matching}

To analyze whether image layers include secrets, we match the selected regular expressions on the images as follows~(we will open-source our tool on acceptance of this paper):
We download and decompress the image layers and then match our regular expressions on the included files. %
Moreover, we recursively extract archive files up to a depth of 3 and match again.
As API documentations often suggest setting secrets in environment variables and not writing them into files, we analyze set variables.
Since Docker allows downloading the small image configuration containing set variables aside of the image, i.e., potential attackers do not have to download and search through all files to find included secrets, we analyze variables separately:
As such, we only download the image configuration file and iterate our regular expression over set environment variables.
Here, we adapt the API expressions, as some expect a specific term before the secret~(cf.\ Table~\ref{tab:regularexpressions} in Appendix~\ref{sec:regularexpressions}), e.g., the service name as part of a variable name.
As the variable names and values are separated in the configuration file, we also split the according expressions and match them individually.

Table~\ref{tab:matchtargets}~(center) lists for each secret domain how many matches and how many distinct matches we found in both, image content and environment variables.
Notably, while only covering two services, i.e., Facebook and Twitter, the expressions in the Social Media domain matched most often over all domains, which already indicates that API~secrets of this domain are often suspect to leakage.

The high redundancy of the matches, visible as the significant decrement between distinct and non-distinct matches, already hints at invalid matches, e.g., private keys or example API tokens prevalent in unit tests or documentation in several layers.
Indeed, the most reoccurring match \texttt{\Remark{mostreoccurringdata}}~(\SI{\Remark{mostreoccurringnumocc}}{}~times in \SI{\Remark{mostreoccurringnumlayer}}{} different layers), is an example key for \Remark{mostreoccurringrule} from a library documentation which creators usually include in their images.
We thus validate our matches extensively.

\subsubsection{Match Validation}
\label{sec:match-validation}

\begin{figure}[!t]
\centering
\includegraphics[width=\columnwidth]{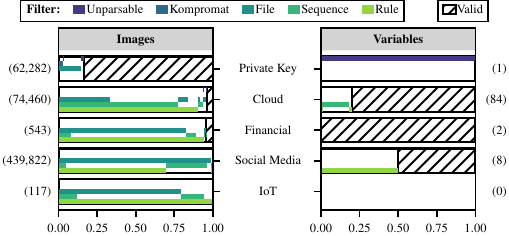}
\vspace{-2em}
\caption{Validation of matches. While most private key matches are valid, API secret matching in Docker images is challenging and requires well-configured filters. Several filters treat a large share of matches in parallel. Absolute number of distinct matches in parentheses.}
\label{fig:plt-filtering}
\vspace{-1.25em}
\end{figure}

To exclude test keys for cryptographic libraries, example API~secrets, and completely invalid matches to get a near lower bound of harmful leaked secrets in Docker images, we use different filters depending on the secret type.
While we show the number of resulting valid secrets in Table~\ref{tab:matchtargets}~(right), Figure~\ref{fig:plt-filtering} details the filtering results separated by the match's origin, i.e., image content or environment variable and domain.

\afblock{Private Keys}
Our regular expressions for private keys match on PEM or XML formatted keys.
Thus, we can first exclude every match that is not parsable~(filter \textit{Unparsable}). %
Figure~\ref{fig:plt-filtering} shows that only a minority of all potential private keys in image layers are unparsable, underlining that image creators include and compromise private keys actually usable in final Docker containers for practical operations.
Contrarily, the single match within the environment variables is only a key fragment and thus not parsable.

Still, we expect a high number of software test keys in Docker images among found keys, as they are part of several libraries creators might include in their images, e.g., OpenSSL.
Since users will most likely not use such keys to secure their deployments, we filter out test keys that are included in kompromat~\cite{kompromat}, a repository listing already compromised secrets~(filter \textit{Kompromat}).
More specifically, we filter keys occurring in RFCs~(\SI{\Remark{kompromatfoundrfcnumdistinct}}{}), libraries for software tests~(\SI{\Remark{kompromatfoundsoftwaretestsnumdistinct}}{}), or as special test vectors~(\SI{\Remark{kompromatfoundtestvectorsnumdistinct}}{}).

To also account for software test keys that are not available in kompromat, we analyze the file paths where respective keys were found~(filter \textit{File}).
While we do not generally exclude all paths containing signal words indicating test or example keys, as users might use such paths also for keys they generated and use in practice, we apply different measures.
For instance, based on locations of test keys identified using kompromat, we deliberately exclude matches in similar locations, i.e., keys within directories where we already detected test keys and all parent directories under which we find more than \(\nicefrac{2}{3}\) test keys.
Last, we exclude file paths typically used by libraries~(cf.\ Appendix~\ref{sec:excludedpaths}), e.g., \texttt{/var/lib/*}, as there is a lower chance that users adapt their keys here.

Figure~\ref{fig:plt-filtering} shows that these filters process the largest share of excluded private key matches.
It further indicates that kompromat only includes a minority of software test keys, i.e., is not directly usable to exclude all false-positive matches.
Still, many of the found keys are not filtered and, thus, most likely, no software test keys.

In total, we found \SI{\Remark{validprivatekeyvalidnumdistinctmatchestotal}}{}~valid private keys potentially in use in practice~(cf.\ Table~\ref{tab:matchtargets}~(right)).
Since all of these keys are located in files, attackers would have to download respective image layers to get access and not only meta information to retrieve environment variables.
Still, since these keys are publicly available and thus compromised, usage in production puts authentication at stake, i.e., attackers can perform impersonation attacks.

\afblock{API Secrets}
Since our ethical considerations deter us from validating API secrets against their service endpoints~(cf.\ Appendix~\ref{sec:ethics}) as applied by TruffleHog~\cite{truffles71:online}, and related methods for false positive detection focus on matches in source code~\cite{Feng-2022,Saha-2020,KallT21,lrnto-icissp21}, which is not prevalent in Docker images, we need alternative measures to filter invalid matches.
By manually supervising our filtering, we ensure that the final set only includes valid-looking API secrets.

Based on invalid matches in GitHub code repositories~\cite{meli-2019}, we expect human-created example keys that contain keywords, e.g., \texttt{EXAMPLE}, or consecutive character sequences, e.g., \texttt{XXXX}, that we must exclude~(filter \textit{Sequence}).
To filter consecutive sequences, we search for segments consisting of ascending, descending~(both with a length of four), and repeating characters~(with a length of three).
Furthermore, we filter matches including sequences that occur unusually often, i.e., we create (\Remark{frequencyngrammin}, \Remark{frequencyngrammax})-character-grams of all matches, exclude grams created over fixed parts of our regular expressions as well as grams only containing digits, and count the number of occurrences over all API matches.
To account for randomly reoccurring grams, we filter all matches that include grams occurring \SI{\Remark{frequencyNgramsTimeFactor}}{}~times more often than the average.
We manually ensured that our filter is not too restrictive but also not to loose leaving often reoccurring grams out.
Figure~\ref{fig:plt-filtering} shows that this filtering excludes a large share of matches.
Interestingly, the most reoccurring gram is \texttt{\Remark{topngramfilter0ngram}}~[sic!], which we could trace back to DNA sequences in images related to bioinformatics underpinning the large variety of different and unexpected file types occurring in Docker images.

Similar to filtering private key matches by their file paths, we also filter API matches occurring in manually selected paths~(filter \textit{File}, cf.\ Appendix~\ref{sec:excludedpaths}).
Essentially, we revisited the location and file types of all matches and excluded paths that most likely do not include any valid secrets compromised by publishing these in Docker images.
Figure~\ref{fig:plt-filtering} indicates that the filtered paths often also include matches filtered by our sequence filter and thus that libraries include strings similar to secrets, e.g., in their documentation.

Still, after manual revision of the remaining matches, we conclude that rules which match on a fixed term before the secret, e.g., the service name, and then allow a specific length of characters are too ambiguous for usage on files in Docker images as they match on arbitrary content, e.g., on hashes with the service name in front.
We thus decide to exclude matches of these rules from our further analysis (gray in Table~\ref{tab:matchtargets}~(left)), i.e., consider these matches invalid, to ensure the integrity of our further results.
Still, a minority of these matches might be valid, potentially enabling attackers to compromise production services or access confidential data.

Comparing the filter results of API secret matches in files and environment variables, the share of valid matches in variables is significantly higher than in files indicating that image creators less likely include secret placeholders in variables.
Still, as Table~\ref{tab:matchtargets}~(right) shows, most secrets are located within the images.
Thus, attackers have a higher chance of finding valid secrets when downloading both environment variables and image content.

In total, we found \SI{\Remark{apinumdistinctmatches}}{}~distinct API secrets in Docker images, mostly related to services from the cloud domain~(\Remark{validapicloudvalidnumdistinctmatchestotal}{}~secrets).
Although we cannot prove the functionality of these secrets, the occurrence of \SI{\Remark{apicloud1numdistinctmatches}}{}~secrets for the \Remark{apicloud1rule} or \Remark{apicloud2numdistinctmatches}~secrets for the \Remark{apicloud2rule} indicate that attackers might be able to reconfigure cloud services maliciously, e.g., by editing DNS or VM options.
Additionally, we found evidence for secrets allowing attackers to access private data from social media (\SI{\Remark{validapisocialmediavalidnumdistinctmatchestotal}}{} secrets), or even access financial services (\SI{\Remark{validapifinancialvalidnumdistinctmatchestotal}}{} secrets, most matches: \Remark{apifinancial0rule}).
Notably, although we focused our image search partly on IoT terms, we found no valid secrets from selected IoT services.

\subsubsection{Secrets Owned by Single Users}
\label{sec:sourcelayers}

\begin{figure}[!t]
\centering
\includegraphics[width=\columnwidth]{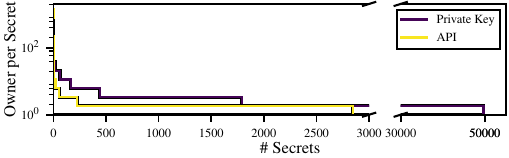}
\vspace{-2.5em}
\caption{Total secrets found in images of a specific number of owners. Most were found in images of single owners.}
\label{fig:src-layerhist}
\vspace{-1.25em}
\end{figure}

Based on findings over leaked secrets found on GitHub~\cite{meli-2019}, we expect most valid secrets to residing in images of single users (as users do not share their secrets intentionally).
Contrarily, invalid matches, e.g., library test keys, would mainly reside in images of multiple owners.

Thus, to check whether the matches we identified as valid secrets are located in images of single users, we analyze the number of different owners that include a specific secret in their images.
To this end, for images from Docker Hub, we consider the repository owner~(embedded in the repository name) as the owner of a secret.
For private registries, we consider the registry's IP address as the owner~(assuming that owners only run a single registry and neglecting that registries might use different (dynamic) IP addresses).

Figure~\ref{fig:src-layerhist} shows that the largest share of valid secrets indeed occurs in images of single owners.
\SI{\Remark{validmatchmultiuserprivatekeyFalsepct}}{\percent} of private keys (\SI{\Remark{validmatchmultiuserprivatekeyFalsenum}}{}~keys) and \SI{\Remark{validmatchmultiuserapiFalsepct}}{\percent} of API secrets (\SI{\Remark{validmatchmultiuserapiFalsenum}}{}~secrets) reside in images of single owners underpinning that these should be protected.
Moreover, we can trace \SI{\Remark{validmatchmultiuserlayer0privatekeyTruenum}}{}~private keys and \SI{\Remark{validmatchmultiuserlayer0apiTruenum}}{}~API secrets of multiple owners back to inheritance.
These secrets were already included in the base image, but w.r.t.\ to the overall occurrence, we conclude that secret spread due to inheritance is no major problem.

To responsibly inform image creators about leaked secrets in their images, we reach out to them whenever possible~(\SI{\Remark{numemaildisclosure}}{}~extractable and valid e-mail addresses) and also contacted the operator of Docker Hub~(cf.\ Appendix~\ref{sec:ethics}).
Early on, we received notifications of creators that removed found secrets from their images.

\takeaway{%
\SI{\Remark{totalvalidmatches}}{} found secrets show that image creators publish confidential information in their publicly available Docker images.
As attackers have access to these secrets relying authentication and other security mechanisms are futile, potentially leading to compromised servers or leaked privacy-sensitive data.
}

\subsection{Origin of Leaked Secrets}
\label{sec:origin-analysis}

Next, we analyze where the validated secrets stem from to see whether specific images are more affected and why.
To this end, we examine the distribution of affected images and compare between private registries and Docker Hub, as well as IIoT specific and Standard images.
Moreover, we evaluate which operation in the original Dockerfile led to the insertion of secrets and inspect the file paths where they reside to get an intuition for their usage.

\subsubsection{Docker Hub Leads Before Private Registries}
\label{sec:sourceregistry}
We already discovered that private registries include potentially sensitive images.
However, until now, it remains unclear whether images on these registries are more often subject to secret leakage than images from Docker Hub, e.g., due to creators believing that these are unavailable for the public.
Thus, we analyze whether leaked secrets occur more often in images from Docker Hub or from private registries.

While we found that \SI{\Remark{numaffectedimages}}{}~images (\SI{\Remark{pctaffectedimages}}{\percent} of images analyzed) contain valid secrets, \SI{\Remark{pctaffectedimagesdockerhub}}{\percent}~of images from Docker Hub and \SI{\Remark{pctaffectedimagesprivate}}{\percent}~of images from private registries are affected.
Thus, creators upload secrets to Docker Hub more often than to private registries indicating that private registry users may have a better security understanding, maybe due to a deeper technical understanding required for hosting a registry.
Yet, both categories are far from being leak-free.

For Docker Hub, besides the increased fraction of leaked secrets, we see an issue for others, i.e., other users can easily deploy containers based on these images.
Thus, there is a higher chance their containers rely their security on included and compromised secrets.
For example, a shared certificate private key could lead to an impersonation attack.
In case of shared API secrets, all deployed containers might use the same API token leading to exhausted rate limits in the best case, but maybe also to overwritten or insufficiently secured private data.
As a single API token does not allow fine-granular exclusions, i.e., it is either valid or revoked for all users, a revocation would also interfere with benign users.

Independent of their origin, attackers could equally misuse the secrets we found to leverage authentication or access privacy- or security-sensitive data.
As such, both user groups of Docker Hub and private registries leak sensitive information, be it through unawareness or a deceptive feeling of security.

\subsubsection{Domains are Similarly Affected}
\label{sec:sourcegroup}

For our image selection on Docker Hub, we specifically included search terms relating to the IIoT, as past research has shown significant security shortcomings in this area.
However, until now it is open whether found images of a certain domain are suspect to revealed secrets more frequently than other images.
To answer this question, we trace images that include secrets back to the query group that led to their inclusion.

We discovered that \SI{\Remark{affectedstandardrepositorypct}}{\percent} of the images only found using queries from the Standard query group and \SI{\Remark{affectediiotrepositorypct}}{\percent} of images only from the IIoT group include valid secrets\footnote{Images found by both query groups are not included.}.
Thus, in case of secret leakage via Docker images and based on our selected search terms, the IIoT domain does not perform worse than our Standard domain.
However, it underpins that the problem of secret leakage in Docker images is a prominent issue for all domains.

\subsubsection{Fresh Private Keys and Copied API Secrets}
\label{sec:history}

\begin{figure}[!t]
\centering
\includegraphics[width=\columnwidth]{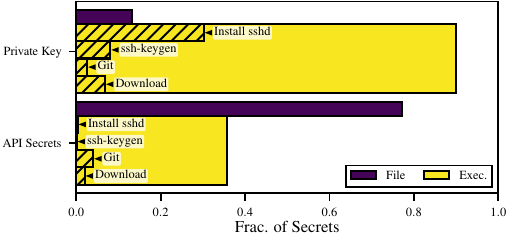}
\vspace{-2.5em}
\caption{Operations that include secrets. Most API secrets are inserted by file and private keys by execution commands.}
\label{fig:src-operation}
\vspace{-1.25em}
\end{figure}

To find countermeasures against secret leakage in Docker images, it is important to understand how these leaked secrets became part of Docker images.
More specifically, for private keys, it is unclear whether creators execute commands in the Dockerfile to create fresh keys, which are then published in images, or whether they manually add them, i.e., using \texttt{ADD} or \texttt{COPY} in a Dockerfile.
Additionally, both, private keys and API secrets, could be indirectly included through other means, e.g., by cloning Git repositories or downloading further data.

Figure~\ref{fig:src-operation} shows that while most API secrets are typically inserted by file operations~(\textit{File}), e.g., copied from the image creator's host system, private keys are predominantly included by executing a command within the Dockerfile~(\textit{Exec.})\footnote{Secrets can be associated with both, File and Exec.\ operations, e.g., when first \texttt{ADD}ed to the image and then copied or moved internally using \texttt{cp} or \texttt{mv}.}.
Thus, private keys might be either downloaded or generated during the creation process.

To further trace the insertion of secrets in Exec.\ layers back to the responsible executed commands, we analyze these commands.
Since image creators often concatenate several bash commands whose output is then included in a single layer without any opportunity to associate files~(and thus secrets) to a specific command, we count each of the commands related to the leakage of a secret.
We show the most prominent of all \SI{\Remark{validmatchnumdistinctcommands}}{}~commands associated with secret leakage in Figure~\ref{fig:src-operation}.
In fact, \SI{\Remark{privatekeyinstsshdpct}}{\percent} of private keys were generated in layers where image creators installed the OpenSSH server.
Since the installation triggers \texttt{ssh-keygen} to generate a fresh host key pair, it is automatically included in the image.

While the procedure of automatic key generation is beneficial on real hardware, i.e., users are not tempted to reuse keys on different hosts, in published Docker images it automatically leads to compromised keys and thus puts the authenticity of all containers relying on this image in danger.
Further \SI{\Remark{privatekeysshkeygenpct}}{\percent} of found private keys were generated by a direct call of \texttt{ssh-keygen}, e.g., to generate fresh SSH client key material, implying the planned usage in production of generated but compromised key material.

Given the massive secret leakage on GitHub~\cite{meli-2019}, we also expect secrets to be included in images by cloning Git repositories.
However, only a minority of secrets can be associated with Git, suggesting that the sets of users leaking secrets via Docker and GitHub are distinct. Furthermore, only a minority of secrets were downloaded~(using \texttt{wget} or \texttt{curl}) both indicating that the secrets we found were most likely exclusively leaked in Docker images and underpinning that they are actually worth being protected.

\subsubsection{File Paths Indicate Usage}
\label{sec:filepath}

\begin{figure}
\begin{subfigure}{0.45\linewidth}
  \centering
  \includegraphics[width=\linewidth]{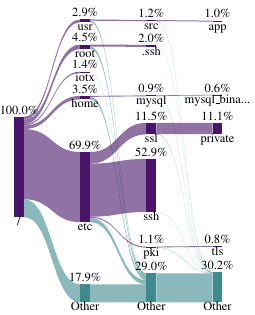}
  \vspace{-1.5em}
  \caption{Private Key}
\end{subfigure}%
\begin{subfigure}{0.45\linewidth}
  \centering
  \includegraphics[width=\linewidth]{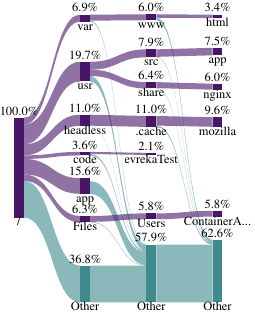}
  \vspace{-1.5em}
  \caption{API Secrets}
\end{subfigure}%
\vspace{-1em}
\caption{Most frequent file paths where we found secrets. While the major location of found private keys focuses on a few file paths, i.e., most private keys are stored where SSH host keys reside, API secrets are spread further.}
\label{fig:src-filepath}
\vspace{-1.25em}
\end{figure}

To further reason about the usage of our found secrets, we analyze their file paths within the images assessing where secrets stem from and how services apply them.
Separated by private keys and API secrets, Figure~\ref{fig:src-filepath} shows the distribution of secrets throughout the directory structure of all images and focuses on the top seven paths.

We found the majority of private keys in \texttt{/etc/ssh} underpinning a high prevalence of compromised SSH host keys.
Another large share occurs in \texttt{/etc/ssl} suggesting compromised keys used for host authentication via TLS.
This path is also the location for TLS default~(\textit{``snakeoil''}) keys that are used if no other information is provided.
They are auto-generated when the \texttt{ssl-cert} package is installed such that every host possesses a unique default key-pair.
However, when installed during the creation of Docker images, the key is included in the image and, thus,  compromised when shared.
Based on the key's filename, indeed, we found \SI{\Remark{numsnakeoiletcssl}}{}~of such keys which are potentially used to offer TLS services with broken authenticity to the public Internet.

Even more alarming, we found keys lying in \texttt{/etc/pki}, indicating that included keys are associated with a Public Key Infrastructure~(PKI), and thus potentially destined to offer services to a higher number of users.
Furthermore, \texttt{/iotx} contains private keys used in relation to the IoT and, as per the repository names, for authentication using IoT protocols like CoAP and MQTT.
Thus, attackers possessing these private keys can leverage the authentication of all connections users establish to each container created based on these images.
In fact, attackers then can access or alter transmitted confidential information, e.g., privacy-sensitive user data or commands of IoT services potentially impacting cyber-physical systems.
In addition, we found keys in \texttt{/root/.ssh}, i.e., a location where SSH client key pairs typically reside.
Hence, these keys might enable attackers to take over SSH servers, trusting these keys and having access to confidential data.

Contrarily, found API secrets are distributed more evenly through the directory structure.
We found the largest share in \texttt{/app}, which is the example folder for including own applications in Docker images~\cite{docker-file-bestpractice}, underlining that image creators compromise their own application's API secrets.
While similar holds for \texttt{/usr/src/app}, another large share of secrets resides in \texttt{/headless/.cache/mozilla} stemming from Firefox profiles containing Google Service API secrets in cached JavaScript files.
Although these secrets are most likely usable in combination with Google Maps or Google Analytics and thus meant to be shared with website visitors, this leakage implies privacy issues:
An attacker could retrace the creator's browsing history, which apparently exists due to the cache being filled, which could show potentially sensitive information.

In addition, we found a large share of Google API secrets (both Cloud and Services) in \texttt{/code/evrekaTest}.
Since we do not use API tokens for further validation~(cf.\ Appendix~\ref{sec:ethics}), we cannot be entirely sure whether these secrets are usable or only generated for testing purposes.
However, manual supervision of the matches and including files suggest that they could be actually in use.

\takeaway{
\SI{\Remark{pctaffectedimages}}{\percent} of analyzed images contain and thus leak secrets.
While the majority stems from public Docker Hub images regardless of their domain, also private registries leak a significant number of secrets.
Notably, associated file paths and commands imply their production use and that various authentication mechanisms are futile.
}

\subsection{Compromised Certificates}
\label{sec:accmaterial}
To further understand the severity of potentially compromised systems, we now focus on found certificates as they provide various information on their relations and use cases.
Thus, we research the trust chain, validity, and usage parameters of \SI{\Remark{knowncompromizedcerts}}{}~compromised certificates occurring in Docker images.

\afblock{Trust Anchors}
While self-signed certificates indicate the usage of certificates in controlled environments, i.e., clients need a safelist with all certificates they can trust, CA-signed certificates imply the usage at larger scale as these are trusted by all clients having a corresponding root certificate installed.
We consider certificates where the issuer and common name are similar as self-signed and CA-signed otherwise.
For CA-signed certificates, we consider those which we can validate against widespread root stores\footnote{Stores from Android, iOS/MacOS, Mozilla NSS, OpenJDK, Oracle JDK, and Windows.} as signed by a public~CA, and otherwise signed by a private~CA.

We discovered that the majority of found compromised certificates~(\SI{\Remark{selfsignedcertspct}}{\percent}) are self-signed, but also \SI{\Remark{privatecacerts}}{} private~CA-signed and \SI{\Remark{casignedcerts}}{} public~CA-signed certificates.
While all systems relying on these certificates open the door for impersonation attacks, the occurrence of CA-signed certificates is especially alarming as such certificates are typically planned to provide authenticity to many clients/users and are universally accepted.
Thus, knowing these certificates' private key not only allows attackers to perform Man-in-the-Middle attacks but also enable them to sign malicious software to compromise other's systems.

\afblock{Validity}
As a countermeasure against key leakage, the certificate's lifetime enforces service operators to request new certificates from time to time, as clients should reject outdated certificates.
Notably, \SI{\Remark{casignedvalidondownload}}{}~public-CA, \SI{\Remark{privatecavalidondownload}}{}~private-CA, and \SI{\Remark{selfsignedvalidondownload}}{}~self-signed certificates were valid when we downloaded their containing image layer, showing that the authenticity of relying services is at stake, i.e., the lifetime does not help in these cases of key leakage.

Interestingly, \SI{\Remark{casignedvalidonhistory}}{}~public-CA, \SI{\Remark{privatecavalidonhistory}}{}~private-CA, and \SI{\Remark{selfsignedvalidonhistory}}{}~self-signed certificates were valid when added to their Docker image~(as per the image's history timestamp). %
While these larger numbers show that the limited lifetime of certificates helps to mitigate leaked private keys, they also indicate that key leakage in images is tedious, i.e., more and more private keys are leaked.

\afblock{Usages}
The usage attributes of certificates can optionally indicate the practical use-case of CA-signed certificates and, thus, further help to understand the severity of the private key leakage.
While all public-CA-signed certificates allow for authentication~(digital signatures), and \SI{\Remark{casignedparsedFindingextensionsextendedkeyusageserverauth}}{} are explicitly declared for server authentication, \SI{\Remark{casignedparsedFindingextensionsextendedkeyusagecodesigning}}{}~(private-CA:~\SI{\Remark{privatecaparsedFindingextensionsextendedkeyusagecodesigning}}{}) allow for code-signing.
Thus, knowing the private key of these certificates, does not only allow attackers to perform Man-in-the-Middle attacks, but also enable to sign malicious software to compromise others systems.

\takeaway{%
\SI{\Remark{knowncompromizedcerts}}{} found compromised certificates show that leaked private keys can have extensive influence on the authenticity of services and software.
Thus, attackers can impersonate services, decrypt past communications, or sign malware to infect production systems.
}

\section{Secret Usage in the Wild}
\label{sec:usageinternet}
Until now, it is open whether the found compromised secrets are used in practice and, if so, to what extent, i.e., whether a single compromised secret is reused due to several Docker containers stemming from the same image.
While we cannot check the validity of API~secrets by using them against their destined endpoint due to our ethical guidelines~(cf.\ Appendix~\ref{sec:ethics}), we can investigate whether hosts on the Internet use found private keys for authentication.

To assess whether Internet-reachable hosts can be suspect to impersonation attacks due to secret leakage in Docker images, we check for TLS- and SSH-enabled hosts relying their authentication on compromised private keys by using the Censys database, i.e., 15 months of active Internet-wide measurement results~\cite{durumeric2015search}.
Here, we search for hosts presenting a public key, i.e., as SSH host key or within a TLS certificate, matching to one of the found compromised keys.
More specifically, we match the fingerprint of public keys in the Censys database on ones extracted from found private keys.

\begin{figure}[!t]
\centering
\includegraphics[width=\columnwidth]{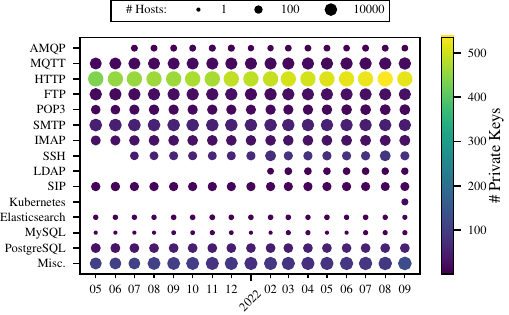}
\vspace{-2.5em}
\caption{Alarming number of hosts~(dot size) relying their authenticity on several compromised keys~(dot color) over time~(x-axis). Used protocols~(y-axis) imply sensitive services.}
\vspace{-0.5em}
\label{fig:plt-censys}
\end{figure}

In Figure~\ref{fig:plt-censys}, we detail how many hosts rely their authenticity on found compromised private keys and how often these keys are reused.
While the total number of hosts relying on compromised keys is worrying on its own~(\SI{\Remark{20220901numuniquehosts}}{}~hosts in Oct.~2022), their protocols, even worse, imply sensitive services.
As such, in October~2022, we find \SI{\Remark{MQTT20220901numuniquehosts}}{}~MQTT and \SI{\Remark{AMQP20220901numuniquehosts}}{}~AMQP~hosts, potentially transferring privacy-sensitive~((I)IoT)~data.
Moreover, \SI{\Remark{FTP20220901numuniquehosts}}{}~FTP, \SI{\Remark{PostgreSQL20220901numuniquehosts}}{}~PostgreSQL, \SI{\Remark{Elasticsearch20220901numuniquehosts}}{}~Elasticsearch, and \SI{\Remark{MySQL20220901numuniquehosts}}{}~MySQL instances serve potentially confidential data.
Regarding Internet communications, we see \SI{\Remark{SIP20220901numuniquehosts}}{}~SIP hosts used for telephony as well as \SI{\Remark{SMTP20220901numuniquehosts}}{}~SMTP, \SI{\Remark{POP320220901numuniquehosts}}{}~POP3, and \SI{\Remark{IMAP20220901numuniquehosts}}{}~IMAP servers used for email.
Since these hosts are susceptible to impersonation attacks due to their leaked private keys, attackers can eavesdrop, relay, or alter the sensitive data transmitted here.

Aggravatingly, we also find services with administrative relevance:
\SI{\Remark{SSH20220901numuniquehosts}}{}~SSH servers rely on \SI{\Remark{SSH20220901corrFingerprint}}{}~compromised host keys and \SI{\Remark{Kubernetes20220901numuniquehosts}}{}~Kubernetes instances use leaked keys opening doors for attacks which can lead to remote-shell access, extension of botnets or further data access.
The comparably low number of compromised keys used (compared to \SI{\Remark{knowncompromizedhostkeys}}{}~found SSH host keys) is probably due to a missing need for SSH servers in Docker containers as other mechanisms, e.g., \texttt{docker exec}, already allow shell access.
Furthermore, we see \SI{\Remark{LDAP20220901numuniquehosts}}{}~LDAP instances relying on leaked secrets.
As LDAP is used as a base for user authentication on attached systems, the integrity of unknown many other clients is at stake.
For instance, attackers could grant themselves root access to a myriad of systems.

The number of actually used keys is low compared to the number of hosts which rely on them indicating that a few Docker images lead to various compromised container deployments.
Thus, the simplicity of Docker to deploy services based on ready-to-use images puts the authenticity of several instances most likely operated by different users under threat.
In this regard, HTTPS hosts stand out in particular.
\SI{\Remark{HTTP20220901numuniquehosts}}{}~HTTPS hosts use \SI{\Remark{HTTP20220901corrFingerprint}}{}~different compromised private keys showing that the reuse of these keys is rampant for Web services.
Thus, attackers can perform Man-in-the-Middle attacks to alter webpages on their delivery or data sent to the server.

Figure~\ref{fig:plt-censys} also underpins that the key usage of compromised keys is long-lasting and rising, i.e., over the complete available period the number of compromised systems grew from \SI{\Remark{20210501numuniquehosts}}{}~(relying on \SI{\Remark{20210501corrFingerprint}}{}~compromised keys) to \SI{\Remark{20220901numuniquehosts}}{}~hosts~(\SI{\Remark{20220901corrFingerprint}}{}~keys) indicating that container images with compromised certificates or SSH host keys included are increasingly used.
Thus, the authenticity of more and more systems is futile, offering an ever-growing attack surface.

While our study is significantly driven by initially found compromised keys in Docker images in the area of the IIoT, Censys does not identify secured IIoT protocols other than AMQP and MQTT via TLS.
Thus, we perform own Internet-wide measurements for a deeper inspection of whether IIoT services also use compromised certificates, e.g., for authentic communication via OPC~UA.
To this end, we select ten secure IIoT protocols from recent literature~\cite{dahlmanns-2022} and mimic its proposed measurement strategy.
Our results show that besides the already large number of compromised AMQP and MQTT hosts, only 2~CoAP hosts use 2~different leaked keys from Docker containers.
That we do not find substantially more compromised hosts using other IIoT protocols underlines that the issue of key leakage is not an IIoT specfic hotspot but a general problem.

\takeaway{
\SI{\Remark{20220901numuniquehosts}}{}~hosts use \SI{\Remark{20220901corrFingerprint}}{}~compromised private keys found in Docker images for authentication on the Internet and encompass deployments using, i.a., MQTT, SMTP, and PostgreSQL.
This widespread usage allows attackers to eavesdrop on confidential or alter sensitive information, e.g., from the IoT, webpages, or databases.
}

\section{Discussion, Limitations \& Mitigations}
The outcome of our work has different aspects.
We have seen that numerous private keys are compromised by image creators publishing their images via Docker registries and shown that security relies on these secrets in practice.
Still, future work could investigate the \textit{limitations} of our approach or implement the derived \textit{mitigation opportunities} from our results.

\afblock{View on Available Images}
Due to rate and computation-time limits and comprehensive ethical considerations~(cf.\ Appendix~\ref{sec:ethics}), we could not analyze all available images on Docker Hub and private registries.
Thus, we might have missed secrets included in single layers or complete images that were not subject to our study.
In this light, the absolute number of found secrets is already very alerting.
Also, in relative numbers, our results should be representative for the selected groups due to our sampling.
Yet, the selected groups, i.e., our Docker Hub search terms, might lead to skewed results overestimating the overall population.
For instance, images that are not targeted at protocols might have been created with fewer secrets.
Thus, we opted for a broad body of terms based on, i.a., public polls~\cite{stackoverflow-survey-2021} to avoid any bias.
Moreover, our private registry analysis has not been targeted but included randomly sampled layers, and we still found a similar share of affected images as on Docker Hub.
As such, we believe that our relative results are---at least in their magnitude---representative for the overall population of Docker images publicly available.

\afblock{Missing Methods to Check API Secrets}
While relying on Internet-wide measurements was a suitable measure to assess the usage of compromised private keys for the authenticity of Internet-reachable services, we could not check whether found API secrets are functional.
The only option would be to contact the corresponding API's endpoint to check for the acceptance of found credentials.
However, due to our ethical considerations, we must not use found secrets as such usage might influence other systems or services.
Thus, we cannot validate them against their respective endpoint.
Still, the number of found secrets is worrying and looking at the usage of compromised private keys, we are convinced that many API secrets are also functional.

\afblock{Causes \& Mitigation Opportunities}
We have seen both creators \textit{actively} copying secrets from their local file system into the image, e.g., most of the API secrets but also private keys, incl.\ certificates, and \textit{passively} generating key material during the image creation process, e.g., by installing an OpenSSH server.
Both behaviors lead to compromised secrets and affect the security of both image creators and users basing their containers on an image and already included secrets.
Most likely, creators and users are unaware of compromising or using compromised foreign secrets.
In fact, compared to GitHub, which provides a graphical interface to browse published files and potentially notice a mistakenly uploaded secret, files in Docker images and containers cannot be browsed easily, i.e., users barely get an overview on included files.
Furthermore, while Git repositories only include manually added files, images of Docker containers contain a complete system directory tree.
Thus, files with included secrets cannot be identified.

The mitigation of these problems must be two-fold.
On the one hand, image creators must be warned that they are uploading their secrets to~(publicly reachable) Docker registries.
On the other hand, when deploying containers based on downloaded images, users should be informed that included secrets, especially private keys, might already be compromised, putting the authentication of deployed services at stake.
To this end, credential-finding tools such as TruffleHog~\cite{truffles71:online} or SecretScanner~\cite{deepfenc70:online} can be integrated on both sides of the Docker paradigm.
When uploading or downloading an image, these tools could then scan all layers of the image for included secrets.
To reduce the number of false positives, for potential API secrets, the tool can also check the secret's function against the respective endpoint~(we think this is also ethically correct on the user's side who downloaded the image).
For private keys, the tools could maintain a list of test keys that are usually included in libraries.
Increasing the image creator's awareness regarding the leakage of such secrets should decrease their number in uploaded images.
Additionally, performing a second check at the user deploying a container based on a downloaded image should further decrease the number of services relying on already compromised secrets.
An additional help could be an API + graphical view for images on Docker Hub, which shows the included files.
This API could also enable third-party solutions similar to those for GitHub~\cite{truffles71:online, GitGuard70:online, michenri37:online} to easily search for known secret file paths.

\section{Conclusion}
Containerization allows integrating applications and their dependencies in self-containing and shareable images making software deployment easy.
However, when focusing on security, sharing of secrets or using already compromised secrets breaks promises, e.g., authenticity or access control.
Thus, cryptographic secrets must not be included in publicly available container images.

Our analysis of \SI{\Remark{numnonemptyimages}}{}~images from Docker Hub and \SI{\Remark{privatemeasurementnumtotalmax}}{}~private registries revealed that, however, \SI{\Remark{pctaffectedimages}}{\percent} include secrets that should not be leaked to the public.
More specifically, we found a near-lower bound of \SI{\Remark{validprivatekeyvalidnumdistinctmatchestotal}}{}~private keys and \SI{\Remark{apinumdistinctmatches}}{}~API secrets.
\SI{\Remark{validapicloudvalidnumdistinctmatchestotal}}{}~API secrets belonging to cloud providers, e.g., \Remark{apicloud1rule}~(\SI{\Remark{apicloud1numdistinctmatches}}{} secrets), or \SI{\Remark{validapifinancialvalidnumdistinctmatchestotal}}{}~secrets to financial services, e.g., \Remark{apifinancial0rule}~(\SI{\Remark{apifinancial0numdistinctmatches}}{}~secrets), show that attackers can cause immediate damage knowing these secrets.
Focusing on the leaked private keys, we find that these are also in use in practice: \SI{\Remark{20220901numuniquehosts}}{}~TLS and SSH hosts on the Internet rely their authentication on found keys, thus being susceptible to impersonation attacks.
Notably, many private keys automatically generate when installing packages during image creation.
While beneficial when running on real hardware where every computer generates its own key, in container images, this process automatically leads to compromised secrets and potentially a sheer number of containers with compromised authenticity.

We further discover that especially private registries serve images with potentially sensitive software, most likely not intended to be publicly shared.
Additionally, these registries might not prevent write access enabling attackers to add malware to images.

Our work shows that secret leakage in container images is a real threat and not neglectable.
Especially the proven usage of leaked private keys in practice verifies numerous introduced attack vectors.
As a countermeasure, the awareness of image creators and users regarding secret compromise must be increased, e.g., by integrating credential search tools into the Docker paradigm.

\begin{acks}
Funded by the German Federal Ministry for Economic Affairs and Climate Action (BMWK) --- Research Project VeN\textsuperscript{2}uS --- 03EI6053K.
Funded by the Deutsche Forschungsgemeinschaft (DFG, German Research Foundation) under Germany's Excellence Strategy --- EXC-2023 Internet of Production --- 390621612.
\end{acks}

\bibliographystyle{ACM-Reference-Format}
\bibliography{paper}

\clearpage

\appendix

\begin{table*}[htpb]
\caption{%
Search queries and derived spellings to receive corresponding Docker repositories from Docker Hub of our Standard and (Industrial) IoT query group.
}%
\vspace{-0.75em}
\scriptsize
\raggedright
{
\raggedright
\normalsize
\textbf{Standard:}
\textit{Trending protocols and technologies}.
\hfill{}
}

\setlength{\tabcolsep}{10pt}
\begin{tabular}{llllllllll}
\makecell[l]{$\cdot$ tls} &
\makecell[l]{$\cdot$ ipp} &
\makecell[l]{$\cdot$ css} &
\makecell[l]{$\cdot$ imap} &
\makecell[l]{$\cdot$ html} &
\makecell[l]{$\cdot$ mysql} &
\makecell[l]{$\cdot$ oracle} &
\makecell[l]{$\cdot$ mariadb} &
\makecell[l]{$\cdot$ memcached} &
\makecell[l]{$\cdot$ elasticsearch} \\
\makecell[l]{$\cdot$ ssh} &
\makecell[l]{$\cdot$ vpn} &
\makecell[l]{$\cdot$ sql} &
\makecell[l]{$\cdot$ pptp} &
\makecell[l]{$\cdot$ java} &
\makecell[l]{$\cdot$ mssql} &
\makecell[l]{$\cdot$ heroku} &
\makecell[l]{$\cdot$ ansible} &
\makecell[l]{$\cdot$ terraform} &
\makecell[l]{$\cdot$ c++ $\rightarrow{}$ c+, c, c } \\
\makecell[l]{$\cdot$ dns} &
\makecell[l]{$\cdot$ irc} &
\makecell[l]{$\cdot$ php} &
\makecell[l]{$\cdot$ xmpp} &
\makecell[l]{$\cdot$ bash} &
\makecell[l]{$\cdot$ redis} &
\makecell[l]{$\cdot$ docker} &
\makecell[l]{$\cdot$ xamarin} &
\makecell[l]{$\cdot$ postgresql} &
\makecell[l]{$\cdot$ ibm db2 $\rightarrow{}$ ibmdb2, ibm+db2} \\
\makecell[l]{$\cdot$ ftp} &
\makecell[l]{$\cdot$ aws} &
\makecell[l]{$\cdot$ quic} &
\makecell[l]{$\cdot$ yarn} &
\makecell[l]{$\cdot$ ipmi} &
\makecell[l]{$\cdot$ shell} &
\makecell[l]{$\cdot$ puppet} &
\makecell[l]{$\cdot$ firebase} &
\makecell[l]{$\cdot$ kubernetes} &
\makecell[l]{$\cdot$ unity 3d $\rightarrow{}$ unity+3d, unity3d} \\
\makecell[l]{$\cdot$ rdp} &
\makecell[l]{$\cdot$ gcp} &
\makecell[l]{$\cdot$ http} &
\makecell[l]{$\cdot$ deno} &
\makecell[l]{$\cdot$ samba} &
\makecell[l]{$\cdot$ proxy} &
\makecell[l]{$\cdot$ pulumi} &
\makecell[l]{$\cdot$ dynamodb} &
\makecell[l]{$\cdot$ javascript} &
\makecell[l]{$\cdot$ ibm cloud $\rightarrow{}$ ibmcloud, ibm+cloud} \\
\makecell[l]{$\cdot$ vnc} &
\makecell[l]{$\cdot$ git} &
\makecell[l]{$\cdot$ smtp} &
\makecell[l]{$\cdot$ chef} &
\makecell[l]{$\cdot$ rsync} &
\makecell[l]{$\cdot$ telnet} &
\makecell[l]{$\cdot$ python} &
\makecell[l]{$\cdot$ cassandra} &
\makecell[l]{$\cdot$ typescript} &
\makecell[l]{$\cdot$ node.js $\rightarrow{}$ node js, node+js, nodejs} \\
\makecell[l]{$\cdot$ smb} &
\makecell[l]{$\cdot$ k8s} &
\makecell[l]{$\cdot$ pop3} &
\makecell[l]{$\cdot$ flow} &
\makecell[l]{$\cdot$ ipsec} &
\makecell[l]{$\cdot$ sqlite} &
\makecell[l]{$\cdot$ mongodb} &
\makecell[l]{$\cdot$ couchbase} &
\makecell[l]{$\cdot$ powershell} &
\makecell[l]{$\cdot$ ibm watson $\rightarrow{}$ ibm+watson, ibmwatson} \\
\makecell[l]{$\cdot$ ipp} &
\makecell[l]{$\cdot$ css} &
\makecell[l]{$\cdot$ imap} &
\makecell[l]{$\cdot$ html} &
\makecell[l]{$\cdot$ mysql} &
&
&
&
&
\\
\end{tabular}

\vspace{1em}

{
\raggedright
\normalsize
\textbf{(Industrial) IoT:}
\textit{Industrial protocols subject to recent research}.
\hfill{}
}

\begin{tabular}{lllll}
\makecell[l]{$\cdot$ atg} &
\makecell[l]{$\cdot$ mqtt} &
\makecell[l]{$\cdot$ codesys} &
\makecell[l]{$\cdot$ ff-hse $\rightarrow{}$ ff hse, ff+hse, ffhse} &
\makecell[l]{$\cdot$ iec-61850 $\rightarrow{}$ iec+61850, iec61850, iec 61850} \\
\makecell[l]{$\cdot$ dnp3} &
\makecell[l]{$\cdot$ cspv4} &
\makecell[l]{$\cdot$ proconos} &
\makecell[l]{$\cdot$ fl-net $\rightarrow{}$ fl net, flnet, fl+net} &
\makecell[l]{$\cdot$ zigbee-ip $\rightarrow{}$ zigbeeip, zigbee ip, zigbee+ip} \\
\makecell[l]{$\cdot$ srtp} &
\makecell[l]{$\cdot$ bacnet} &
\makecell[l]{$\cdot$ ethercat} &
\makecell[l]{$\cdot$ hart-ip $\rightarrow{}$ hart+ip, hartip, hart ip} &
\makecell[l]{$\cdot$ ansi c12.22 $\rightarrow{}$ ansi c12 22, ansi+c12+22, ansic1222} \\
\makecell[l]{$\cdot$ iccp} &
\makecell[l]{$\cdot$ modbus} &
\makecell[l]{$\cdot$ profinet} &
\makecell[l]{$\cdot$ iec-104 $\rightarrow{}$ iec 104, iec104, iec+104} &
\makecell[l]{$\cdot$ ethernet/ip $\rightarrow{}$ ethernet+ip, ethernetip, ethernet ip} \\
\makecell[l]{$\cdot$ amqp} &
\makecell[l]{$\cdot$ siemens} &
\makecell[l]{$\cdot$ pc worx $\rightarrow{}$ pcworx, pc+worx} &
\makecell[l]{$\cdot$ omron fins $\rightarrow{}$ omron+fins, omronfins} &
\makecell[l]{$\cdot$ red lion crimson v3 $\rightarrow{}$ redlioncrimsonv3, red+lion+crimson+v3} \\
\makecell[l]{$\cdot$ coap} &
\makecell[l]{$\cdot$ tridium} &
\makecell[l]{$\cdot$ opc-ua $\rightarrow{}$ opcua, opc+ua, opc ua} &
\makecell[l]{$\cdot$ melsec-q $\rightarrow{}$ melsecq, melsec q, melsec+q} &
\makecell[l]{$\cdot$ automatic tank gauge $\rightarrow{}$ automatic+tank+gauge, automatictankgauge} \\
\\
\end{tabular}
\vspace{-1.75em}
\label{tab:searchterms}
\end{table*}

\section{Ethical Considerations}
\label{sec:ethics}

Our research curates a comprehensive archive of leaked security secrets in Docker images on Docker Hub and private registries whose leakage is again a threat to security.
Moreover, to find private registries and deployments relying their security on leaked secrets, we leverage Internet-wide measurements that can have unintended implications, e.g., high load on single network connections impacting stability or alerting sysadmins due to unknown traffic.
Thus, we base our research on several ethical considerations.

First, we take well-established guidelines~\cite{dittrich_menlo-report_2012} and best practices of our institution as base for our research.
We handle all collected data with care and inform image creators and Docker~Inc., to responsibly disclose our findings~(cf.\ Appendix~\ref{subsec:ethics:handling}).
Moreover, we comply with recognized measurement guidelines~\cite{durumeric-zmap-2013} for our Internet-wide measurements reducing their impact (cf.\ Appendix~\ref{subsec:ethics:impact}).

\subsection{Handling of Data \& Responsibilities}
\label{subsec:ethics:handling}
During our research, we always only collect and request publicly available data, i.e., our access is limited to publicly available image repositories.
At no time do we bypass access control, e.g., by guessing passwords. 
We, thus, cannot download private images.
Still, we revealed that many of the public images contain sensitive security secrets~(cf.\ Section~\ref{sec:layer-analysis}) which we stored for further analysis.
All found secrets are stored on secured systems.
Furthermore, we refrain from releasing our dataset including these secrets or image names, to not provide an archive of leaked secrets or pinpoints for potential attackers.
While this restriction prevents others from independently reproducing our results, we consider this decision to constitute a reasonable trade-off to protect affected users.

\afblock{Responsible Disclosure}
To further support affected users in removing their secrets from publicly available Docker images, we target to responsibly disclose our findings.
To this end, we extract e-mail addresses from maintainer variables set in Dockerfiles and furthermore derive addresses from Gravatar accounts linked to affected Docker Hub accounts.
In this regard, we identified \SI{\Remark{numemaildisclosure}}{}~e-mail addresses we contacted to notify about our possible findings.
Already after a few hours, we received >30 answers of owners appreciating our efforts, fixing their images or informing us that the image at hand is not used anymore.
A handful informed us that no secrets were leaked helping us to refine our filtering.
Moreover, we decided to reach out to the operator of Docker Hub, i.e., Docker~Inc., to discuss potential further disclosure to unidentifiable creators.

\subsection{Reducing Impact of Measurements}
\label{subsec:ethics:impact}

To reduce the impact of our active Internet scans, we follow widely accepted Internet measurement guidelines~\cite{durumeric-zmap-2013}.

\afblock{Coordination}
We coordinate our measurements with our Network Operation Center to reduce the impact on the Internet and to react correspondingly.
Abuse emails are handled informing about the intent of our measurements and how to opt-out of our measurements.
As part of this opt-out process, we maintain a blocklist to exclude IPs from our measurements.

\afblock{External Information}
For giving external operators information about our research intent, we provide rDNS records for all our scan IPs and transmit contact information in the HTTP header of each request to the registries.
Moreover, we host a webpage on our scan IPs, which gives further information on our project and how to opt-out.
Over time, also due to other measurements, we excluded 5.8\,M IP addresses (0.14\% of the IPv4 address space).

\afblock{Limiting Load}
To limit load and stress on all systems involved (along the path and the end-host), we deliberately reduce our scan-rate.
Our scans are stretched over the course of one day and use \texttt{zmap}'s address randomization to spread load evenly.
We further limit the load on single private registries when downloading available images.
While we paid to increase the existing rate limiting for image downloads on Docker Hub~(cf.\ Appendix~\ref{sec:imageselectionhub}), private registries typically do not implement any rate limiting.
Hence, to prevent our scanner from overloading registries running on resource-constrained hardware or connected via slow or volume-billed Internet connections, we decide to only download image layers randomly until their size sums up to at most \SI{250}{\mega\byte}.
Additionally, we shuffle the downloads of layers of different registries to further distribute the load.

\subsection{Overall Considerations}
Without taking our goals into account, summarizing the sensitive nature and the impact of our measurements can quickly lead to the conclusion that our measurements are not beneficial.
However, we consider it public interest and fundamental for improving security to know about potential security issues and how widespread these are.
The Docker paradigm does not include any mechanisms to prevent image creators from (accidentally) adding security secrets to their images and no mechanisms exist that warns users relying on already compromised security secrets.
Hence, we consider it essential to know whether secrets are widely included in publicly available Docker images and whether these are in use at scale to steer future decisions for counter-measures.
To answer this question, we carefully weighed the impact of our measurements against their benefit and have taken sensible measures to reduce the risks of building a large archive of leaked security secrets and risks introduced by active Internet measurements.

\section{Image Download from Docker Hub}
\label{sec:imageselectionhub}
The limit of image manifest downloads from Docker Hub depends on the booked plan, e.g., free users are allowed to pull only \SI{800}{}~images per day.
Hence, for a faster analysis of images on Docker Hub, we purchased two Pro accounts, that allow \SI{5000}{}~image downloads per day each.
Still, we are required to perform our analysis on a subpart of available images as the download of one image of every of the \SI{9321726}{}~available repositories would require \SI{933}{}~days under best conditions.
Thus, we decided to limit our analysis on two categories:
\begin{inparaenum}[(i)]
	\item a context of standard protocol and frequently used technologies, and 
	\item an (Industrial) IoT context for comparison.
\end{inparaenum}
Both categories have communication in common as here security can be affected on an Internet scale.

\afblock{Standard Context}
To generate a wide view on secret leakage in Docker images, we create a list of search queries comprising standard protocols~\cite{schumann-2022}, and frequently used technologies~\cite{stackoverflow-survey-2021}.
To find related images, we employ Docker Hub's API to perform searches over all available images and retrieve results users would retrieve when using the \texttt{docker search} CLI command or Docker Hub's web interface.
To ensure that different handling of special characters in technology and protocol names does not exclude any images, we include different spelling variants in our query list, i.e., we include terms as they are, but also replace non-alpha-numeric characters by \texttt{+} and \textit{space}.
Table~\ref{tab:searchterms}~(top) shows our constructed search queries for the standard context.

\afblock{(Industrial) IoT Context}
We extend our analysis on images in the (Industrial) IoT context, as deployments in this area showed massive security deficits in past~\cite{dahlmanns-2022,2020-dahlmanns-imc-opcua,nawrocki-passics-20,mirian-icsmes-2016,leverett-shodanclassification-2011,hansson-analysisshodan-2018}, in single cases traced back to security secret leakage via GitHub and Docker images~\cite{dahlmanns-2022}.
As search terms, we take (Industrial) IoT protocol names that were subject to recent research~\cite{dahlmanns-2022}.
We proceed similar as in the standard context, i.e., include derived spellings of these terms, and show our constructed search query of this context in Table~\ref{tab:searchterms}~(bottom).

\section{Regular Expressions}
\label{sec:regularexpressions}
Following already established procedures to find security secrets in code repositories~\cite{meli-2019,truffles71:online}, we build our secret detection in Docker Images on regular expressions, i.e., we try to match regular expressions derived from secrets on the content of included files. %
Table~\ref{tab:regularexpressions} shows our composed list of regular expressions covering a variety of secrets, i.e., asymmetric private keys and API keys, as well as accompanying material we use for our analysis, i.e., public keys and certificates.
We orientate our expressions towards related work~\cite{meli-2019} and TruffleHog~\cite{truffles71:online}, an established tool to find secrets in various sources, i.e., the local file system, Git repositories, S3 storages, and syslogs.
Specifically, we inherit Meli et al.'s~\cite{meli-2019} regular expressions to allow comparisons between the occurrence of leaked secrets in GitHub repositories at scale and our findings.
Furthermore, they composed their expressions comprehensibly, i.e., they included API~keys for certain services by the occurrence of service domains in Alexa's Top 50 Global and United States lists in combination with a list of well-known APIs manually filtered for services with a high risk on key leakage and keys with a distinctive signature~(to reduce the number of false-positives).
For private keys they focus on the most prevalent types and form to store, i.e., RSA, elliptic curve keys, PGP, and general keys in PEM format.

To spread our analysis and align our expressions to the scope of our search queries~(cf.\ Appendix~\ref{sec:imageselectionhub}), we adapt our expression for private keys to match every type of private key in PEM format and, furthermore, extend the list of expressions to also match private key blocks, keys in PKCS7 format, and keys stored in XML format~(due to their unambiguous signature).
Regarding API~secrets to match, we extend our list with expressions from TruffleHog~\cite{truffles71:online} on basis of services being currently trending under developers~\cite{stackoverflow-survey-2021} or having a high risk for misuse and the regular expressions including a unique signature~(also to reduce the number of false positives).
For some services we found more than one type of secret, i.e., secrets for different API~versions~(GitHub v1 and v2), or different types of keys~(Stripe).
Our final list contains \SI{48}{}~expressions which we match on the content of every file in the images part of our study.

\begin{figure}[!b]
\vspace{-3.4em}
\begin{subfigure}{0.45\linewidth}
  \centering
  \includegraphics[width=\linewidth]{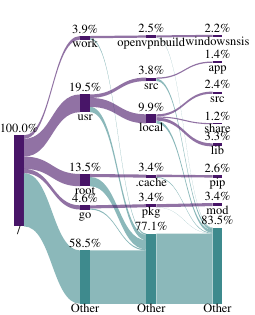}
  \vspace{-1.5em}
  \caption{Private Key}
\end{subfigure}%
\begin{subfigure}{0.45\linewidth}
  \centering
  \includegraphics[width=\linewidth]{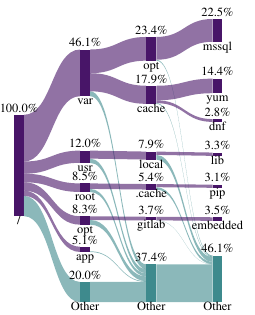}
  \vspace{-1.5em}
  \caption{API Secrets}
\end{subfigure}%
\vspace{-1em}
\caption{Most frequent file paths suspect to our filtering based on file path or extension.}
\label{fig:filtering-filepath}
\end{figure}

\vspace{-20em}
\begin{table*}
\renewcommand{\arraystretch}{1.2}
\caption{%
Regular expressions we matched on each file's content in the layers and environment variables of selected images.
}%
\vspace{-0.75em}
\scriptsize
\centering
\arrayrulecolor{black}
\begin{tabu}{cccclc}
\multicolumn{2}{c}{\textbf{Domain}}                                                                             & \textbf{Name}                    & \textbf{Subordinate}  & \multicolumn{1}{c}{\textbf{Expression}}                                                                                                                                                                                                                                                                                                                                                                                                                                                                                                                                                                      & \textbf{Source}              \\ 
\hline\hline
\multicolumn{2}{c}{\multirow{3}{*}{Private Key}}                                                                & PEM Private Key                  & ~                     & \begin{tabular}[c]{@{}l@{}}(?i)-----\textbackslash{}s*?BEGIN[ A-Z0-9\_-]*?PRIVATE KEY\textbackslash{}s*?-----[a-zA-Z0-9\textbackslash{}/\textbackslash{}n\textbackslash{}r=+]*\\-----\textbackslash{}s*?END[ A-Z0-9\_-]*? PRIVATE KEY\textbackslash{}s*?-----\end{tabular}                                                                                                                                                                                                                                                                                                                                   & \multirow{3}{*}{own}         \\
                                                                                &                               & PEM Private Key Block            &                       & \begin{tabular}[c]{@{}l@{}}(?i)-----\textbackslash{}s*?BEGIN[ A-Z0-9\_-]*?PRIVATE KEY BLOCK\textbackslash{}s*?-----[a-zA-Z0-9\textbackslash{}/\textbackslash{}n\textbackslash{}r=+]*\\-----\textbackslash{}s*?END[ A-Z0-9\_-]*? PRIVATE KEY BLOCK\textbackslash{}s*?-----\end{tabular}                                                                                                                                                                                                                                                                                                                       &                              \\
                                                                                &                               & PEM PKCS7                        &                       & (?i)-----\textbackslash{}s*?BEGIN PKCS7\textbackslash{}s*?-----[a-zA-Z0-9\textbackslash{}/\textbackslash{}n\textbackslash{}r=+]*-----\textbackslash{}s*?END PKCS7\textbackslash{}s*?-----                                                                                                                                                                                                                                                                                                                                                                                                                    &                              \\
                                                                                &                               & XML Private Key                  & ~                     & (?i)\textless{}(RSAKeyValue\textbar{}DSAKeyValue\textbar{}ECKeyValue)\textgreater{}(.\textbar{}{[}\textbackslash{}n\textbackslash{}r])+\textless{}\textbackslash{}/RSAKeyValue\textbar{}\textbackslash{}/DSAKeyValue\textbar{}\textbackslash{}/ECKeyValue)\textgreater{}                                                                                                                                                                                                                                                                                                                                     &                              \\ 
\hline\hline
\parbox[t]{2mm}{\multirow{45}{*}{\rotatebox[origin=c]{90}{\textbf{API}}}}       & \multirow{20}{*}{Cloud}       & Alibaba                          &                       & \textbackslash{}b(LTAI[a-zA-Z0-9]\{17,21\})[\textbackslash{}"';\textbackslash{}s]*                                                                                                                                                                                                                                                                                                                                                                                                                                                                                                                           & \cite{truffles71:online}                   \\
                                                                                &                               & Amazon AWS                       & \multicolumn{1}{l}{~} & \textbackslash{}b((?:AKIA\textbar{}ABIA\textbar{}ACCA\textbar{}ASIA)[0-9A-Z]\{16\})\textbackslash{}b                                                                                                                                                                                                                                                                                                                                                                                                                                                                                                         & \cite{truffles71:online}                   \\
\rowfont{\color{gray}}
                                                                                &                               & Azure                            &                       & (?i)(client\_secret\textbar{}clientsecret).\{0,20\}([a-z0-9\_\textbackslash{}.\textbackslash{}-\textasciitilde{}]\{34\})                                                                                                                                                                                                                                                                                                                                                                                                                                                                                     & \cite{truffles71:online}                   \\
\rowfont{\color{gray}}
                                                                                &                               & DigitalOcean                     &                       & (?i)(?:digitalocean)(?:.\textbar{}{[}\textbackslash{}n\textbackslash{}r])\{0,40\}\textbackslash{}b([A-Za-z0-9\_-]\{64\})\textbackslash{}b                                                                                                                                                                                                                                                                                                                                                                                                                                                                    & \cite{truffles71:online}                   \\
                                                                                &                               & Github                           &                       & \textbackslash{}b((?:ghp\textbar{}gho\textbar{}ghu\textbar{}ghs\textbar{}ghr)\_{[}a-zA-Z0-9]\{36,255\})\textbackslash{}b                                                                                                                                                                                                                                                                                                                                                                                                                                                                                     & \cite{truffles71:online}                   \\
\rowfont{\color{gray}}
                                                                                &                               & \multirow{2}{*}{\textcolor{black}{Gitlab}}          & v1                    & (?i)(?:gitlab)(?:.\textbar{}{[}\textbackslash{}n\textbackslash{}r])\{0,40\}\textbackslash{}b([a-zA-Z0-9\textbackslash{}-=\_]\{20,22\})\textbackslash{}b                                                                                                                                                                                                                                                                                                                                                                                                                                                      & \multirow{2}{*}{\cite{truffles71:online}}  \\
                                                                                &                               &                                  & v2                    & \textbackslash{}b(glpat-[a-zA-Z0-9\textbackslash{}-=\_]\{20,22\})\textbackslash{}b                                                                                                                                                                                                                                                                                                                                                                                                                                                                                                                           &                              \\
                                                                                &                               & Google Cloud                     &                       & \textbackslash{}\{[\textasciicircum{}\{]+auth\_provider\_x509\_cert\_url[\textasciicircum{}\}]+\textbackslash{}\}                                                                                                                                                                                                                                                                                                                                                                                                                                                                                            & \cite{meli-2019}    \\
                                                                                &                               & Google Services                  &                       & \textbackslash{}bAIza[0-9A-Za-z\textbackslash{}-\_]\{35\}\textbackslash{}b                                                                                                                                                                                                                                                                                                                                                                                                                                                                                                                                   & \cite{meli-2019}    \\
                                                                                &                               & Heroku                           &                       & (?i)(?:heroku)(?:.\textbar{}{[}\textbackslash{}n\textbackslash{}r])\{0,40\}\textbackslash{}b([0-9Aa-f]\{8\}-[0-9a-f]\{4\}-[0-9a-f]\{4\}-[0-9a-f]\{4\}-[0-9a-f]\{12\})\textbackslash{}b                                                                                                                                                                                                                                                                                                                                                                                                                       & \cite{truffles71:online}                   \\
                                                                                &                               & IBM Cloud Identity Services      &                       & (?i)(?:ibm)(?:.\textbar{}{[}\textbackslash{}n\textbackslash{}r])\{0,40\}\textbackslash{}b([A-Za-z0-9\_-]\{44\})\textbackslash{}b                                                                                                                                                                                                                                                                                                                                                                                                                                                                             & \cite{truffles71:online}                   \\
\rowfont{\color{gray}}
                                                                                &                               & Login Radius                     & ~                     & (?i)(?:loginradius)(?:.\textbar{}{[}\textbackslash{}n\textbackslash{}r])\{0,40\}\textbackslash{}b([0-9a-f]\{8\}-[0-9a-f]\{4\}-[0-9a-f]\{4\}-[0-9a-f]\{4\}-[0-9a-f]\{12\})\textbackslash{}b                                                                                                                                                                                                                                                                                                                                                                                                                   & \cite{truffles71:online}                   \\ 
                                                                                &                               & MailChimp                        &                       & \textbackslash{}b[0-9a-f]\{32\}-us[0-9]\{1,2\}\textbackslash{}b                                                                                                                                                                                                                                                                                                                                                                                                                                                                                                                                              & \cite{meli-2019}                          \\
                                                                                &                               & MailGun                          &                       & \textbackslash{}bkey-[0-9a-zA-Z]\{32\}\textbackslash{}b                                                                                                                                                                                                                                                                                                                                                                                                                                                                                                                                                      & \cite{meli-2019}                          \\
                                                                                &                               & Microsoft Teams                  &                       & \begin{tabular}[c]{@{}l@{}}(https:\textbackslash{}/\textbackslash{}/[a-zA-Z-0-9]+\textbackslash{}.webhook\textbackslash{}.office\textbackslash{}.com\textbackslash{}/webhookb2\textbackslash{}/[a-zA-Z-0-9]\{8\}-[a-zA-Z-0-9]\{4\}-\\{[}a-zA-Z-0-9]\{4\}-[a-zA-Z-0-9]\{4\}-[a-zA-Z-0-9]\{12\}\textbackslash{}@[a-zA-Z-0-9]\{8\}-[a-zA-Z-0-9]\{4\}-[a-zA-Z-0-9]\{4\}-\\{[}a-zA-Z-0-9]\{4\}-[a-zA-Z-0-9]\{12\}\textbackslash{}/IncomingWebhook\textbackslash{}/[a-zA-Z-0-9]\{32\}\textbackslash{}/[a-zA-Z-0-9]\{8\}-[a-zA-Z-0-9]\{4\}-\\{[}a-zA-Z-0-9]\{4\}-[a-zA-Z-0-9]\{4\}-[a-zA-Z-0-9]\{12\})\end{tabular} & \cite{truffles71:online}                   \\
\rowfont{\color{gray}}
                                                                                &                               & Netlify                          & ~                     & (?i)(?:netlify)(?:.\textbar{}{[}\textbackslash{}n\textbackslash{}r])\{0,40\}\textbackslash{}b([A-Za-z0-9\_-]\{43,45\})\textbackslash{}b                                                                                                                                                                                                                                                                                                                                                                                                                                                                      & \cite{truffles71:online}                   \\ 
                                                                                &                               & Twilio                           & ~                     & \textbackslash{}bSK[0-9a-fA-F]\{32\}\textbackslash{}b                                                                                                                                                                                                                                                                                                                                                                                                                                                                                                                                                        & \cite{meli-2019}                          \\ 
\cline{2-6}
                                                                                & \multirow{15}{*}{Financial}   & Amazon MWS                       &                       & \textbackslash{}bamzn\textbackslash{}.mws\textbackslash{}.[0-9a-f]\{8\}-[0-9a-f]\{4\}-[0-9a-f]\{4\}-[0-9a-f]\{4\}-[0-9a-f]\{12\}\textbackslash{}b                                                                                                                                                                                                                                                                                                                                                                                                                                                            & \cite{meli-2019}                          \\
\rowfont{\color{gray}}
                                                                                &                               & Bitfinex                         &                       & (?i)(?:bitfinex)(?:.\textbar{}{[}\textbackslash{}n\textbackslash{}r])\{0,40\}\textbackslash{}b([A-Za-z0-9\_-]\{43\})\textbackslash{}b                                                                                                                                                                                                                                                                                                                                                                                                                                                                        & \cite{truffles71:online}                   \\
\rowfont{\color{gray}}
                                                                                &                               & Coinbase                         &                       & (?i)(?:coinbase)(?:.\textbar{}{[}\textbackslash{}n\textbackslash{}r])\{0,40\}\textbackslash{}b([a-zA-Z-0-9]\{64\})\textbackslash{}b                                                                                                                                                                                                                                                                                                                                                                                                                                                                          & \cite{truffles71:online}                   \\
\rowfont{\color{gray}}
                                                                                &                               & Currency Cloud                   &                       & (?i)(?:currencycloud)(?:.\textbar{}{[}\textbackslash{}n\textbackslash{}r])\{0,40\}\textbackslash{}b([a-zA-Z0-9\textbackslash{}-=\_]\{20,22\})\textbackslash{}b                                                                                                                                                                                                                                                                                                                                                                                                                                               & \cite{truffles71:online}                   \\
\rowfont{\color{gray}}
                                                                                &                               & Paydirt                          &                       & (?i)(?:paydirtapp)(?:.\textbar{}{[}\textbackslash{}n\textbackslash{}r])\{0,40\}\textbackslash{}b([a-z0-9]\{32\})\textbackslash{}b                                                                                                                                                                                                                                                                                                                                                                                                                                                                            & \cite{truffles71:online}                   \\
\rowfont{\color{gray}}
                                                                                &                               & Paymo                            &                       & (?i)(?:paymoapp)(?:.\textbar{}{[}\textbackslash{}n\textbackslash{}r])\{0,40\}\textbackslash{}b([a-zA-Z0-9]\{44\})\textbackslash{}b                                                                                                                                                                                                                                                                                                                                                                                                                                                                           & \cite{truffles71:online}                   \\
\rowfont{\color{gray}}
                                                                                &                               & Paymongo                         &                       & (?i)(?:paymongo)(?:.\textbar{}{[}\textbackslash{}n\textbackslash{}r])\{0,40\}\textbackslash{}b([a-zA-Z0-9\_]\{32\})\textbackslash{}b                                                                                                                                                                                                                                                                                                                                                                                                                                                                         & \cite{truffles71:online}                   \\
                                                                                &                               & PayPal Braintree                 &                       & \textbackslash{}baccess\_token\textbackslash{}\$production\textbackslash{}\${[}0-9a-z]\{16\}\textbackslash{}\${[}0-9a-f]\{32\}\textbackslash{}b                                                                                                                                                                                                                                                                                                                                                                                                                                                              & \cite{meli-2019}                          \\
                                                                                &                               & Picatic                          &                       & \textbackslash{}bsk\_live\_{[}0-9a-z]\{32\}\textbackslash{}b                                                                                                                                                                                                                                                                                                                                                                                                                                                                                                                                                 & \cite{meli-2019}                          \\
                                                                                &                               & \multirow{2}{*}{Stripe}          & ST                    & \textbackslash{}bsk\_live\_{[}0-9a-zA-Z]\{24\}\textbackslash{}b                                                                                                                                                                                                                                                                                                                                                                                                                                                                                                                                              & \multirow{2}{*}{\cite{meli-2019}}         \\
                                                                                &                               &                                  & RE                    & \textbackslash{}brk\_live\_{[}0-9a-zA-Z]\{24\}\textbackslash{}b                                                                                                                                                                                                                                                                                                                                                                                                                                                                                                                                              &                              \\
                                                                                &                               & \multirow{2}{*}{Square}          & AT                    & \textbackslash{}bsq0atp-[0-9A-Za-z\textbackslash{}-\_]\{22\}\textbackslash{}b                                                                                                                                                                                                                                                                                                                                                                                                                                                                                                                                & \multirow{2}{*}{\cite{meli-2019}}         \\
                                                                                &                               &                                  & OA                    & \textbackslash{}bsq0csp-[0-9A-Za-z\textbackslash{}-\_]\{43\}\textbackslash{}b                                                                                                                                                                                                                                                                                                                                                                                                                                                                                                                                &                              \\
\rowfont{\color{gray}}
                                                                                &                               & Ticketmaster                     &                       & (?i)(?:ticketmaster)(?:.\textbar{}{[}\textbackslash{}n\textbackslash{}r])\{0,40\}\textbackslash{}b([a-zA-Z0-9]\{32\})\textbackslash{}b                                                                                                                                                                                                                                                                                                                                                                                                                                                                       & \cite{truffles71:online}                   \\
\rowfont{\color{gray}}
                                                                                &                               & WePay                            & ~                     & (?i)(?:wepay)(?:.\textbar{}{[}\textbackslash{}n\textbackslash{}r])\{0,40\}\textbackslash{}b([a-zA-Z0-9\_?]\{62\})\textbackslash{}b                                                                                                                                                                                                                                                                                                                                                                                                                                                                           & \cite{truffles71:online}                   \\ 
\cline{2-6}
                                                                                & \multirow{3}{*}{Social Media} & \multirow{2}{*}{Facebook}        & Key                   & \textbackslash{}b([A-Za-z0-9\_\textbackslash{}.]\{69\}-[A-Za-z0-9\_\textbackslash{}.]\{10\})\textbackslash{}b                                                                                                                                                                                                                                                                                                                                                                                                                                                                                                & \cite{truffles71:online}                             \\
                                                                                &                               &                                  & Key                   & \textbackslash{}bEAACEdEose0cBA[0-9A-Za-z]+\textbackslash{}b                                                                                                                                                                                                                                                                                                                                                                                                                                                                                                                                                 & \cite{meli-2019}                          \\
                                                                                &                               & Twitter                          & ~                     & \textbackslash{}b[1-9][0-9]+-[0-9a-zA-Z]\{40\}\textbackslash{}b                                                                                                                                                                                                                                                                                                                                                                                                                                                                                                                                              & \cite{meli-2019}                          \\
\cline{2-6}
\rowfont{\color{gray}}
                                                                                & \multirow{4}{*}{\textcolor{black}{IoT}}          & Accuweather                      &                       & (?i)(?:accuweather)(?:.\textbar{}{[}\textbackslash{}n\textbackslash{}r])\{0,40\}([a-z0-9A-Z\textbackslash{}\%]\{35\})\textbackslash{}b                                                                                                                                                                                                                                                                                                                                                                                                                                                                       & \cite{truffles71:online}                   \\
                                                                                &                               & Adafruit IO                      &                       & \textbackslash{}b(aio\textbackslash{}\_{[}a-zA-Z0-9]\{28\})\textbackslash{}b                                                                                                                                                                                                                                                                                                                                                                                                                                                                                                                                 & \cite{truffles71:online}                   \\
\rowfont{\color{gray}}
                                                                                &                               & OpenUV                           &                       & (?i)(?:openuv)(?:.\textbar{}{[}\textbackslash{}n\textbackslash{}r])\{0,40\}\textbackslash{}b([0-9a-z]\{32\})\textbackslash{}b                                                                                                                                                                                                                                                                                                                                                                                                                                                                                & \cite{truffles71:online}                   \\
\rowfont{\color{gray}}
                                                                                &                               & Tomtom                           & ~                     & (?i)(?:tomtom)(?:.\textbar{}{[}\textbackslash{}n\textbackslash{}r])\{0,40\}\textbackslash{}b([0-9Aa-zA-Z]\{32\})\textbackslash{}b                                                                                                                                                                                                                                                                                                                                                                                                                                                                            & \cite{truffles71:online}                   \\ 
\hline\hline
\multicolumn{2}{c}{\multirow{8}{*}{\begin{tabular}[c]{@{}c@{}}Accompanying\\Material\end{tabular}}}             & PEM Certificate                  & ~                     & (?i)-----\textbackslash{}s*?BEGIN CERTIFICATE\textbackslash{}s*?-----[a-zA-Z0-9\textbackslash{}/\textbackslash{}n\textbackslash{}r=+]*-----\textbackslash{}s*?END CERTIFICATE\textbackslash{}s*?-----                                                                                                                                                                                                                                                                                                                                                                                                        & \multirow{8}{*}{own}         \\
                                                                                &                               & PEM Certificate Request          &                       & \begin{tabular}[c]{@{}l@{}}(?i)-----\textbackslash{}s*?BEGIN CERTIFICATE REQUEST\textbackslash{}s*?-----[a-zA-Z0-9\textbackslash{}/\textbackslash{}n\textbackslash{}r=+]*\\-----\textbackslash{}s*?END CERTIFICATE REQUEST\textbackslash{}s*?-----\end{tabular}                                                                                                                                                                                                                                                                                                                                              &                              \\
                                                                                &                               & PEM Public Key                   &                       & \begin{tabular}[c]{@{}l@{}}(?i)-----\textbackslash{}s*?BEGIN[ A-Z0-9\_-]*?PUBLIC KEY\textbackslash{}s*?-----[a-zA-Z0-9\textbackslash{}/\textbackslash{}n\textbackslash{}r=+]*\\-----\textbackslash{}s*?END[ A-Z0-9\_-]*? PUBLIC KEY\textbackslash{}s*?-----\end{tabular}                                                                                                                                                                                                                                                                                                                                     &                              \\
                                                                                &                               & PEM Public Key Block             &                       & \begin{tabular}[c]{@{}l@{}}(?i)-----\textbackslash{}s*?BEGIN[ A-Z0-9\_-]*?PUBLIC KEY BLOCK\textbackslash{}s*?-----[a-zA-Z0-9\textbackslash{}/\textbackslash{}n\textbackslash{}r=+]*\\-----\textbackslash{}s*?END[ A-Z0-9\_-]*? PUBLIC KEY BLOCK\textbackslash{}s*?-----\end{tabular}                                                                                                                                                                                                                                                                                                                         &                              \\
                                                                                &                               & SSH Host Key                     & ~                     & \textbackslash{}bssh-[0-9a-zA-Z]+ AAAA\textbackslash{}S+ \textbackslash{}S+\textbackslash{}b                                                                                                                                                                                                                                                                                                                                                                                                                                                                                                                 &                              \\ 

\hline\hline
\end{tabu}
\label{tab:regularexpressions}
\vspace{-1.75em}
\end{table*}
\vspace{20em}

\section{Filtering Based on Filepaths}
\label{sec:excludedpaths}

After matching our regular expressions on arbitrary file content available in Docker images, extensive filtering is required to exclude false positive matches, i.e., matches that do not contain any secret.
Our \textit{File} filter bases on file paths derived from matches our \textit{Kompromat} filter excluded, i.e., all parent directories under which we find more than \(\nicefrac{2}{3}\) test keys known by kompromat~\cite{kompromat} and all directories that include known test keys directly.
Additionally, it takes manually compiled file paths, e.g., where standard libraries reside~(\texttt{/var/lib/*}) or package managers store their downloads~(e.g., \texttt{*/.cache/pip/*}) and extensions of database files~(e.g., \texttt{db} and \texttt{dbf}) into account which we selected after manually revisit all matches as these produced a high number of false positives.

\newpage

Figure~\ref{fig:filtering-filepath} shows the seven most prevalent file paths that contain matches excluded by our \textit{File} filter.
Indeed, most of the exclusions are matches included in folders belonging to package managers and thus most likely test secrets.
The massive filtering of API secret matches in \texttt{/var/opt/mssql} is due to the high number of false positives of the Twitter regular expressions on database files.

\end{document}